\documentclass[a4paper,fleqn,usenatbib]{mnras}

\usepackage[T1]{fontenc}
\usepackage{ae,aecompl}

\usepackage{graphicx}	% Including figure files
\usepackage{amsmath}	% Advanced maths commands
\usepackage{amssymb}	% Extra maths symbols
\usepackage{longtable,lscape}
\usepackage{multirow}
\title[NGC 6067: a young populous open cluster]{NGC 6067: a young and massive open cluster with high metallicity\thanks{Based on observations collected with the MPG/ESO 2.2-meter
Telescope operated at the La Silla Observatory (Chile) jointly by the Max Planck Institute for Astronomy and the European Organisation for Astronomical Research in the 
Southern Hemisphere under ESO programme 087.D-0603(A). }}

\author[J. Alonso-Santiago et al.]{
J. Alonso-Santiago,$^{1}$\thanks{E-mail: javier.alonso@ua.es}
I. Negueruela,$^{1}$
A. Marco,$^{1,2}$
H. M. Tabernero,$^{1,3}$
\newauthor C. Gonz\'alez-Fern\'andez,$^{4}$ and N. Castro$^{5}$
\\
$^{1}$Dpto de F\'{i}sica, Ingenier\'{i}a de Sistemas y Teor\'{i}a de la Se\~{n}al, Escuela Polit\'ecnica Superior, Universidad de Alicante, Apdo.99 E-03080, Spain\\
$^{2}$Department of Astronomy, University of Florida, 211 Bryant Space Science Center, Gainesville, FL 32611, USA\\
$^{3}$Dpto de Astrof\'{i}sica, Facultad de CC. F\'{i}sicas, Universidad Complutense de Madrid, E-28040, Madrid, Spain\\
$^{4}$Institute of Astronomy, University of Cambridge, Madingley Road, Cambridge CB3 OHA, UK\\
$^{5}$Department of Astronomy, University of Michigan, 1085 S. University Avenue, Ann Arbor, MI 48109-1107, USA
}

\date{Accepted XXX. Received YYY; in original form ZZZ}

\pubyear{2016}

\begin{document}
\label{firstpage}
\pagerange{\pageref{firstpage}--\pageref{lastpage}}
\maketitle

\begin{abstract}
NGC 6067 is a young open cluster hosting the largest population of evolved stars among known Milky Way clusters in the 50\,--\,150 Ma age range. It thus represents the best laboratory
in our Galaxy to constrain the evolutionary tracks of 5\,--\,$7\:$M$_{\sun}$ stars. 

We have used high-resolution spectra of a large sample of bright cluster members (45), combined with archival photometry, to obtain accurate parameters for the cluster as well as
stellar atmospheric parameters. We derive a distance of 1.78\,$\pm$\,0.12~kpc, an age of $90\pm20\:$Ma and a tidal radius of 14.8$^{+6.8}_{-3.2}$ arcmin. We estimate an initial 
mass above $5\,700\:$M$_{\odot}$, for a present-day evolved population of two Cepheids, two A supergiants and 12 red giants with masses $\approx6\:$M$_{\odot}$.

We also determine chemical abundances of Li, O, Na, Mg, Si, Ca, Ti, Ni, Rb, Y, and Ba for the red clump stars. We find a supersolar metallicity, [Fe/H]=$+0.19\pm0.05$, and a homogeneous
chemical composition, consistent with the Galactic metallicity gradient. The presence of a Li-rich red giant, star 276 with A(Li)=2.41, is also detected. An over-abundance of Ba is found, supporting the 
enhanced $s$-process.
 
The ratio of yellow to red giants is much smaller than one, in agreement with models with moderate overshooting, but the properties of the cluster Cepheids do not seem consistent with current Padova models for 
supersolar metallicity.

\end{abstract}

\begin{keywords}
open clusters and associations: individual: NGC 6067 -- Hertzsprung-Russell and colour-magnitude diagrams -- stars: abundances -- stars: fundamental parameters -- 
stars: late-type -- stars: variables: Cepheids 
\end{keywords}

%%%%%%%%%%%%%%%%%%%%%%%%%%%%%%%%%%%%%%%%%%%%%%%%%%

%%%%%%%%%%%%%%%%% BODY OF PAPER %%%%%%%%%%%%%%%%%%

\section{Introduction}\label{sec:intro}

Stellar clusters are excellent laboratories for the study of stellar evolution because all the stars in a cluster are formed from the same interstellar cloud, at roughly the same time
with a similar chemical composition. Their evolution, thus, will mainly depend on their initial mass. In very populous clusters, such as globular clusters, all the evolutionary stages 
are represented, and stellar evolution can be directly inferred. On the contrary, in young open clusters, generally rather less populous, only glimpses of the evolution of high- 
and intermediate-mass stars are provided. As a consequence, the physics of the most massive intermediate-mass stars %(5\,M$_{\odot}$\,<\,M$_*$\,<\,9\,M$_{\odot}$ ) 
is poorly constrained.

%Type II-P supernovae (SNe), the most common SNe in the Local Universe, come from low-luminosity red supergiants \citep{Smar09_rev}. Observationally, the lower limit for the mass of core-collapse
%SN progenitors is set at $M_{*}=8.5^{+1.0}_{-1.5}\:$M$_{\odot}$ \citep{Smar09,Smar15}, in mild to moderate conflict with stellar models \citep{Ibel13,Jo13}, which predict a lower limit 
%directly proportional to metallicity, always above $8.3\:$M$_{\sun}$ and in excess of $9\:$M$_{\sun}$ at solar metallicity}. The possibility of explosions from stars with masses as low as $7\:$M$_{\sun}$ has lead to renewed interest 
%in the post-MS evolution of stars in this mass range, with emphasis on aspects such as the super-AGB} phase \citep{Po08,Do15} or the detectability of massive Oxygen-Neon white dwarfs \citep[WDs;][]{He05,Sa09,Cu16}. 

NGC~6067 is a young open cluster that occupies a central position in the Norma Cloud, a rich region projected towards the central part of the Galactic disc [$\alpha$(2000)\,=\,16h\:13m\:11s, 
 $\delta$(2000)\,=\,$-54^{\circ}\:13\arcmin\:06\arcsec$; $\ell=329\fdg75$, $b=-2\fdg21$]. It is well known for hosting two classical Cepheids \citep{Eg83,An13}: V340~Nor and
 QZ~Nor. \citet{Th62}, hereinafter Th62, performed the first complete study of this cluster, combining $UBV$ photometry, spectral classification and radial velocity. 
\defcitealias{Th62}{Th62}
 
The age of NGC 6067 is not well known yet and hence, one of the main goals of this work is the determination of a reliable cluster age by using independent methods. Different
studies place it in the 50\,--\,150 Ma range. On the one side, the confirmed membership of V340~Nor (a Cepheid with a period 
of 11.3~d) and the integrated spectrum of the cluster suggest an age around $50\:$Ma \citep{Sa93}. On the other side, according to the spectral classification published by 
\citetalias{Th62}, the most complete to date, NGC 6067 is similar to the Pleiades in age, which at that time  was assumed to be around 150 Ma. Recently, \citet{Ma13} with new photometry obtained a younger age (80 Ma) that 
fits the brightest cluster members (i.e.\ B-type and red giants).  

Already in 1962, \citetalias{Th62} suggested that NGC 6067 looked like a ``young populous cluster''. \citet{Me08} confirmed via radial velocity measurements 
the membership of 14 luminous cool stars, most of which have been classified in the past, based on photographic spectra, as K\,Ib supergiants. The cluster is known to contain a large 
population of evolved stars: 14 B giants, at least two A/F supergiants, two Cepheids (F/G supergiants) and 12 late-G/early-K giants. Despite this richness, only one paper contains
stellar atmospheric parameters and chemical abundances for a few members \citep{Lu94}. In this paper we fill this gap by determining parameters and chemical abundances for several dozen
stars.  In particular, we want to characterise those stars located in the upper main sequence as well as the more evolved population, as a way to evaluate agreement between atmospheric models developed to 
study hot stars and those usually applied to cool stars. Moreover, the high number of evolved stars in NGC~6067 allows its use as a probe to test different theoretical stellar evolutionary models with a large, homogeneous population.

Given the moderately short distance to NGC 6067 (2.1~kpc according to \citetalias{Th62}) and the high number of members, a very accurate {\em Gaia} distance is expected in the near future. With its young age and large population of evolved members, 
NGC 6067 may become a reference point for the study of intermediate-mass stars. In this paper, we set out to derive accurate stellar and cluster parameters and explore this
potentiality. The paper is organised as follows: in the second section we describe the data used, both the photometric and the spectroscopic observations. In Sect.~\ref{sec:3} we
explain the methodology followed to perform a complete study of the main parameters of the cluster, as well as the atmospheric parameters and chemical abundances for the observed stars.
We also present in this section the results obtained, whereas in Sect.~\ref{sec:4} these are put in context and compared to those of previous authors. Finally, in Sect.~\ref{sec:5}
we summarize the main results and conclusions of our work.

\section{Observations and data}

\subsection{Spectroscopy}

\begin{figure*}  %usando el * ocupa las dos columnas, también para tablas / puedo usar para localizar la figura t(op), b(otton) o h(ere)
  \centering         
%el pdflatex no compila .eps, pero sí funciona si lo compilo a mano acabo pasando epstopdf
  \includegraphics[width=13cm]{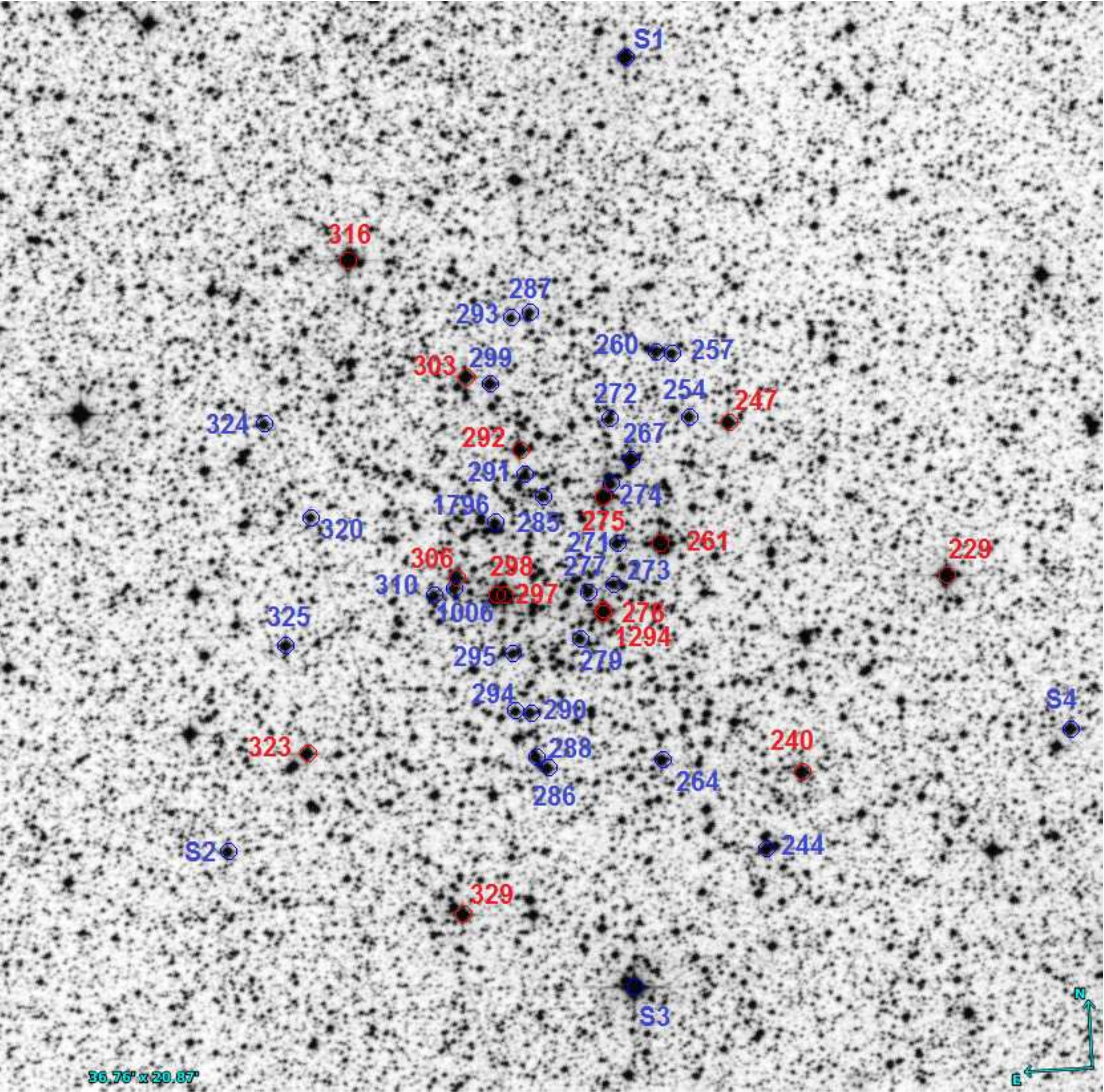}   %posición {h} me da problemas
  \caption{Finding chart for stars with spectroscopy in the field of NGC~6067. The chart is a $21\arcmin\times21\arcmin$ POSS2 Red image. Stars are marked, according to the text, as blue (blue stars) 
  or red circles (cool stars) depending on their spectral type. The number of each star corresponds to the numbering shown in the 
  WEBDA database for this cluster. For stars without the WEBDA numbering we used other designation: S1 (HD\,145304), S2 (CPD~$-53$ 7467), S3 (HD\;145324) and S4 (HD\;145139).} 
  \label{6067}  % ref para cuando la nombre, da igual lo que ponga, el programa las ordena fig1, fig2, ... pero es útil por si cambio la posición
\end{figure*}

We obtained high-resolution $\acute{e}chelle$ spectra using FEROS (Fiber Extended Range Optical Spectrograph) mounted on the ESO/MPG 2.2-metre telescope at the La Silla Observatory, 
in Chile. FEROS \citep{Kau99} covers the wavelength range from 3500 to 9200\,\AA{}, providing a resolving power of $R=48\,000$. This spectral region is covered in 39 orders, with 
small gaps between the orders appearing only at the longest wavelengths. The spectra were taken during four consecutive nights in 2011 May, 10\,--\,13 under programme 
087.D-0603(A). The exposure times ranged from 600 to 4800 seconds to achieve a typical signal-to-noise ratio of around $S/N\approx70$\,--\,80 for blue stars in the classification 
region (4000\,--\,5000\,\AA{}) and $S/N\approx90$\,--\,100 for the cool stars (6000\,--\,7000\,\AA{}). Later, in May 2015, under programme 095.A-9020(A), we observed the star HD\;145175
(star 229 in Table~\ref{SpT}) with the same instrument. To complete our sample, we took from the ESO Archive Raw Data spectra of QZ~Nor, observed originally in 2007 under programme 
060.A-9120(B). In total, we have spectra for 48 stars. Figure~\ref{6067} displays all the stars observed on a chart of the cluster. 

The spectroscopic reduction was performed using the {\scshape feros-drs} pipeline based on {\scshape midas} routines comprising the usual steps of: bad pixel and cosmic
correction, bias and dark current subtraction, removal of scattered light, optimum order extraction, flatfielding, wavelength calibration using Th-Ar lamps exposures, 
rectification and merging of the $\acute{e}chelle$ orders. 

%con !htbp parece que lo pone AQUI aunque no es perfecto es mejor que antes

\subsection{Photometry}\label{sec:2.2}
In order to complement the spectroscopic data we used the $BV$ photometry of \citet{An07}, downloading it from the WEBDA database\footnote{Available at \url{http://univie.ac.at/webda/}} \citep{Webda},
from which we adopted the numbering of the stars. \citet{An07} provide photometry for the highest number of stars, 1070, as they compiled values from \citetalias{Th62} and
\citet{Wa85fot} putting them on the same scale, with the latter as reference.

We completed our dataset with $JHK_{\textrm{S}}$ photometry from the 2MASS catalogue \citep{2MASS}. We selected stars with good-quality photometric flags (i.e.\ without any ``$U$''
flag). Stars inside a wide circle of radius 3$0\arcmin$ around the cluster centre were taken with the aim of determining the size of the cluster. All photometric data for stars with spectroscopy are displayed in Table~\ref{Ph}, in the appendix.

\section{Results}\label{sec:3}

\subsection{Cluster membership}

Throughout Sect.~\ref{sec:3} we carry out an analysis of our observations together with archival photometry. We mainly rely on radial velocities (Sect.~\ref{sec:rv}) 
to evaluate cluster membership. Then, according to the positions in the colour-magnitude diagrams (Sect.~\ref{sec:CMD}) and chemical abundances (Sect.~\ref{sec:abund}), we confirm 
or discard likely members among the spectroscopically observed stars (see Table~\ref{SpT}).

\subsection{Spectral classification}

We obtained high-resolution spectra for the brightest stars in the cluster. We selected likely blue members from \citetalias{Th62} and confirmed evolved members from \citet{Me08}.
To make sure that we were not leaving out member stars of comparable brightness, we applied the criteria described in \citet{Be55} over the 2MASS data to select more targets. We 
chose early-type stars using their $Q_{\textrm{IR}}$ index \citep{Ne07}. We identified possible red luminous members by combining their $Q_{\textrm{IR}}$ index with their positions in the
$K_{\textrm{S}}/(J-K_{\textrm{S}})$ diagram. In total, we observed 48 objects, most of which are B-type stars. For the analysis, we divided the sample in two groups: the blue stars
(i.e.\ stars with B and A spectral types) and the cool stars (the late-type stars, with G and K types). All the targets are listed in Table~\ref{SpT} together with their main
characteristics: name (NGC\,6067, when possible following the WEBDA numbering), equatorial coordinates referred to epoch J2000.0 (R.A., Dec.), spectral type, exposure time 
($t_{\textrm{exp}}$), signal-to-noise ratio (SNR), radial velocity ($V_{\textrm{rad}}$), cluster membership (Member, $y$es or $n$ot) and peculiarity notes for some stars (Notes).

\begin{table*}
\caption{Log of the observed stars. Spectral types and radial velocities are also included.\label{SpT}}
\begin{center}
\begin{tabular}{lcccccccc}   %longtable es para poner primero landscape
\hline\hline
\multirow{2}{*}{NGC 6067}  & R.A.     & Dec.     & \multirow{2}{*}{Spectral type}  & $t_{\textrm{exp}}$ & \multirow{2}{*}{SNR} & $V_{\textrm{rad}}$    & \multirow{2}{*}{Member} & \multirow{2}{*}{Notes}\\
                           & (J2000.0) & (J2000.0) &                                & (s)                &                      & (km\,s$^{-1}$)       &                         & \\ 
\hline
HD\;145139 & 243.019331 & -54.284729 &  B9.5\,III          & 1200 &  79 &     5.3 $\pm$ 0.9  & n & \\
229        & 243.082762 & -54.354095 &  K0\,III            & 1800 &  96 &     3.32 $\pm$ 0.02  & n & \\
240        & 243.165896 & -54.294746 &  K2\,II             &  600 &  92 & $-$41.91 $\pm$ 0.02  & y & \\
244        & 243.187149 & -54.318604 &  B7\,III-IV         & 2100 &  72 & $-$40.6 $\pm$ 2.9  & y & \\
247        & 243.197358 & -54.183491 &  G8\,Ib-II          &  600 &  92 & $-$38.82 $\pm$ 0.04  & y & \\
254        & 243.218388 & -54.181019 &  B7\,IV             & 1200 &  60 & $-$38.7 $\pm$ 7.9  & y & \\
257        & 243.226108 & -54.160816 &  B6\,IV             & 4800 &  57 & $-$35.7 $\pm$ 6.3  & y & \\
260        & 243.234543 & -54.160004 &  B6\,IV             & 2400 &  71 & $-$40.6 $\pm$ 7.2  & y & \\
261        & 243.237130 & -54.220406 &  K2\,Iab-Ib         &  600 & 108 & $-$39.42 $\pm$ 0.03  & y &  \\
264        & 243.241291 & -54.288956 &  B4\,V              & 2400 &  74 & $-$34.0 $\pm$ 4.0  & y & BSSc \\
HD\;145304 & 243.244057 & -54.066616 &  B2\,III            & 1200 & 120 & $-$41.1 $\pm$ 5.7  & y & BSSc \\
267        & 243.251027 & -54.193668 &  BN2.5\,III         &  900 & 145 & $-$43.3 $\pm$ 5.7  & y & BSSc \\
271        & 243.260353 & -54.219982 &  B8\,III + B8\,V    & 1200 &  64 & $-$18.0 $\pm$ 5.2  & y & SB2\\
272        & 243.261746 & -54.180622 &  B8\,IV shell       & 1200 &  54 & $-$42.5 $\pm$ 15.5 & y & \\
273        & 243.263526 & -54.232758 &  B7\,IV             & 1200 &  53 & $-$41.2 $\pm$ 8.4  & y & \\
274        & 243.262250 & -54.201130 &  B8\,III            & 4800 &  59 & $-$66.6 $\pm$ 7.6  & y & Binary?\\
275        & 243.266812 & -54.205208 &  K3\,Ib-II          &  600 & 105 & $-$39.87 $\pm$ 0.03  & y & \\
276        & 243.269432 & -54.240875 &  K4\,II             &  600 &  92 & $-$40.65 $\pm$ 0.02  & y & \\
1294       & 243.270122 & -54.241905 &  K2\,II             &  600 &  92 & $-$35.68 $\pm$ 0.02  & y & \\
277        & 243.277234 & -54.234901 &  B9\,III\,Si        & 1200 &  47 & $-$40.3 $\pm$ 1.4  & y & \\
279        & 243.282726 & -54.249516 &  B8\,III            & 4800 &  61 & $-$39.2 $\pm$ 1.4  & y & \\
285        & 243.299324 & -54.204128 &  B7\,IV             & 4800 &  52 & $-$37.8 $\pm$ 4.9  & y & \\
286        & 243.303496 & -54.290005 &  B7\,IIIe           & 1200 &  79 & $-$39.2 $\pm$ 9.1  & y & \\
287        & 243.302170 & -54.145798 &  B7\,IV             & 1200 &  55 & $-$41.7 $\pm$ 9.8  & y & Binary?\\
288        & 243.309679 & -54.286213 &  B7\,III-IV         & 2400 &  78 & $-$40.0 $\pm$ 2.6  & y & \\
290        & 243.311429 & -54.272305 &  B5\,shell          & 1200 &  52 & $-$39.8 $\pm$ 3.7  & y & \\
291        & 243.308498 & -54.196980 &  B9\,III\,Si        & 1500 &  23 & $-$36.7 $\pm$ 4.5  & y & \\
292        & 243.310704 & -54.189117 &  K0\,Ib-II          &  600 &  91 & $-$38.65 $\pm$ 0.02  & y & \\
293        & 243.312194 & -54.147339 &  B7\,V              & 3000 &  46 & $-$39.0 $\pm$ 13.5 & y & \\
294        & 243.319414 & -54.271442 &  B8\,IVe            & 1200 &  55 & $-$39.8 $\pm$ 12.8 & y & \\
295        & 243.319466 & -54.253189 &  B7\,IV             & 3600 &  48 & $-$39.8 $\pm$ 4.2  & y & \\
297 (V340 Nor)& 243.322565 & -54.234879 &  G2\,Iab         &  600 & 228 & $-$32.05 $\pm$ 0.13  & y & Cepheid\\
298        & 243.326993 & -54.234463 &  A5\,II             & 1100 & 173 & $-$35.9 $\pm$ 12.5 & y & \\
299        & 243.325234 & -54.167809 &  B7\,III + B8?      & 2400 &  78 & $-$28.7 $\pm$ 8.3  & y & SB2\\
303        & 243.337821 & -54.165279 &  G8\,II             & 1200 & 154 & $-$39.49 $\pm$ 0.05  & y & \\
1006       & 243.349542 & -54.231850 &  A3\,V              & 1200 &  59 & $-$31.7 $\pm$ 10.0 & n & \\
306        & 243.348307 & -54.228519 &  K2\,II             &  600 & 100 & $-$38.69 $\pm$ 0.02  & y & \\
310        & 243.360811 & -54.233719 &  B8\,III            & 1200 &  57 & $-$39.4 $\pm$ 1.5  & y & \\
316        & 243.398675 & -54.126804 &  K2\,Ib + B         & 2500 & 124 & $-$40.22 $\pm$ 0.03  & y & SB2\\
320        & 243.425241 & -54.207333 &  B6\,V              & 1500 &  44 & $-$35.5 $\pm$ 7.7  & y & \\
323        & 243.433354 & -54.281738 &  G8\,II             &  600 & 170 & $-$38.91 $\pm$ 0.03  & y & \\
324        & 243.448136 & -54.177147 &  B7\,IV             & 1500 &  57 & $-$45.1 $\pm$ 7.6  & y & \\
325        & 243.442158 & -54.247295 &  B8\,IIIe           & 2400 &  74 & $-$41.5 $\pm$ 5.9  & y & \\
CPD-53 7467& 243.478922 & -54.311527 &  B8\,IIIp           & 2400 &  74 & $-$38.0 $\pm$ 2.0  & y & \\
QZ Nor     & 243.835322 & -54.354095 &  G1\,Iab            & 1200 &  82 & $-$33.16 $\pm$ 0.05  & y & Cepheid\\
HD\;145324 & 243.261693 & -54.360264 &  A5\,Ib-II          &  300 & 199 & $-$43.3 $\pm$ 2.1  & y? & \\
329        & 243.352939 & -54.334885 &  K0\,Ib             &  600 &  90 & $-$39.09 $\pm$ 0.02  & y & \\
1796       & 243.325923 & -54.211670 &  B6\,Ve             & 1200 &  51 & $-$38.9 $\pm$ 13.5 & y & Binary?\\
\hline
\end{tabular}
\end{center}
%\begin{list}{}{}
% \end{list}
\end{table*}

\subsubsection{Blue stars}\label{sec:blue_stars}

\begin{figure}  %usando el * ocupa las dos columnas, también para tablas / puedo usar para localizar la figura t(op), b(otton) o h(ere)
  \centering         
%el pdflatex no compila .eps, pero sí funciona si lo compilo a mano acabo pasando epstopdf
  \includegraphics[width=\columnwidth]{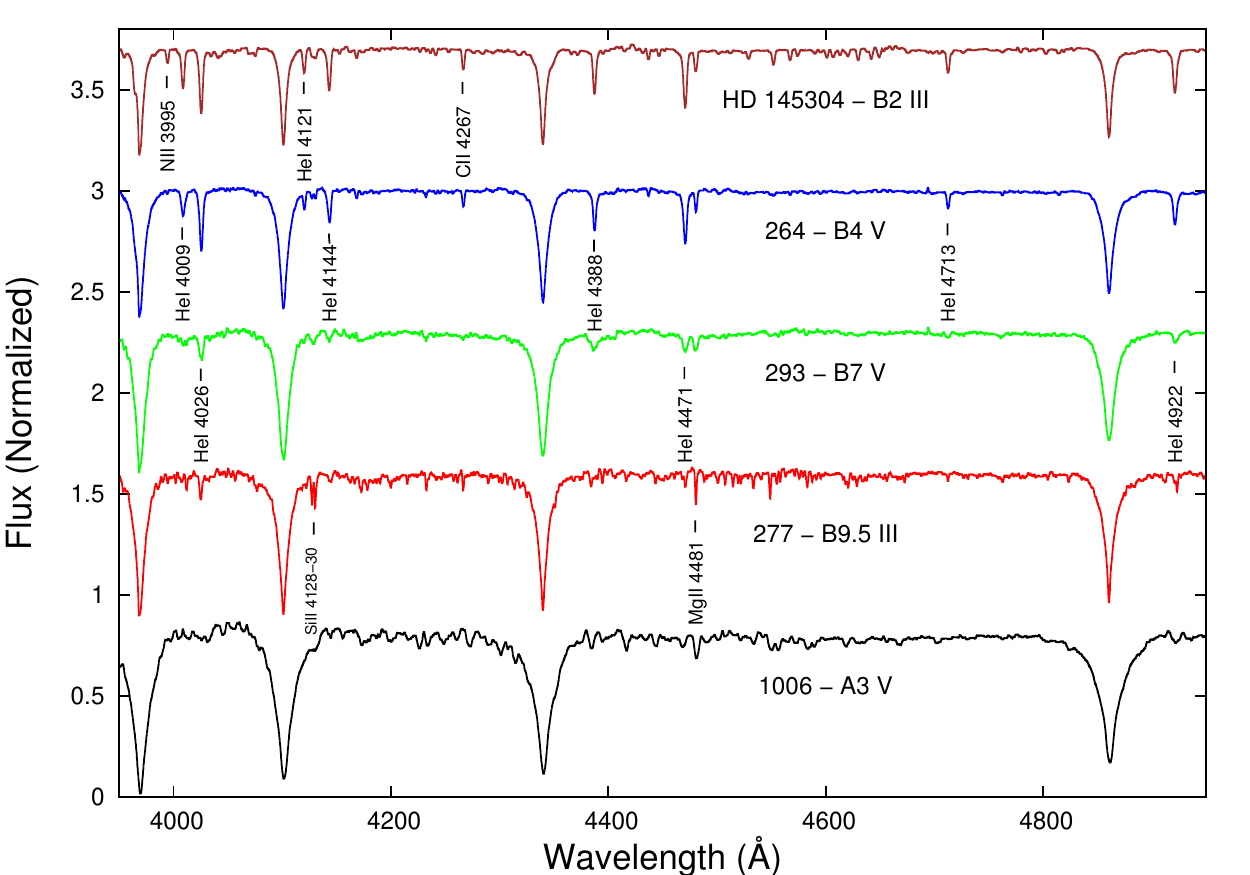}   %posición {h} me da problemas
  \caption{Typical spectra (degraded to $R=4\,000$ for a proper comparison with the spectral-type templates) of the blue stars in NGC~6067, covering almost the entire B spectral 
  class. The most important lines are marked (except the Balmer lines, the four deepest).\label{secuencia}} 
   % ref para cuando la nombre, da igual lo que ponga, el programa las ordena fig1, fig2, ... pero es útil por si cambio la posición
\end{figure}

We took spectra of the blue stars in the field of NGC~6067 to study the upper main sequence and the main sequence turnoff point (MSTO). We classified them by comparison with high-quality standards from the IACOB spectroscopic
database\footnote{\url{http//www.iac.es/proyecto/iacob/}} \citep{iacob1,iacob2,iacob3}. We followed classical criteria of classification in the optical wavelength range (4000\,--\,5000\,\AA{}) according to \citet{Ja87}. For most of our stars, with mid- and late-B types, the ratios \ion{Si}{ii}\,$\lambda$4128--30/\ion{He}{i}\,$\lambda$4144 and 
\ion{Mg}{ii}\,$\lambda$4481/\ion{He}{i}\,$\lambda$4471 are the main classification criteria, together with the profiles of the Balmer lines, which depend on effective gravity, and
so can be used to evaluate luminosity class. For the earliest stars, with types around B2\,--\,B3, we used instead the ratios \ion{Si}{ii}~$\lambda$4128--30/ \ion{Si}{iii}~$\lambda$4553,
\ion{Si}{ii}~$\lambda$4128--30/\ion{He}{i}~$\lambda$4121, \ion{N}{ii}~$\lambda$3995/\ion{He}{i}~$\lambda$4009, and \ion{He}{i}~$\lambda$4121/\ion{He}{i}~$\lambda$4144 \citep{Wa90}. We found 30 B-type and three A-type (one of which is not a member) stars. The objects observed cover almost the whole B spectral type, although most of them lie in the B7\,--\,B8 range (see Fig.~\ref{secuencia}). 

We find the MSTO close to spectral type B6. Three of the  stars studied (see Table~\ref{resumen_blue}) have earlier spectral types and lie brighter and bluewards
of the MSTO (Figs.~\ref{BV} and~\ref{JHK}). These objects are good candidates to blue-straggler stars (BSS).

\begin{figure}  %usando el * ocupa las dos columnas, también para tablas / puedo usar para localizar la figura t(op), b(otton) o h(ere)
  \centering         
%el pdflatex no compila .eps, pero sí funciona si lo compilo a mano acabo pasando epstopdf
  \includegraphics[width=\columnwidth]{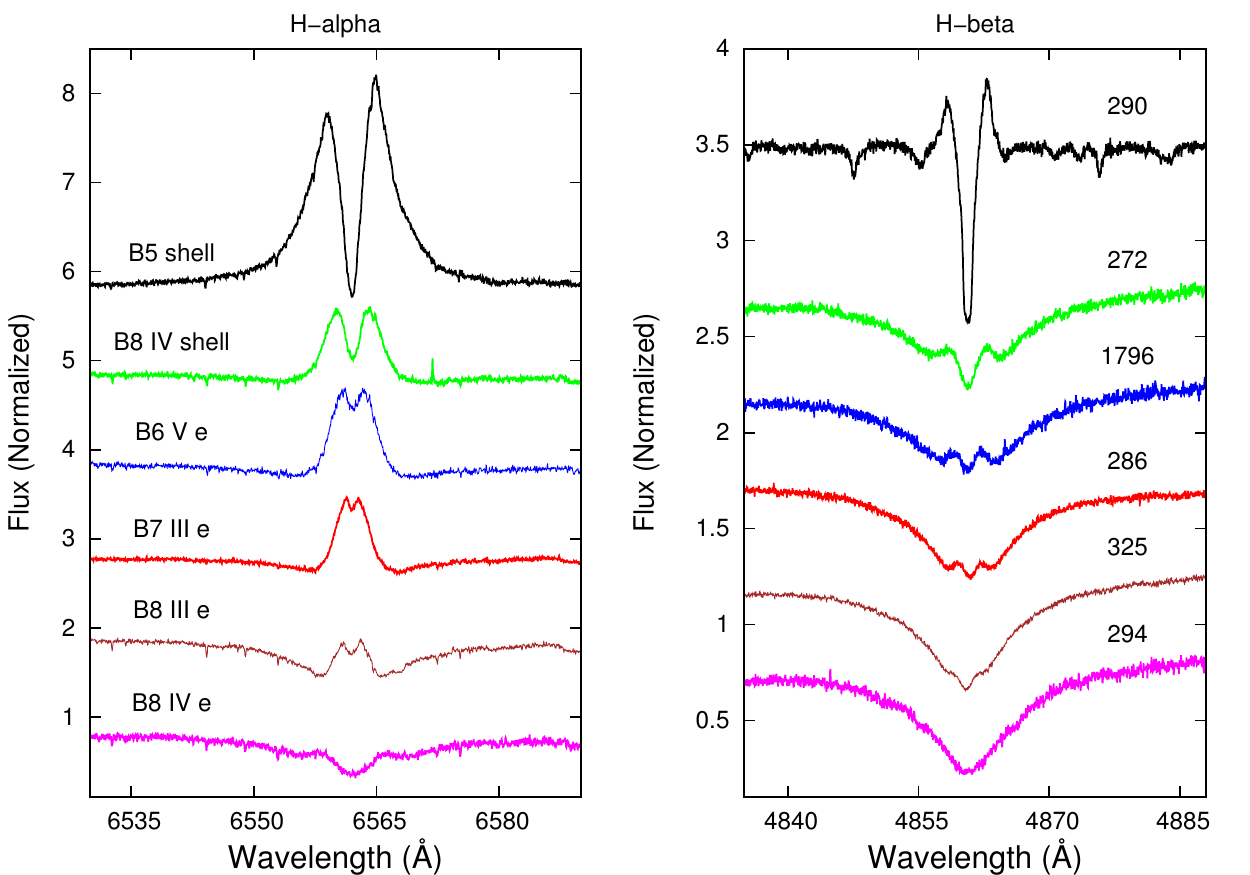}   %posición {h} me da problemas
  \caption{Spectra around H$\alpha$ (left) and H$\beta$ (right) of the six Be stars in NGC~6067. Each star is displayed with the same colour in both panels. Spectral types are shown in the left plot, while identifiers are given in the right panel. Different emission profiles are shown. Notice the strong emission of star 290 (black). On the contrary, the emission of star 294 (magenta) is barely discernible in this figure.
  \label{emision}}  % ref para cuando la nombre, da igual lo que ponga, el programa las ordena fig1, fig2, ... pero es útil por si cambio la posición
\end{figure}

Six of the B-type stars observed are emission-line stars (Be); two of them present a shell profile (see Table~\ref{resumen_blue}). This number represents a fraction of Be stars to total (B+Be) of about 21 per cent. 
Especially remarkable is star 290. It exhibits strong emission that will greatly hinder subsequent analyses, since the stellar atmosphere modelling may be severely affected by the emission characteristics due to circumstellar material. In Fig.~\ref{emision}, we show the H$\alpha$ and H$\beta$ line profiles for all the emission stars.

\begin{figure} %usando el * ocupa las dos columnas, también para tablas / puedo usar para localizar la figura t(op), b(otton) o h(ere)
  \centering         
%el pdflatex no compila .eps, pero sí funciona si lo compilo a mano acabo pasando epstopdf
  \includegraphics[width=\columnwidth]{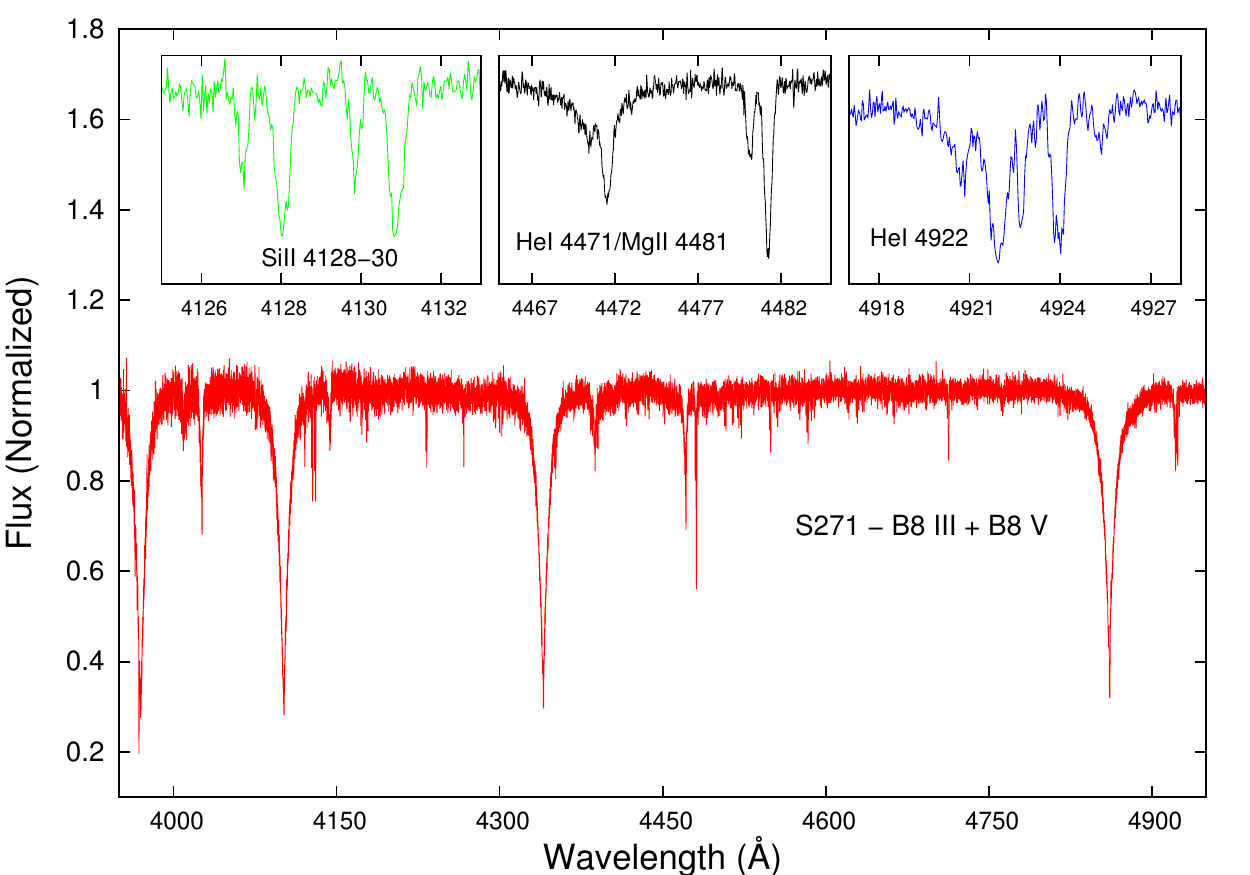}   %posición {h} me da problemas
  \caption{High-resolution spectrum of star 271, a new spectroscopic binary. Three regions are shown in detail at the top: \ion{Si}{ii}~$\lambda$4128\,--\,30 (left), \ion{He}{i}~$\lambda$4471 and \ion{Mg}{i}~$\lambda$4481 (middle) and \ion{He}{i}~$\lambda$4922 (right). As seen in the inset, and especially noticeable in the \ion{Si}{ii} region, lines appear double, and thus it is possible to observe both components.\label{doble}}  % ref para cuando la nombre, da igual lo que ponga, el programa las ordena fig1, fig2, ... pero es útil por si cambio la posición
\end{figure}

In addition, we found three spectroscopic binaries (see Fig.~\ref{doble} and Table~\ref{resumen_blue}), in which we can identify both companions (i.e.\ SB2). Two of them consist of two late B stars, while the third pair is formed by one blue and one red star. Given the asymmetric shape of their lines, stars 274, 287 and 1796 are likely binaries too, but we cannot confirm their nature with just one spectrum.
Finally, star 1006, with a spectral type A3\,V, is as bright as the B-type giants, and therefore cannot be a cluster member, but rather is a foreground star.

\begin{table}
\caption{Spectral types for remarkable blue stars in NGC~6067.}
\begin{center}
\begin{tabular}{lc}   %longtable es para poner primero landscape
\hline\hline
NGC 6067 & Sp T\\
\hline
\multicolumn{2}{c}{Blue-straggler candidates}\\    
\hline
HD\;145304 & B2\,III\\ 
264 & B4\,V\\
267 & BN2.5\,III\\
\hline
\multicolumn{2}{c}{Be stars}\\
\hline
272 & B8\,IV\,shell\\
286 & B7\,IIIe\\
290 & B5\,shell\\
294 & B8\,IVe\\
325 & B8\,IIIe\\
1796 & B6\,Ve\\
\hline
\multicolumn{2}{c}{Binaries}\\
\hline
271 & B8\,III + B8\,V\\
299 & B7\,III + B8?\\
316 & K2\,Ib + B\\
\hline
\end{tabular}
\label{resumen_blue}
\end{center}
\end{table}

\subsubsection{Red stars}
There are spectra for 15 red stars, most of which are late-G and early-K (super)giants. To classify them we focused on the near-infrared wavelenght range (8400\,--\,8900\,\AA{}), around 
the \ion{Ca}{ii} triplet. The triplet weakens towards later spectral types and lower luminosity classes \citep{Ja87}, but many other classification criteria are available in this range. 
In addition, the \ion{Mg}{i} triplet (5167, 5173, 5184\,\AA{}) is very useful for the stars in our sample
because of its sensitivity to luminosity \citep{Ja87,Ca15}. The two Cepheids were also observed, presenting spectral types G2\,Iab (V340~Nor) and G1\,Iab (QZ~Nor).

\subsection{Rotational and radial velocities}\label{sec:rv}

We estimated the projected rotational velocity ($v \sin\,i$) by using the {\scshape iacob-broad} code \citep{iacob_broad}, based on the Fourier transform method. This method allows the separation of rotational broadening from other broadening mechanisms, such as the macroturbulent velocity ($\zeta$). For the cool stars, we used six lines of \ion{Fe}{i} and \ion{Ni}{i}, 
while for B-type stars, which have very few strong isolated metallic lines, we could only rely on the \ion{Mg}{ii} line at 4481\,\AA{}. For each line we carried out three measurements. 
The errors listed reflect the scatter among measurements, in terms of rms. The results, projected rotational and macroturbulent velocities, are shown together with stellar parameters in Tables~\ref{Par_cal} and~\ref{Par} for blue and red stars, respectively.
 
Radial velocities, referred to the heliocentric reference frame of rest were obtained through Fourier cross-correlation. For blue stars we used the {\scshape fxcor} task within the {\scshape iraf} 
package\footnote{{\scshape iraf} is distributed by the National Optical Astronomy Observatories, which are operated by the Association of Universities for Research in Astronomy, Inc., under
cooperative agreement with the National Science Foundation.}. The spectrum of each star was correlated against a template spectrum and the correlation peak was fitted with a Gaussian function. 
%For the red stars we employed as a template the \emph{Kurucz Solar Atlas} \citep{sunatlas}, using the 5000\,--\,7000\,\AA{} range. For the B-type stars, we used instead a 
As a template we used a previously generated grid of theoretical stellar spectra, which were convolved with rotational and instrumental (Gaussian) profiles using the {\scshape fortran} routine {\scshape 
rotin3} in the {\scshape synspec} software \citep{rotin3}. We restricted our attention to the range between 4000 and 5000\,\AA{}. For the fast rotators the H$\beta$ and H$\gamma$ lines were masked in
order to improve the correlation. Given the low reddening, diffuse interstellar bands are not visible in the spectra, and so masking them was not necessary. There are no telluric lines in this
spectral range either.
For red stars, we employed instead the {\scshape iSpec} software \citep{ispec}, especially designed for the study of cool stars, computing the cross-correlation against a list of atomic
line masks from asteroids observed with the NARVAL spectrograph.

The radial velocity (RV) obtained for all the stars observed is displayed in Table~\ref{SpT}. In the case of cool stars we are reaching the instrumental limit ($\approx20-30\:$m\,s$^{-1}$) thanks
to the carefully selected mask list. These errors are computed following \citet{Zu03}. In contrast, errors for B-type stars are larger because they have high rotational velocities and the error in RV is proportional to the width of the cross-correlation peak, which depends on the (projected) rotational velocity. In fact, for fast rotators
($v\sin\,i>200\:$km\,s$^{-1}$), the errors obtained in the correlation exceed $10\:$km\,s$^{-1}$ (see Table~\ref{SpT}).

%We employed two different methods to calculate the average radial velocity of the cluster. On the one hand, we used the variance of each velocity as weight to derive the mean velocity.
%The result obtained is $\left< RV\right> =-38.4\pm1.5\:$km\,s$^{-1}$. On the other hand, by using a two-sigma-clipping statistics, we calculate a similar value, $\left< RV\right> 
%= -39.5\pm0.9\:$ km\,s$^{-1}$. We determine an average (heliocentric) radial velocity for the cluster, $\left< RV \right> =-39.5\pm0.9\:$km\,s$^{-1}$. This value is compatible with those 
%presented in other studies but more reliable, as we used a much larger sample. \citetalias{Th62}, by using only four stars, give a value of $\left< RV \right> =-39.8\pm0.8\:$km\,s$^{-1}$ whereas
%\citet{Me08} obtained $\left< RV \right> =-39.4\pm1.0\:$km\,s$^{-1}$ with very accurate measurements of ten stars. We do not find any dynamical structure within the cluster. There are no obvious 
%correlations between radial velocity and position, at least up to the distance covered by our observations, around $10\arcmin$ from the cluster centre.

In order to calculate the average (heliocentric) radial velocity of the cluster, we used a two-sigma-clipping statistics obtaining a $\left< RV \right> =-39.5\pm0.9\:$km\,s$^{-1}$ and a velocity 
dispersion, $\sigma_{\textrm{rad}}$\,=\,1.5\:km\,s$^{-1}$. This value is compatible with those presented in other studies but more reliable, as we used a much larger sample. \citetalias{Th62}, by using only four stars, give a value 
of $\left< RV \right> =-39.8\pm0.8\:$km\,s$^{-1}$ whereas \citet{Me08} obtained $\left< RV \right> =-39.4\pm1.0\:$km\,s$^{-1}$ with very accurate measurements of ten cool stars. We do not find any dynamical
structure within the cluster. There are no obvious correlations between radial velocity and position, at least up to the distance covered by our observations, around $10\arcmin$ from the cluster centre.

\begin{figure}  %usando el * ocupa las dos columnas, también para tablas / puedo usar para localizar la figura t(op), b(otton) o h(ere)
  \centering         
%el pdflatex no compila .eps, pero sí funciona si lo compilo a mano acabo pasando epstopdf
  \includegraphics[width=\columnwidth]{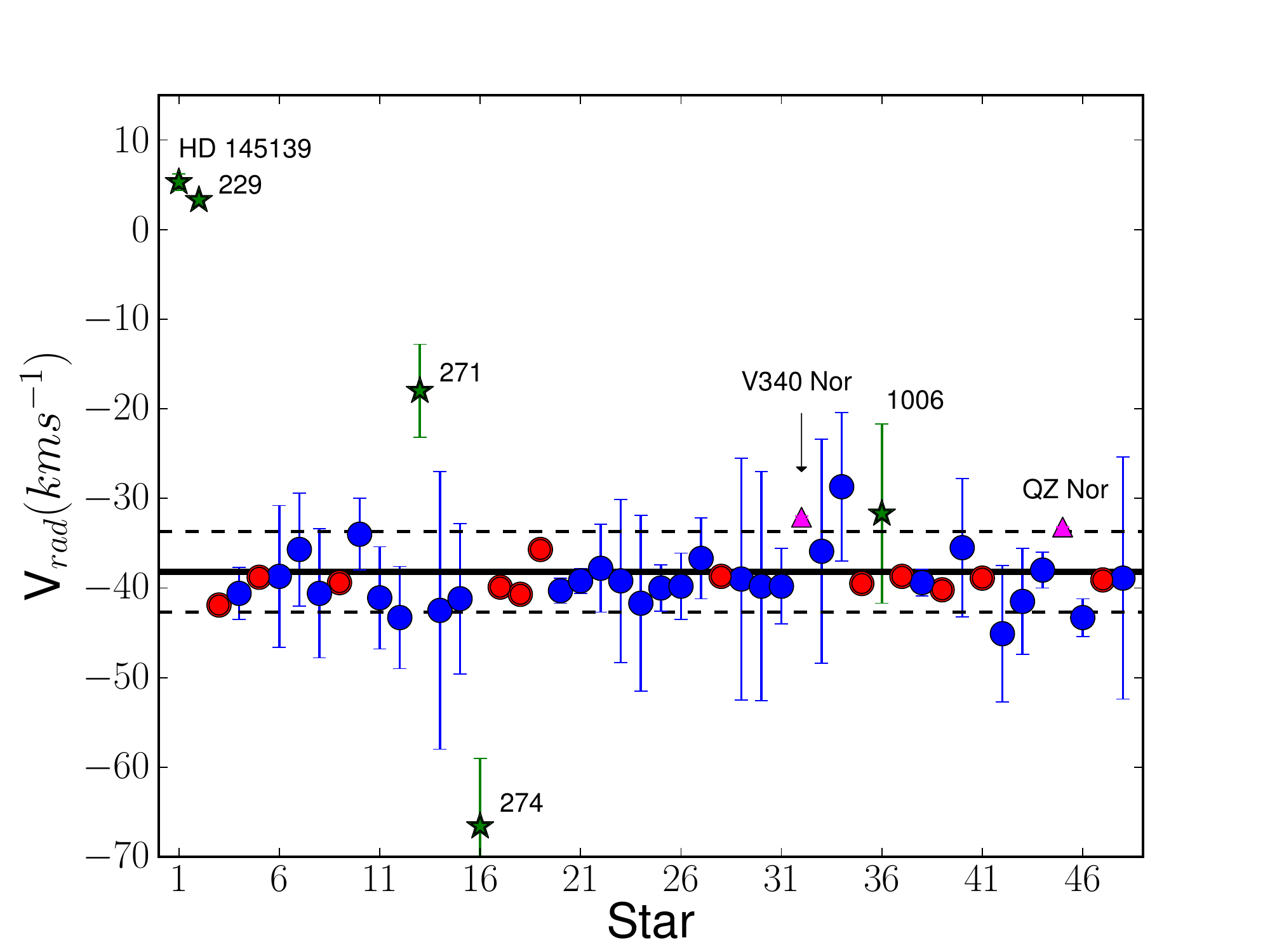}   %posición {h} me da problemas
  \caption{Radial velocity members. For all the stars contained in Table~\ref{SpT} the radial velocity is plotted. Blue circles represent blue stars, red circles are cool stars and green stars
  show the outliers. Cepheids, showing the velocity measured in this work and not their mean, are highlighted by magenta triangles. The black line is the $\left< RV \right>$ and the dashed ones delimit
  the $3\sigma$ confidence intervals. Stars with radial velocities compatible within $3\sigma$ with the cluster mean are considered cluster members, with the exception of star 1006, which is a foreground
  star.\label{vrad2}} %The particular cases of the two Cepheids and stars 271 and 274 are discussed as well.\label{vrad2}}
 \end{figure}

We have considered as members those stars whose radial velocities are within $3\sigma$ of the cluster mean (see Fig.~\ref{vrad2}). According to this, we can only discard the membership 
of stars HD\;145139 and 229 (in addition to star 1006, mentioned in Sect.~\ref{sec:blue_stars}). Star 271 presents a discrepant velocity, below $-30\:$km\,s$^{-1}$, but this is an SB2. 
Another outlier is star 274, with a more negative radial velocity, $v_{\textrm{rad}}=-66.6\:$km\,s$^{-1}$. However, as mentioned, this star is also likely a binary. We have selected
all the remaining stars (45), as likely members, and used them to estimate the main cluster parameters.  

Cepheids, as is well known, are variable stars, showing variations in their spectral type and radial velocity. However, \citet{Ma13} provide, based on data from the literature, the whole velocity 
curve for both cluster Cepheids, deriving their mean radial velocities, $\left< RV \right> =-40.3\pm0.2\:$km\,s$^{-1}$ for QZ Nor $-$\,which further, according to \citet{An14}, presents a modulated
RV curve\,$-$ and $\left< RV \right> = -39.3\pm0.1\:\textrm{km}\,\textrm{s}^{-1}$ for V340 Nor. 
These velocities  match well the average value for the cluster, thus confirming the membership of both. Obviously, the velocities measured for the Cepheids in this study have not been used for
calculating the cluster average.

\subsection{HR diagram: Reddening and determination of distance and age}\label{sec:CMD}

\subsubsection{Reddening}

From the spectral types and the observed $(B-V)$ colours of B-type likely members without emission lines or companions, 13 stars in total, we obtain the colour excess for individual stars. 
The intrinsic colours (i.e.\ $(B-V)_0$), as a function of spectral type, were adopted from \citet{Fi70} 
%and the observed colours, from \citet{An07}
. In Table~\ref{reddening} we display the results obtained, which correspond to a mean colour excess for the cluster of $E(B-V)=0.35\pm0.04$ (the uncertainty is expressed in terms of the rms; see Table~\ref{red&dist} for a comparison with results by other authors). This mean value will be used for dereddening the colour-magnitude diagram (CMD) in Sect.~\ref{sec:isochrones}.
%This value will be used, in Sect.~\ref{sec:isochrones}, for dereddening the colour-magnitude diagram (CMD). As \citetalias{Th62} noted, the reddening is rather homogeneus across
%the cluster. We used the standard extincion law \citep{Ca89} to obtain a total visual absorption, $A_{V}=1.09\pm0.13$~mag.

\begin{table}
  \caption{Optical colour excesses for B-type cluster members.\label{reddening}}
\begin{center}
\begin{tabular}{ccccc}
\hline\hline
NGC 6067 & Sp T & $(B-V)$ & $(B-V)_0$ & $E(B-V)$\\
\hline
254 & B7\,IV     & 0.26 & $-$0.13 & 0.39\\
260 & B6\,IV     & 0.24 & $-$0.14 & 0.38\\
264 & B4 V       & 0.13 & $-$0.18 & 0.31\\
267 & BN2.5\,III & 0.18 & $-$0.22 & 0.40\\
273 & B7\,IV     & 0.22 & $-$0.13 & 0.35\\
274 & B8\,III    & 0.27 & $-$0.10 & 0.37\\
277 & B9\,III Si & 0.21 & $-$0.08 & 0.29\\
279 & B8\,III    & 0.19 & $-$0.10 & 0.29\\
285 & B7\,IV     & 0.26 & $-$0.13 & 0.39\\
288 & B7\,III-IV & 0.25 & $-$0.12 & 0.37\\
291 & B9\,III Si & 0.24 & $-$0.08 & 0.32\\
295 & B7\,IV     & 0.21 & $-$0.13 & 0.34\\
310 & B8\,III    & 0.30 & $-$0.10 & 0.40\\
\hline 
\end{tabular}
\end{center}
\end{table}

We performed analogous calculations with photometry from 2MASS and the calibration by \citet{St09}, specific for 2MASS photometry. The resulting mean colour excess, derived from 19 stars, both hot and cool, is $E(J-K_{\textrm{S}})=0.21\pm0.05$. For a standard reddening law, we should expect $E(J-K_{\textrm{S}})$ = 0.546 $E(B-V)\simeq0.19$, fully compatible with our value.

\subsubsection{Fitting isochrones}\label{sec:isochrones}

We employed the isochrone fitting method to determine simultaneously the age and distance of the cluster. We fit isochrones by eye to the dereddened CMD, that is, the observational CMD corrected for reddening and absorption. Since we find an iron abundance [Fe/H]\,=\,+0.19 (see later; Sect.~\ref{sec:abund}), we resorted to Padova isochrones \citep{Marigo}, which are available for super-solar metallicity. The isochrones are computed using a Kroupa initial mass function (IMF) corrected for unresolved binaries \citep{Kroupa}. We choose the value of $Z$ with the approximation [M/H]\,=\,$\log$\,($Z$/Z$_{\odot}$), with
Z$_{\odot}$\,=\,0.019 for \citet{Marigo} tracks. We identify as the best-fitting isochrone that which can describe best the MSTO and the position of the red giants, paying special attention to likely members determined via radial velocity. 

\begin{figure}  %usando el * ocupa las dos columnas, también para tablas / puedo usar para localizar la figura t(op), b(otton) o h(ere)
  \centering         
%el pdflatex no compila .eps, pero sí funciona si lo compilo a mano acabo pasando epstopdf
  \includegraphics[width=\columnwidth]{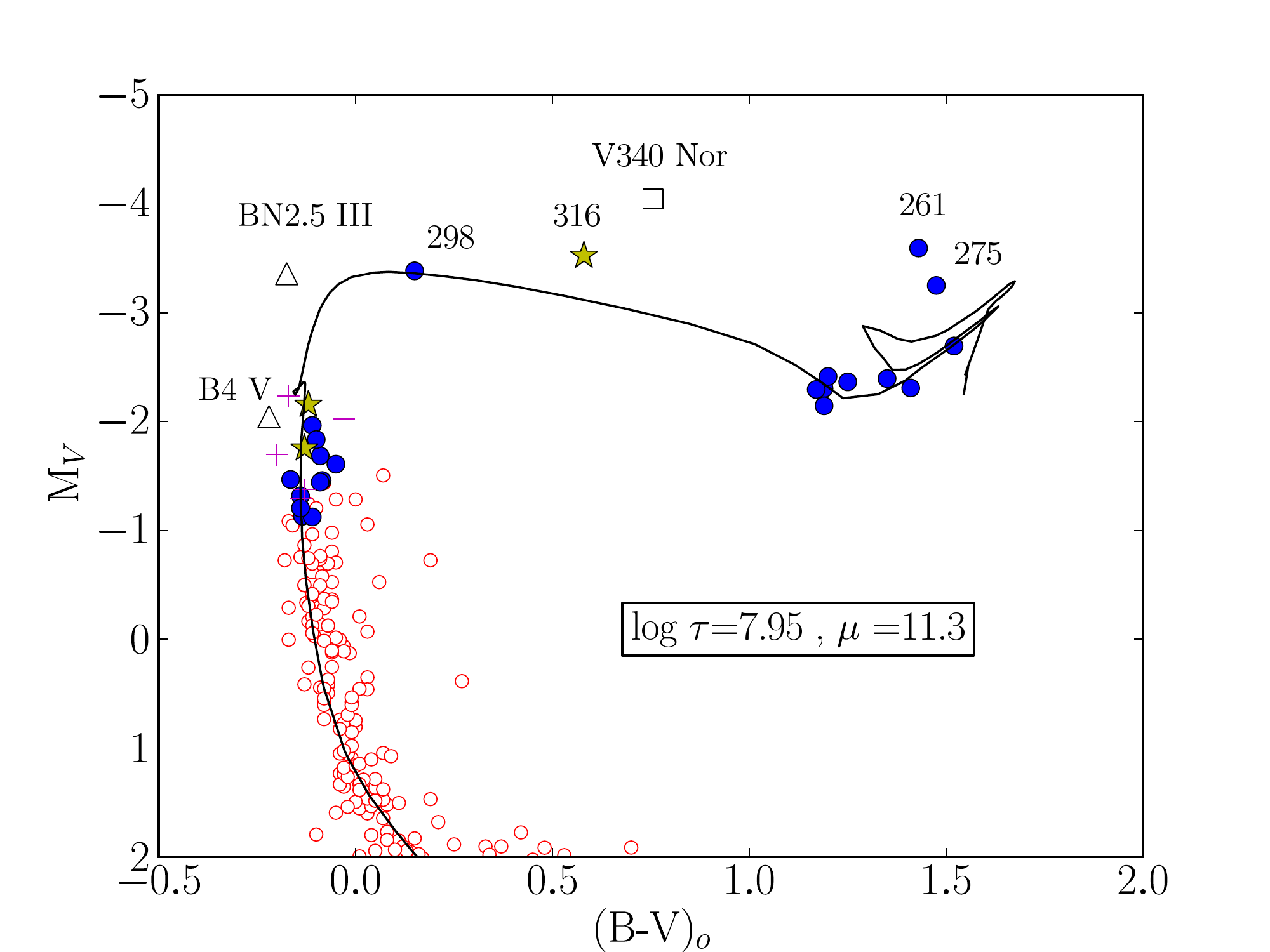}   %posición {h} me da problemas
  \caption{Dereddened $M_V$/$(B-V)_0$ diagram.  The open
  square represents the Cepheid V340~Nor (QZ~Nor lies outside the area covered by the photometry), pluses are Be stars, binaries appear as star-symbols (note especially star 316), BSSCs are represented by 
  triangles and all the other stars with spectra are filled circles. Open circles are stars with photometry from \citet{An07}, but no spectroscopic observations. The black line 
  is the best-fit Padova isochrone, corresponding to log $\tau$=7.95 and distance modulus=11.3.} 
  \label{BV}  % ref para cuando la nombre, da igual lo que ponga, el programa las ordena fig1, fig2, ... pero es útil por si cambio la posición
\end{figure}

In Fig.~\ref{BV} we show the $M_V$/$(B-V)_0$ diagram and the best-fit isochrone. 
%The photometry is taken from \citet{An07}. 
There are 33 of our likely members with photometric
values in \citet{An07}. Most of them lie very close to the isochrone, but there are a few exceptions: the Cepheid V340~Nor (see the latter discussion, Sect.~\ref{sec:Cepheids}), the
binary star 316 (because of the very different colours of its two components: K2\,Ib + B), and the two most luminous red stars -- star 275 (with spectral type K3\,Ib-II) and especially 
star 261 (K2\,Iab-Ib) that lies significantly above the isochrone. All these stars have RVs compatible with membership. From this fit we can derived a $\log\,\tau=7.95\pm0.10$ 
and a distance modulus $\mu=11.30\pm0.15$. These errors show the range of isochrones that give a good fit.

\begin{figure} %usando el * ocupa las dos columnas, también para tablas / puedo usar para localizar la figura t(op), b(otton) o h(ere)
  \centering      
%el pdflatex no compila .eps, pero sí funciona si lo compilo a mano acabo pasando epstopdf
  \includegraphics[width=\columnwidth]{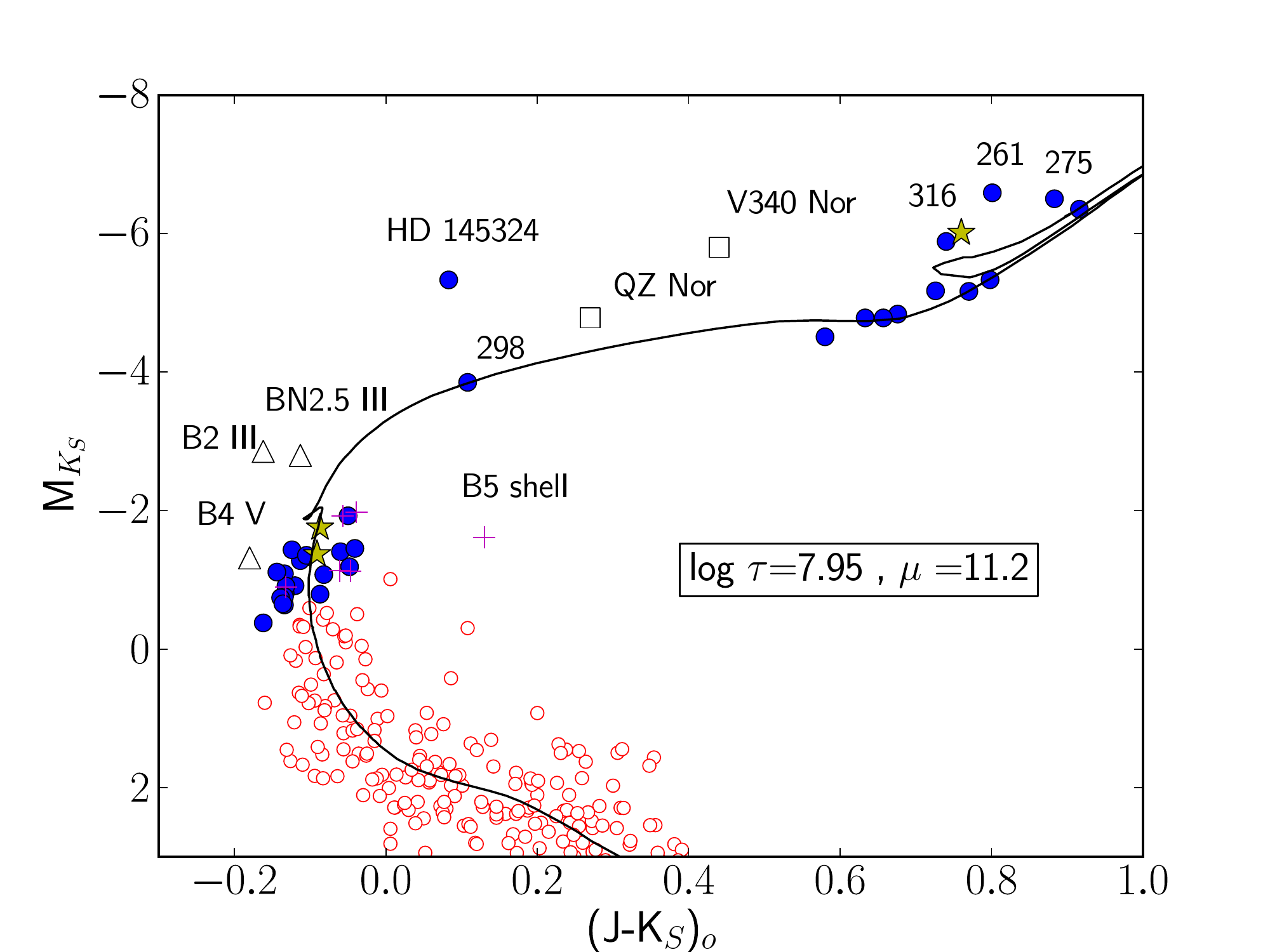}   %posición {h} me da problemas
  \caption{Same as Fig.~\ref{BV} but in the dereddened 2MASS CMD. Open circles are stars from 2MASS within 2.5$\arcmin$ of the cluster core.} 
  \label{JHK}  % ref para cuando la nombre, da igual lo que ponga, el programa las ordena fig1, fig2, ... pero es útil por si cambio la posición
\end{figure}

Figure~\ref{JHK} shows the $M_{K_{\textrm{S}}}$/$(J-K_{\textrm{S}})_0$ diagram with photometry from 2MASS. Optical data are restricted to a smaller field of view centred on the cluster core, whereas 2MASS is all-sky. We thus show all 45 likely members (solid circles in Fig.~\ref{JHK}) for which we have spectra, including the 12 red luminous members.
The position of these evolved stars in the red clump and the blue loop fits the isochrone quite well. The position of the MSTO is in good agreement as well, and only the 
Cepheids and the stars HD\;145324 (A5\,Ib-II, too bright for the isochrone), 290 (B5\,shell) and 261, again, fall away from the isochrone. The best fit corresponds to the values $\log \tau=7.95\pm0.10$ 
and $\mu =11.20\pm\,0.15$ (where errors are estimated as in the previous fit.)

Since the results of the optical and infrared photometry are compatible within the errors, we take the average, and thus derive an age of 
$\log \tau = 7.95\pm0.10$ (corresponding to $\tau=90\pm20\:$Ma) and a distance modulus $\mu=11.25\pm0.15$ (i.e.\ $d=1.78\pm0.12$~kpc). It is important 
to point out that in both CMDs the isochrone reproduces very well the position of the red stars (equivalent to the clump), but not the Cepheids. We will come back to this 
in~Sect.~\ref{sec:Cepheids}.

Distance estimates for NGC 6067 range between 940~pc \citep{Trumpler} and 2.1~kpc \citepalias{Th62}. \citet{Wa85fot} provide the deepest $BV$ colour-magnitude diagram for the cluster
(including photometry for 1019 stars), from which they determined a distance of 1.6~kpc. Based partly on 2MASS, \citet{Tu10} obtains a distance of 1.7~kpc, which agrees well with
that for the Cepheid V340~Nor established by \citet{Sto11} via the infrared surface brightness technique. More recently, \citet{Ma13} using new $BVJH$ photometry obtain a similar 
value ($d=1.8$~kpc), a result that matches Wesenheit distances computed for both cluster Cepheids (See Table~\ref{red&dist}).

Our value for the cluster age is compatible with all previous determinations. \citetalias{Th62}, comparing the CMD of 
NGC~6067 with those of other galactic clusters, suggested that its age was similar to that of the Pleiades. \citet{Me811} based on the shape of the CMD and the presence of BSSs, 
Cepheids and bright red giants with luminosity class II, also included NGC 6067 in the Pleiades age group. \citet{Me812} estimated for this group a turnoff point at spectral type
B6, as we have found for NGC~6067 in this work. \citet{Sa93} measuring the $EWs$ of the Balmer lines and, more recently, \citet{Ma13} using new $BVJH$ and Padova isochrones 
obtained a similar value ($\log\,\tau=7.89\pm0.15$), even if the age of the Pleiades is now believed to be $\approx120$~Ma old. With an age just short of 100~Ma, NGC~6067 is 
the oldest cluster containing K0\,--\,K2 Ib-II supergiants \citep[see the discussion in][]{Be55}. 

\begin{table}
\caption{Comparison of the reddening and distance for NGC 6067 derived in this work with respect to results found in the literature. \label{red&dist}}
\begin{center}
\begin{tabular}{lcc}   %longtable es para poner primero landscape
\hline\hline
Reference & $E(B-V)$ & d (kpc)\\
\hline
\citet{Th62}   & 0.33 $\pm$ 0.07 & 2.10 $\pm$ 0.30 \\
\citet{Wa85res}& 0.35 $\pm$ 0.01 & 1.62 $\pm$ 0.07 \\
\citet{An07}   & 0.34 $\pm$ 0.03 & 1.61 $\pm$ 0.06 \\
\citet{Ma13}   &     ---         & 1.75 $\pm$ 0.10 \\
This work      & 0.35 $\pm$ 0.04 & 1.78 $\pm$ 0.12 \\
\hline
\end{tabular}
\end{center}

\end{table}

\subsection{Centre and cluster radius}

To determine the cluster centre and radius, we made use of the 2MASS photometry, because of its uniformity and spatial coverage. This allows us to obtain reliable data on the 
projected distribution of stars for a large extension around the nominal centre of the cluster. Since the centre is the location of maximum stellar density, it was found 
by fitting two separate Gaussians to the profiles of star counts, one in right ascension and the other in declination \citep{Tad2}. The estimated centre lies at 
$\alpha$\,=\,243.2925$\pm$0.008$^{\circ}$ and \,\, $\delta$\,=\,$-54$.2424$\pm$0.008$^{\circ}$. Our result presents a small offset from the nominal value (this work\,$-$\,nominal) of
$\Delta\alpha=-0.2\pm0.5\arcmin$ and $\Delta\delta=-1.4\pm0.5\arcmin$.

\begin{figure}  %usando el * ocupa las dos columnas, también para tablas / puedo usar para localizar la figura t(op), b(otton) o h(ere)
  \centering         
%el pdflatex no compila .eps, pero sí funciona si lo compilo a mano acabo pasando epstopdf
  \includegraphics[width=\columnwidth,scale=0.8]{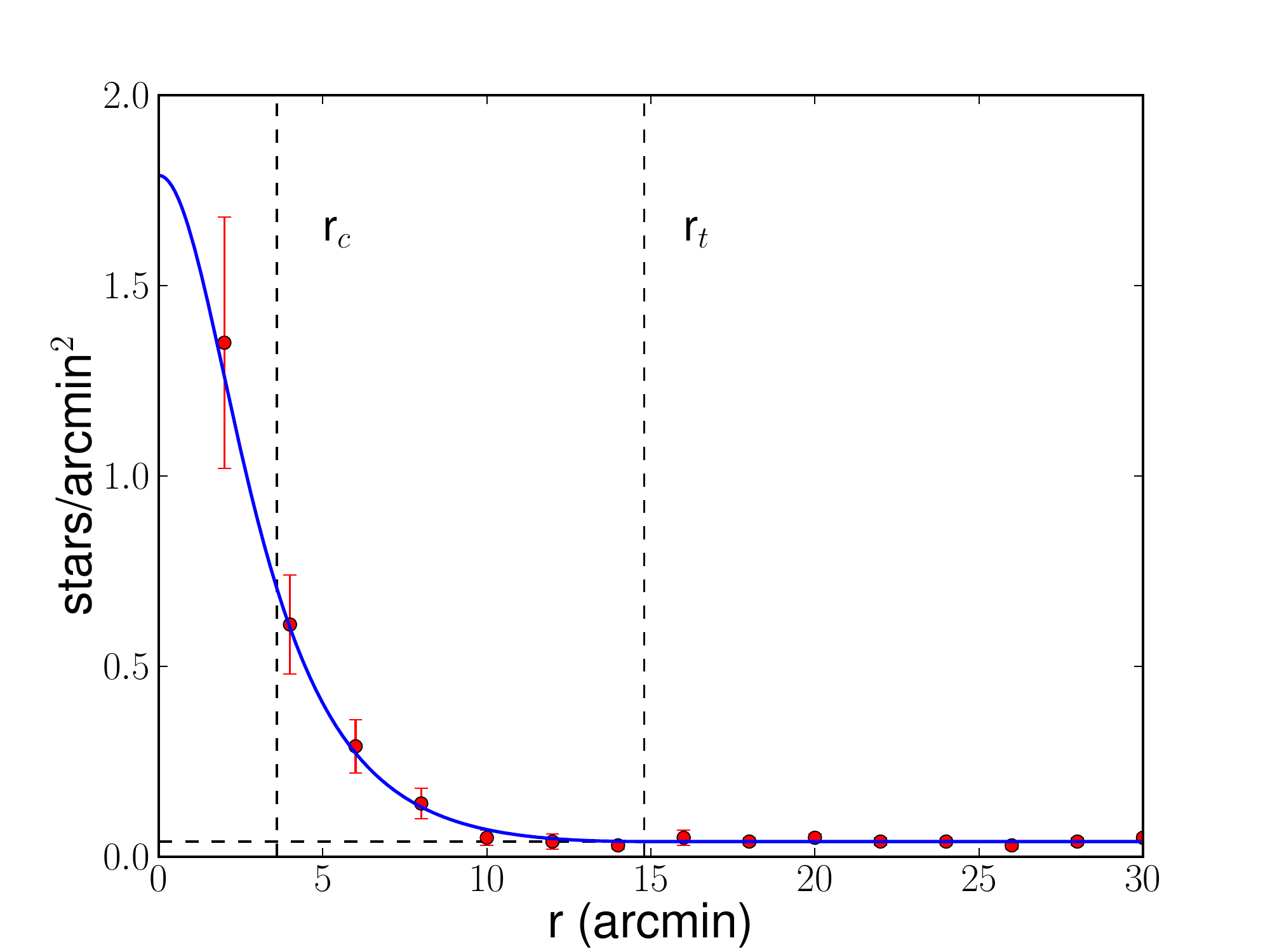}   %posición {h} me da problemas
  \caption{Radius determination of NGC~6067 using the projected density distribution of late-B stars. The red circles are the observed values with their Poisson errors whereas 
the solid blue line represents the fitted King profile from which we obtained a tidal radius of 14.8 arcmin. The dashed lines place the positions of the core (r$_{\textrm{c}}$) and
tidal (r$_{\textrm{t}}$) radii respectively. The background density (0.04 stars/arcmin$^2$) is represented by the black horizontal line.} 
  \label{king}  % ref para cuando la nombre, da igual lo que ponga, el programa las ordena fig1, fig2, ... pero es útil por si cambio la posición
\end{figure}

Traditional values for the cluster diameter vary from $15\arcmin$ to $32\arcmin$ \citepalias{Th62}. \citet{Ma13}, confirmed the membership in the cluster of QZ Nor at a position 
some $20\arcmin$ from the centre, suggesting the existence of a larger halo. We evaluated the stellar density profile, $\rho(r)$, from direct star counts in concentric annuli 
around the cluster centre up to a reasonable distance of 30 arcmin (see Fig.~\ref{king}). The cluster 
is situated in a very rich region, and distinguishing it properly from the background is difficult. When simply counting stars, the density is almost constant at 26 
stars/arcmin$^2$. To solve this difficulty, we used bright B-type stars as tracers of the cluster extent. The selection was made by choosing the stars that complied with these 
conditions: 9.5 < $K_{\textrm{S}}$ < 11.5, $-0.1 < (J-K_{\textrm{S}})_0 < 0.2$ and $Q_{\textrm{IR}}<0.125$ \citep{Ne07}. To this end, we used the average colour excess and the 
distance modulus calculated in the previous subsection. Now the cluster is clearly enhanced and we can fit the density profile to a three-parameters King-model \citep{king}:
\begin{equation}
 $ $\rho(r) = \rho_0 \left\{\displaystyle\frac{1}{\sqrt{1+(r/r_{\textrm{c}})^2}} - \displaystyle\frac{1}{\sqrt{1+(r_{\textrm{t}}/r_{\textrm{c}})^2}} \right\}^2$ $
\end{equation} 
where $\rho_0$ is a constant, $r_{\textrm{c}}$ is the cluster core radius and $r_{\textrm{t}}$ is the tidal radius. The core radius is defined as the radial distance at which the
value of density becomes half of the central density. The tidal radius is the distance at which the cluster disappears in its environment. We performed a Bayesian curve-fitting,
obtaining these angular values: r$_{\textrm{c}}$\,=\,3.6$^{+2.1}_{-1.1}$ \,\,arcmin and r$_{\textrm{t}}$\,=\,14.8$^{+6.8}_{-3.2}$ arcmin. The uncertainties correspond to the 5 and 
95 percentiles of the posterior distribution of the parameters provided. These values correspond to physical sizes $r_{\textrm{c}}$\,=\,1.9$^{+1.2}_{-0.7}$~pc and 
$r_{\textrm{t}}$\,=\,7.7$^{+4.0}_{-2.2}$~pc respectively. This radius is slightly larger, but compatible within errors, than the $12.3\pm1.3$~arcmin 
calculated by \citet{Pi08}. All the stars observed in the present work are inside this radius. The Cepheid QZ~Nor is located $\approx20$~arcmin from the cluster centre, and so it lies within
the tidal radius upper limit, possibly in the halo, as suggested by previous works \citep{An07,Ma13,An13}. 

\subsection{Luminosity function and total mass}\label{sec:lum_mass}

\begin{figure}
  \centering         

  \includegraphics[width=\columnwidth]{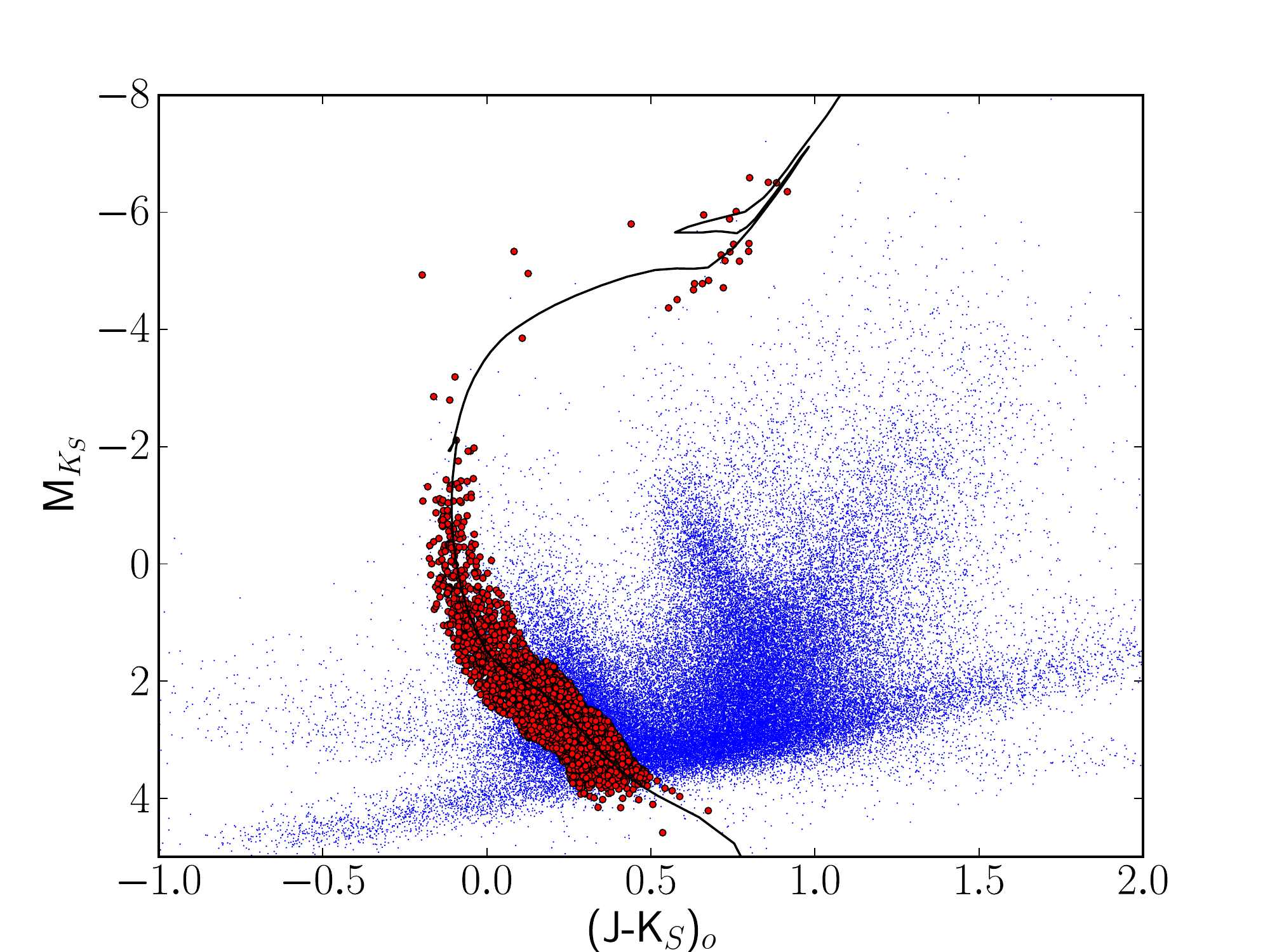}  
  \caption{Dereddened 2MASS CMD showing the best-fitting Padova isochrone. All the 2MASS sources inside the cluster tidal radius are represented as blue dots. We have selected as
  likely members (filled red circles) those stars whose position on the CMD is closer than $(J-K_{\textrm{S}})_0$=\,$\pm0.1$ to the isochrone.}
  \label{seleccion}  
\end{figure}

Once we fixed the cluster tidal radius, we chose all the 2MASS sources inside this radius. On a 2MASS CMD, see Fig.~\ref{seleccion}, we selected as likely cluster members those 
stars close enough to the expected location on the CMD. On average, we chose a value of $(J-K_{\textrm{S}})_0$=\,$\pm0.1$ from the Marigo isochrone used in 
Sect.~\ref{sec:isochrones}. After this, we counted the stars selected as a function of the absolute magnitude $J$ using magnitude bin intervals with a size of $\Delta J$ = 0.5~mag.

\begin{figure} 
  \centering         

  \includegraphics[width=\columnwidth]{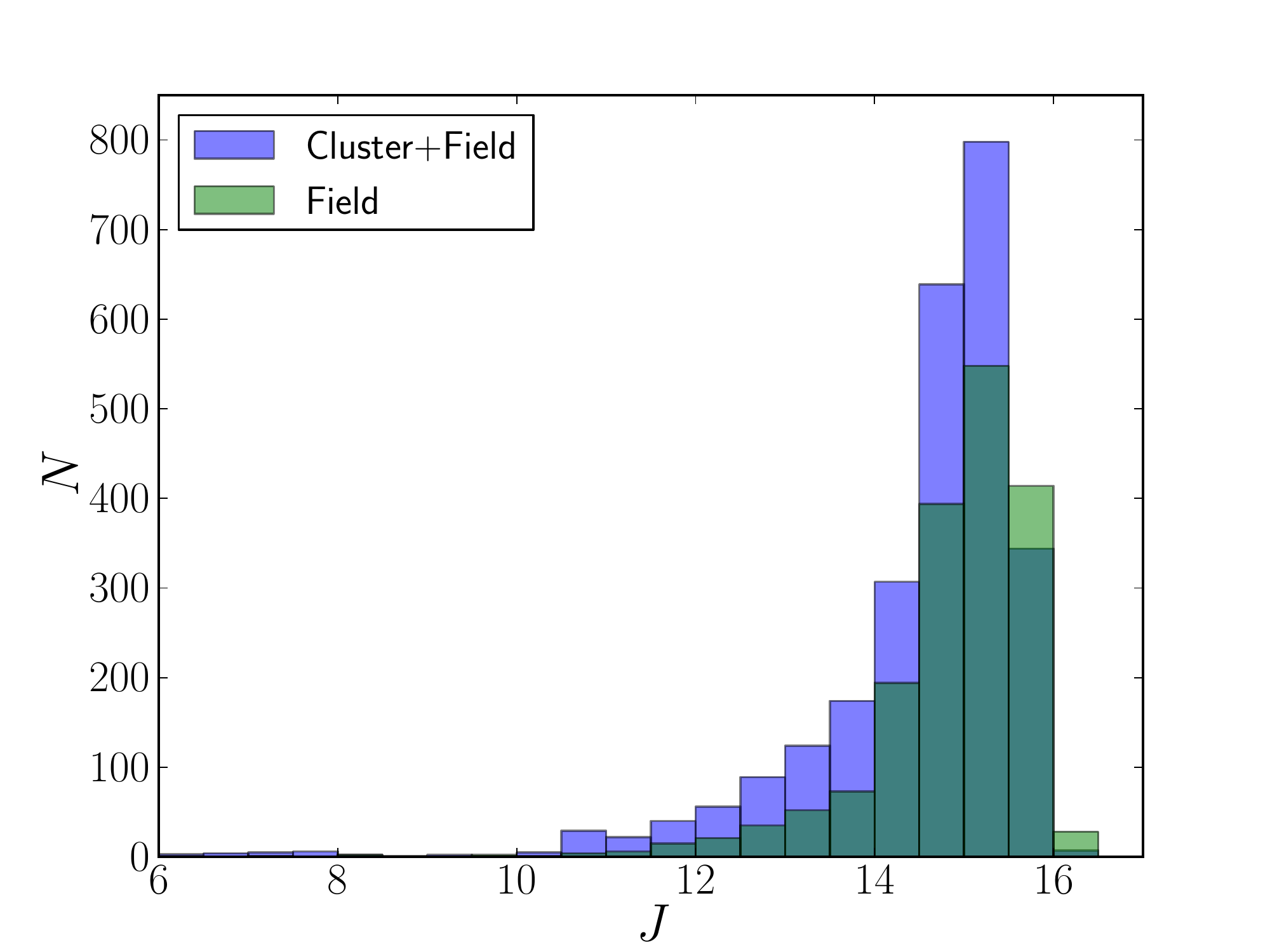}  
  \caption{$J$-band luminosity function, i.e. number of stars ($N$) in each $J$ magnitude bin, for the field of NGC~6067 inside the tidal radius of the cluster (blue) compared to that of a surrounding annulus sampling the same area used to estimate the field contamination (green). The maximum lies at $J=14.8\:$mag. At fainter magnitudes, the number of field stars dominates the total count. The 
    completeness limit is reached at $J=14.7$ according to the criterium explained in the text.}
  \label{JPL}  
\end{figure}

Before analysing this luminosity function we needed to correct for the field contamination.  
%to ensure that our study is performed on the cluster stars. In order to estimate the field
%contamination in the cluster region
For this, we studied the population within an outer annulus with the same area, sampling the surrounding field. The inner limit of this annulus was set at $25\arcmin$ from the centre of the cluster, sufficiently far from the halo (our upper limit on the tidal radius is $21\farcm6$) to avoid including 
potential cluster members. In Fig.~\ref{JPL} we show the luminosity function in the $J$ band for both the cluster region and the external annulus. The maximum lies at $J$\,=\,14.8~mag;
at fainter magnitudes, the contamination is more important than the cluster itself. We applied field decontamination and then, to prevent observational biases in the data, we 
corrected for completeness. Following \citet{Maria}, we estimated a limiting magnitude at $J=$14.7~mag with a completeness limit around 90 per cent. The
limiting magnitude was computed as the mean of the magnitudes at the peak star count bin and its two adjacent bins, on both sides, weighted by the number of stars in 
each bin.

Given the luminosity function we can estimate the total mass of the cluster by multiplying the number of stars in each bin by the mean mass of the bin. To do this, a stellar 
mass-luminosity relation is needed. We took it from the Padova isochrone used in the CMD (see Fig.~\ref{JHK}). With this conversion, we found a value of 
$1\,900^{+1100}_{-600}\:$M$_{\odot}$. We compute the errors by calculating the cluster mass within the upper and lower radius limits. We have also considered the Poissonian errors when counting the number of stars in each bin. This 
estimation does not take into account the effects of unresolved binarity. Following \citet{Tadross} in his study of NGC~1193, we assumed a 
binary frecuency of 50 per cent \citep{Ja70} and an average mass ratio of binary systems of 1.3 \citep{Allen}. With these assumptions, the correction adds $\approx730\:$M$_{\odot}$, giving a total (present-day) cluster mass of $\approx2\,600^{+1\,500}_{-800}\:$M$_{\odot}$ down to the completeness limit (i.e. $M_{J}\approx+3.3$). This 
limit roughly corresponds to a G0\,V spectral type, and so, when integrating the IMF, it accounts for $\approx$52\% of the total cluster mass (for a standard Kroupa IMF). We should thus expect a total mass 
for NGC\,6067\,$\approx5000\:$M$_{\odot}$). 
%As already discussed above, it is hard to separate low-mass cluster stars from field ones so this
%last result for the cluster mass is probably overestimated and it should be considered as an upper limit. 
%Given the very large size of the cluster, it is very difficult to define a comparison field that could give an adequate estimate of the field population.

We can also estimate the cluster mass using a parametrised initial mass function (IMF). We used the multiple-part power law IMF defined by \citet{Kroupa}. Firstly, we set the free parameter of the IMF by counting the stars within a certain mass range. We selected the stars located at the top of the main sequence, from spectral type B6\,V to B9\,V, where a separation from the field population is much easier. The
selection was made by performing a cut-off in colour and magnitude. The cut in magnitude corresponds to $-0.5 \leq M_{J}\leq +0.9$. For the colour, we select $-0.10 \leq (J-K_{\textrm{S}})_0\leq -0.06$ \citep{Wi97}, but we allow an 
uncertainty of $\pm0.05$~mag, driven by the dispersion seen in the values of $E(J-K_{{\textrm S}})$. According to the calibration of \citet{St92}, this range of spectral types covers stars between 2.5 and $4.1\:$M$_{\sun}$. With
these criteria, and after subtracting for field contamination as in the previous paragraph, we find $98\pm12$ stars within $r_{{\textrm t}}$. We checked the validity of this normalisation by calculating the number of stars predicted 
in the 4.5\,--\,$6.0\:$M$_{\sun}$ range, which is approximately $33\pm4$. This is compatible, though slightly smaller than the 45 stars for which we have spectra; the agreement is good when we consider only the number 
of stars above the MSTO. Then, by correcting for binarity and integrating the IMF, we obtain a present cluster {\em total} mass of $4\,000^{+1\,200}_{-600}\:$M$_{\odot}$, roughly in agreement with the previous determination. 
We note that the uncertainties do not take into account possible variations in the slope of the IMF. By integrating the IMF up to $150\:$M$_{\sun}$, we find a lower limit to the total initial mass of $5\,700^{+1\,700}_{-900}\:$M$_{\odot}$.

\subsection{Mass segregation and virial mass}

Mass segregation in open clusters has been known for a long time \citep{Sp69}. For a dynamically relaxed cluster, during the approach towards energy equipartition, the more massive
stars segregate into the cluster core as they lose kinetic energy, while the lower-mass stars reside in the outer region of the cluster \citep{Mo02}. The result of this process is
observed in many clusters: there is an overdensity of massive stars in the core. In addition, within this denser region, mergers are favoured leading to the formation of objects that appear rejuvenated \citep{Sch16}.

\begin{figure}  %usando el * ocupa las dos columnas, también para tablas / puedo usar para localizar la figura t(op), b(otton) o h(ere)
  \centering         
%el pdflatex no compila .eps, pero sí funciona si lo compilo a mano acabo pasando epstopdf
  \includegraphics[width=\columnwidth]{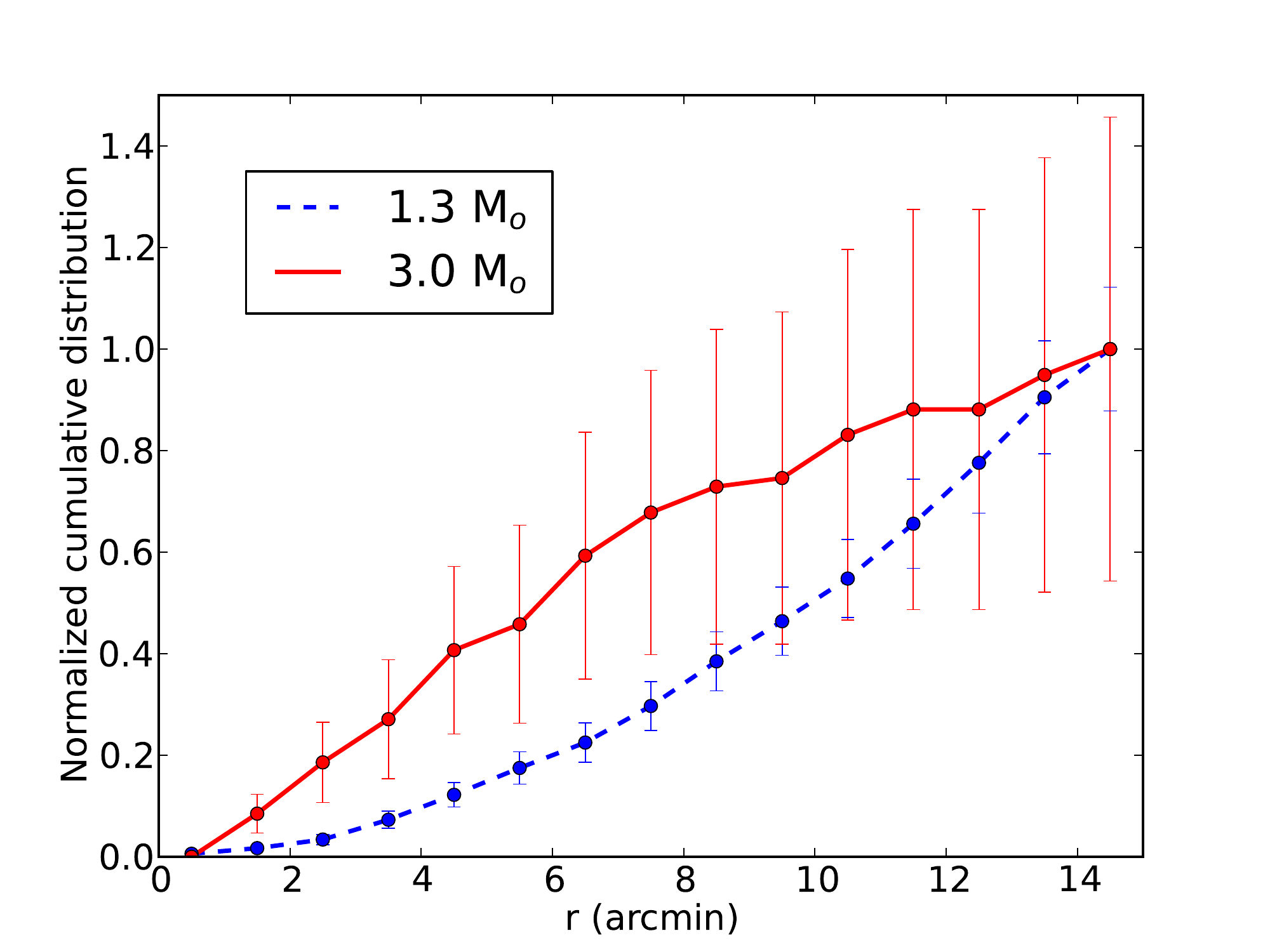}   %posición {h} me da problemas
  \caption{Mass segregation in NGC~6067. The cumulative fraction is represented against the distance from the cluster centre. Two different mass ranges centred on 1.3 and 
  $3\:$M$_{\odot}$ respectively are shown together with their Poisson errors. The more massive stars are more strongly concentrated towards the cluster centre. \label{segreg}}  
\end{figure}

We tried to find any evidence of mass segregation in NGC~6067 by counting stars. We plot (see Fig.~\ref{segreg}) the normalized cumulative distribution of stars as a function of 
the distance to the cluster centre. We selected the stars by doing a cut-off in magnitude: $14.2 < J \leq 14.7$ (centred on $1.25\:$ M$_{\odot}$) and 
$11.7 < J\leq 12.2$ ($3.0\:$M$_{\odot}$). We find that more massive (brighter) stars accumulate more quickly with radius than less massive stars, inferring the existence of mass 
segregation in the cluster.

Since we found mass segregation in NGC~6067, and therefore it is dynamically relaxed, we can use the virial theorem to estimate the cluster mass. We need to make two assumptions:
i) the cluster presents spherical symmetry, and ii) the motions of the stars in the cluster are isotropic. In this case, the three components of the velocity dispersion are equal,
and so:
\begin{equation}
 $ $ \langle v^2 \rangle \approx 3 \sigma^2_{\textrm{rad}}      $ $ 
\end{equation} 
and finally, after operating, we inferred the mass according to this expression:
\begin{equation}
 $ $   M_{\textrm{vir}}\, = \frac{3\, \sigma^2_{\textrm{rad}}\, R\,}{2\, G}\: ,    $ $ 
\end{equation} 
where $G$ is the gravitational constant ($G$\,=\,4.302\,$\cdot$10$^{-3}$ (km\,s$^{-1}$)$^2$ pc\,/M$_{\odot}$\,), $R$ is the cluster radius (tidal radius, in pc) and
$\sigma_{\textrm{rad}}$ is the velocity dispersion calculated in Sect.~\ref{sec:rv}. The result is a 
virial mass, M$_{\textrm{vir}}=6\,100^{+3\,200}_{-1\,700}$~M$_{\odot}$, compatible, within errors, with the two previous determinations. The errors reflect the uncertainties
in the determination of the cluster radius.

\subsection{Stellar atmospheric parameters}

As previously mentioned, we divided the sample in two groups: the early stars and the late stars. We tried to ensure a self-consistent spectroscopic analysis whenever possible. We did not include any of the SB2 found, as we cannot separate the two components with a single spectrum.

\subsubsection{Blue stars}

We employed the technique described by \citet{Ca12}, also in \citet{Le07}. The stellar atmospheric parameters were derived through an automatic $\chi^2$-based algorithm searching for
the set of parameters that best reproduce the main strong lines observed in the range $\approx$\,4000\,--\,5000\,\AA{}. Two different radiative transfer codes were employed to generate 
two partially overlapping synthetic spectral grids: the first one uses KURUCZ atmospheric models from the ATLAS--APOGEE grid from 7\,000\,K to 14\,000\,K. In hotter stars (i.e.\
the earliest B-types), a more complex treatment of the radiative transfer problem, such as non-LTE processes, is needed. Stars above this temperature threshold  were analysed using 
a grid of {\scshape fastwind} synthetic spectra \citep{Si11,Ca12}. The stellar atmosphere code {\scshape fastwind} \citep{Sa97,Pu05} enables non-LTE calculations and assumes a spherical 
geometry. Table~\ref{Par_cal} displays the parameters derived for all the hot single stars.

\begin{table*}
\caption{Stellar atmospheric parameters for the hot stars.\label{Par_cal}}
\begin{center}
\begin{tabular}{lcccc}   %longtable es para poner primero landscape
\hline\hline
Star & $v \sin\,i$ (km\,s$^{-1}$) &  $\zeta$ (km\,s$^{-1}$) & $T_{\textrm{eff}}$ (K) & $\log\,g$\\
\hline
\multicolumn{5}{|c|}{Members}\\
\hline
244      &  90.3 $\pm$ 2.7 & 63.3 $\pm$ 27.4 & 12500 $\pm$ 746   & 3.40 $\pm$ 0.16\\
254      & 184.5 $\pm$ 6.1 & $<\,$88.4       & 12500 $\pm$ 755   & 3.60 $\pm$ 0.16\\
257      & 135.2 $\pm$ 14.9& 64.9 $\pm$ 26.7 & 15000 $\pm$ 1000  & 3.70 $\pm$ 0.14\\
260      & 130.0 $\pm$ 5.4 & 60.6 $\pm$ 33.8 & 12000 $\pm$ 500   & 3.30 $\pm$ 0.10\\
264      &  99.8 $\pm$ 23.9& $<\,$84.3       & 16000 $\pm$ 1000  & 3.70 $\pm$ 0.10\\
267      & 113.7 $\pm$ 2.1 & 65.8 $\pm$ 22.6 & 18000 $\pm$ 1000  & 3.10 $\pm$ 0.10\\
272$^{*}$  & 280.4 $\pm$ 15.3 & $<\,$100.1   & 12000 $\pm$ 889   & 3.30 $\pm$ 0.26\\
273      & 248.9 $\pm$ 11.1 & $<\,$79.9      & 12000 $\pm$ 500   & 3.50 $\pm$ 0.14\\
274      &  60.5 $\pm$ 2.9 & 41.0 $\pm$ 15.0 & 13500 $\pm$ 520   & 3.80 $\pm$ 0.11\\
277      &  42.6 $\pm$ 4.3 & 61.3 $\pm$ 8.8  & 11750 $\pm$ 500   & 3.70 $\pm$ 0.21\\
279      &  23.2 $\pm$ 3.7 & 31.4 $\pm$ 8.5  & 13000 $\pm$ 1000  & 3.60 $\pm$ 0.10\\
285      &  64.6 $\pm$ 11.7& $<\,$98.1       & 14000 $\pm$ 1000  & 4.00 $\pm$ 0.16\\
286$^{*}$ & 130.5 $\pm$ 11.9& $<\,$122.7     & 11750 $\pm$ 500   & 3.30 $\pm$ 0.14\\
287      & 279.7 $\pm$ 24.1& 31.4 $\pm$ 8.5  & 12500 $\pm$ 692   & 3.50 $\pm$ 0.19\\
288      &  67.9 $\pm$ 3.1 & 35.9 $\pm$ 6.1  & 12500 $\pm$ 500   & 3.50 $\pm$ 0.10\\
290$^{*}$  & 29.7 $\pm$ 1.6& $<\,$115.4      & 9000  $\pm$ 500   & 1.70 $\pm$ 0.13\\
291      &   0.4 $\pm$ 0.8 & $<\,$134.1      & 12000 $\pm$ 2120  & 4.40 $\pm$ 0.84\\
293      & 235.8 $\pm$ 31.7& $<\,$118.9      & 12000 $\pm$ 785   & 3.60 $\pm$ 0.20\\
294$^{*}$ & 279.6 $\pm$ 9.1 & 23.7 $\pm$ 19.4 & 12000 $\pm$ 751  & 3.40 $\pm$ 0.19\\
295      &  51.4 $\pm$ 7.5 & 62.2 $\pm$ 12.9 & 13500 $\pm$ 581   & 3.90 $\pm$ 0.18\\
298      &  25.6 $\pm$ 1.7 & 26.7 $\pm$ 9.8  & 7500  $\pm$ 500   & 1.90 $\pm$ 0.10\\
%299      & 12000 $\pm$ 516   & 3.40 $\pm$ 0.12\\  DOBLE
310      &  48.9 $\pm$ 0.6 & 30.3 $\pm$ 7.1  & 12000 $\pm$ 500   & 3.40 $\pm$ 0.12\\
320      & 179.0 $\pm$ 15.2& $<\,$105.3      & 15000 $\pm$ 1000  & 4.00 $\pm$ 0.10\\
324      & 168.6 $\pm$ 11.8& $<\,$138.4      & 12500 $\pm$ 1111  & 3.50 $\pm$ 0.31\\
325$^{*}$ & 120.8 $\pm$ 3.8 & 52.8 $\pm$ 17.8 & 12000 $\pm$ 750  & 3.40 $\pm$ 0.17\\
%1006     & 10250 $\pm$ 651   & 4.00 $\pm$ 0.30\\
1796$^{*}$ & 281.3 $\pm$ 14.5 & 14.4 $\pm$ 11.1& 13000 $\pm$ 744 & 3.70 $\pm$ 0.15\\
7467     &  48.8 $\pm$ 4.4 & 42.2 $\pm$ 10.6 & 13000 $\pm$ 500   & 3.50 $\pm$ 0.15\\
%HD\;145139 & 10750 $\pm$ 515   & 3.50 $\pm$ 0.20\\
HD\,145304 & 135.0 $\pm$ 1.4 & $<\,$81.2     &  18000 $\pm$ 1000  & 3.10 $\pm$ 0.11\\
HD\,145324 &   8.6 $\pm$ 0.7 & 18.0 $\pm$ 4.9 & 7750  $\pm$ 500 & 1.30 $\pm$ 0.10\\
\hline
\multicolumn{5}{|c|}{Non-members}\\
\hline
1006     & 186.6 $\pm$ 21.7 & $<\,$154.7     & 10250 $\pm$ 651   & 4.00 $\pm$ 0.30\\
HD\;145139 &  27.5 $\pm$ 0.8& 33.8 $\pm$ 2.4 & 10750 $\pm$ 515   & 3.50 $\pm$ 0.20\\
\hline
\end{tabular}
\end{center}
\begin{list}{}{}
\item[]$^{*}$ These objects are Be stars. In addition, stars 272 and 290 are shell stars and these parameters do not correspond to the actual star, but to the shell spectrum.
  \end{list}
\end{table*}

\begin{figure*} 
  \centering         

  \includegraphics[width=14cm, height=6cm]{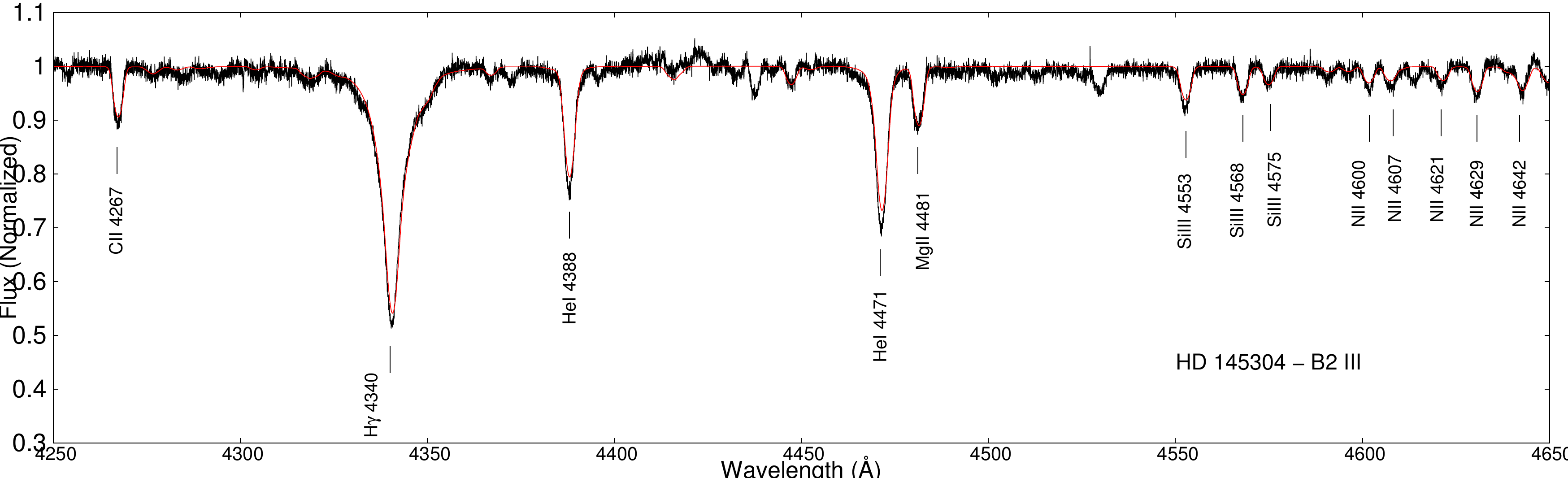}  
  \caption{Best-fitting model for star HD\,145304. The synthetic spectrum, in red, is plotted on the observed one (black). The most prominent lines are labelled.}

  \label{Berto}  
\end{figure*}

\subsubsection{Cool stars}\label{sec:par_frios}

We employed a methodology to derive stellar atmospheric parameters based on a list of 83 \ion{Fe}{i}-\ion{Fe}{ii} features, since Fe lines are numerous as well as very sensitive
to stellar parameters for late-type stars. Their atomic parameters were taken from the VALD database\footnote{\url{http://vald.astro.uu.se/}} \citep{pis95,kup2000}. For the Van der Waals 
damping data, we took the values given by the Anstee, Barklem, and O'Mara theory, when available in VALD \citep[see ][]{bar00}. 

We took KURUCZ LTE plane-parallel stellar atmospheric models from the ATLAS--APOGEE grid\footnote{\url{http://www.iac.es/proyecto/ATLAS--APOGEE/}} 
\citep{mes12}. We expanded the grid by calculating some additional atmospheric models using ATLAS9 \citep{kur93} and the opacity distribution functions ODFs from the ATLAS--APOGEE
web-page. Effective temperature (T$_{\textrm{eff}}$) ranges from 3\,500\,K to 7\,000\,K with a step of 250\,K, whereas surface gravity ($\log\,g$) varies from 0.0 to 5.0~dex in
0.5~dex steps. However, above 6\,000\,K the grid only reaches $\log\,g=0.5$~dex. Finally, in the case of metallicity (using [Fe/H] as a proxy), the grid covers from $-1.0$ to 
1.0~dex in~0.25 dex steps. With these atmospheric models, we generated a grid of synthetic spectra by using the radiative transfer code {\scshape spectrum} \citep{graco94}. The 
microturbulent velocity ($\xi$) was fixed according to the calibration given in \citet{Ad12}.     

As a starting point, we employed a modified version of the {\scshape stepar} code \citep{Ta12}, adapted to the present problem, that uses stellar synthesis instead of an $EW$ method. 
As optimization method we used the Metropolis--Hastings algorithm. Our method generates a Markov-chain of 20\,000 points starting from an arbitrary point. It then performs a 
statistical analysis on the resulting chain to obtain the final stellar atmospheric parameters. As objective function we used a $\chi^2$ in order to fit the selected iron 
lines. We fixed the stellar rotation to the value previously derived (Sect.~\ref{sec:rv}), and the instrumental broadening to the resolution of the FEROS spectrograph. 
We left the macroturbulent broadening as a free parameter to absorb any residual broadening. In Table~\ref{Par}, we display the parameters obtained for the cool stars: 
effective temperature, surface gravity, and metallicity (in terms of [Fe/H]). 

\begin{table*}
\caption{Stellar atmospheric parameters for the cool stars. \label{Par}}
\begin{center}
\begin{tabular}{lccccc}   %longtable es para poner primero landscape
\hline\hline
Star & $v \sin\,i$ (km\,s$^{-1}$)$^{*}$ &  $\zeta$ (km\,s$^{-1}$) & $T_{\textrm{eff}}$ (K) & $\log\,g$ & [Fe/H]\\
\hline
\multicolumn{6}{|c|}{Members}\\
\hline
%229    & 4158 $\pm$ 97  & 1.56 $\pm$ 0.35  & 0.22 $\pm$ 0.23\\
240    & $<\,5.5$ & 5.84 $\pm$ 1.09 & 4051 $\pm$ 106 & 0.99 $\pm$ 0.36  & 0.24 $\pm$ 0.24\\
247    & $<\,7.8$ & 11.21 $\pm$ 1.59& 4321 $\pm$ 123 & 0.48 $\pm$ 0.40  & 0.28 $\pm$ 0.23\\
261    & $<\,8.9$ & 5.58 $\pm$ 1.83 & 4036 $\pm$ 100 & 0.26$^{**}$       & 0.44 $\pm$ 0.20\\ 
275    & $<\,6.3$ & 6.00 $\pm$ 0.94 & 3782 $\pm$ 80  & 0.00$^{**}$       & 0.10 $\pm$ 0.18\\
276    & $<\,5.8$ & 4.26 $\pm$ 1.35 & 3771 $\pm$ 92  & 0.72 $\pm$ 0.39  & 0.24 $\pm$ 0.26\\
292    & $<\,5.9$ & 5.66 $\pm$ 0.93 & 4233 $\pm$ 97  & 0.49 $\pm$ 0.36  & 0.06 $\pm$ 0.18\\
297    &11.0 $\pm$ 1.9 & 2.00 $\pm$ 3.37 & 5776 $\pm$ 183 & 0.82 $\pm$ 0.36  & 0.09 $\pm$ 0.11\\
303    & $<\,10.3$ & 9.23 $\pm$ 2.17 & 4224 $\pm$ 94  & 0.70 $\pm$ 0.36  & 0.06 $\pm$ 0.18\\
306    & $<\,5.4$ & 6.17 $\pm$ 0.73 & 3898 $\pm$ 76  & 0.60 $\pm$ 0.34  & 0.18 $\pm$ 0.22\\
323    & $<\,6.4$ & 8.32 $\pm$ 0.61 & 4531 $\pm$ 70  & 0.58 $\pm$ 0.25  & 0.20 $\pm$ 0.14\\
329    & $<\,5.9$ & 4.69 $\pm$ 0.99 & 4063 $\pm$ 89  & 0.78 $\pm$ 0.36  & 0.18 $\pm$ 0.23\\
1294   & $<\,5.5$ & 4.77 $\pm$ 0.99 & 4080 $\pm$ 105 & 1.19 $\pm$ 0.39  & 0.16 $\pm$ 0.28\\
QZN    & 7.0 $\pm$ 0.7 & 10.92 $\pm$ 2.10& 6031 $\pm$ 272 & 1.21 $\pm$ 0.79  & 0.65 $\pm$ 0.22\\
\hline
\multicolumn{6}{|c|}{Non-member}\\
\hline
229    & 5.0 $\pm$ 0.6 & 4.89 $\pm$ 0.63 & 4158 $\pm$ 97  & 1.56 $\pm$ 0.35  & 0.22 $\pm$ 0.23\\
\hline
\end{tabular}
\end{center}
\begin{list}{}{}
\item[]$^{*}$ Except for the Cepheids (stars 297 and QZN), these values are upper limits.
\item[]$^{**}$ For stars 261 and 275 the tabulated value of gravity is a lower limit (these stars are close to the limit of the grid used).
  \end{list}
\end{table*}

\begin{figure}
  \centering         
  \includegraphics[width=\columnwidth]{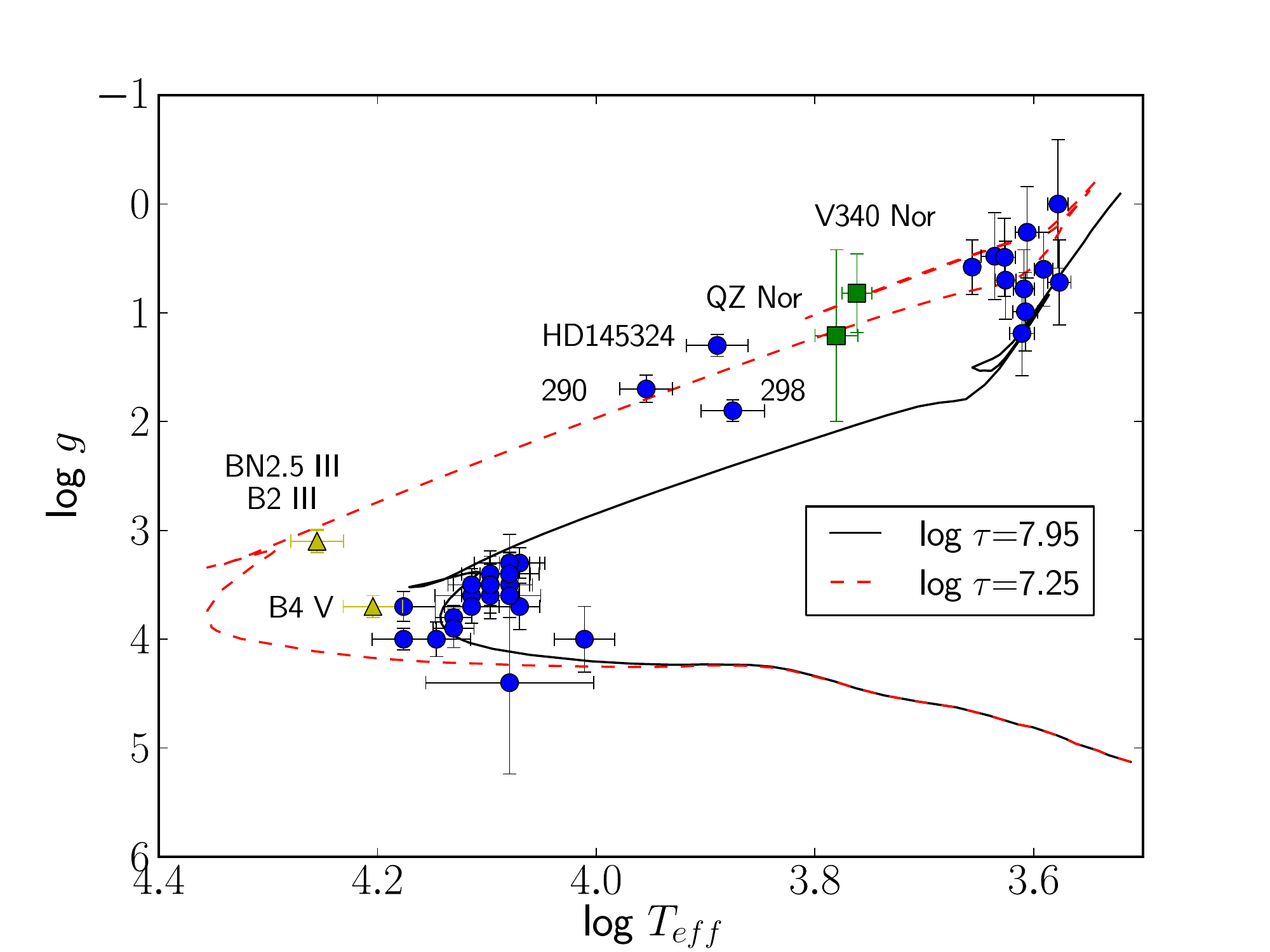}  
  \caption{Kiel diagram. Green squares represent the Cepheids and yellow triangles are BSSs. Two Padova isochrones are drawn: the solid line represents the isochrone providing the
  best fit to the CMDs, while the dashed line is an unrealistic attempt to fit the objects that appear younger than the bulk of the cluster.\label{pHR}}  
\end{figure}

In Fig.~\ref{pHR}, we show a $\log\,g$--$\log\,T_{\textrm{eff}}$ (Kiel) diagram for all the stars with atmospheric parameters. This type of diagram is independent of the distance 
to the cluster, and so provides complementary information to the CMDs. We also plot the isochrone that provides a best fit to the CMDs. However, we see that it is not possible
to fit hot and cool stars with just a single isochrone in the Kiel diagram. Most evolved stars are more luminous than the isochrone and would thus seem to be younger than those at the
top of the main sequence (but note that star 290 is really a shell star, and the parameters displayed correspond to the shell spectrum, and not the underlying star). Since the cluster contains some apparent BSSs, a second younger isochrone is added to test if all these objects could be coeval. A reasonable fit can be 
obtained for an isochrone with an age around 18 Ma. 
%This isochrone would fit two of the BSSs, the two Cepheids, and the stars 290 (B5 shell) and HD\,145324, as well as the brightest red stars. 
However, this apparent fit is unrealistic. The whole unevolved population, including the B4\,V BSSc, fits the older isochrone, and so there is no evidence for a younger unevolved population. 
Finally, many of the evolved stars (the A-type bright giant 298 and some of the red stars) fall in between the two isochrones. Therefore we conclude that there is no real evidence
for multiple populations and discuss possible causes for the poor fit in the Kiel diagram later, in Sect.~\ref{sec:parameters}.

\subsection{Chemical abundances}\label{sec:abund}

We tried to derive chemical abundances for the hot stars, but have been unable to obtain valid results for most of our targets. The A supergiants fall outside the range of
$T_{\textrm{eff}}$ included in the {\sc fastwind} grid. For the late-B giants, the procedure used resulted in errors so large that the measurements were considered unreliable. 
This is due to the weakness of the metallic lines and the lack of the Fe atomic model in the {\sc fastwind} formal solution. In the B4\,--\,B7 stars close to the main sequence, there are no useful metallic lines,
except for \ion{Mg}{ii}~4481\,\AA{}, and so no attempt to derive abundances was made. Only for the two hottest stars have we been able to determine 
abundances for a few elements (see Table~\ref{abund_cal}). Once the stellar parameters are set, in order to obtain the chemical abundaces, we computed a tailored {\scshape fastwind} 
grid of models by varying the abundances of the species under study (C, N, O, Mg, and Si) and performed initially an authomatic $\chi^2$-fitting algorithm and then a visual 
check for discarding weak or blended features. The technique, as well as the lines considered in the analysis ,are described in detail by \citet{Ca12}. The errors obtained 
are larger than or equal to 0.2 dex, the step size of the grid used. 

\begin{table*}
\caption{Chemical abundances, relative to solar abundances by \citet{As05}, measured on the two B2\,III giants.}
\begin{center}
\begin{tabular}{lccccc}   %longtable es para poner primero landscape
\hline\hline
Star & [C/H] & [N/H] & [O/H] & [Mg/H] & [Si/H]\\
\hline
HD\,145304  & $-$0.40 $\pm$ 0.21 & 0.80 $\pm$ 0.20 & 0.00 $\pm$ 0.20 & 0.40 $\pm$ 0.20 & 0.40 $\pm$ 0.20\\ 
267         & $-$1.40 $\pm$ 0.23 & 1.00 $\pm$ 0.20 & 0.00 $\pm$ 0.20 & 0.00 $\pm$ 0.30 & 0.00 $\pm$ 0.21\\
\hline

\end{tabular}
\label{abund_cal}
\end{center}
\end{table*}

For the late-type stars, we employed a method based on $EW$s measured in a semi-automatic fashion using {\scshape tame} \citep{kan12} for Na, Mg, Si, Ca, Ti, Ni, Y, and 
Ba. We also measured $EW$s by hand for two special and delicate cases, oxygen and lithium, using the {\scshape iraf} {\scshape splot} task.

\begin{figure}  %usando el * ocupa las dos columnas, también para tablas / puedo usar para localizar la figura t(op), b(otton) o h(ere)
  \centering         
%el pdflatex no compila .eps, pero sí funciona si lo compilo a mano acabo pasando epstopdf
  \includegraphics[width=\columnwidth]{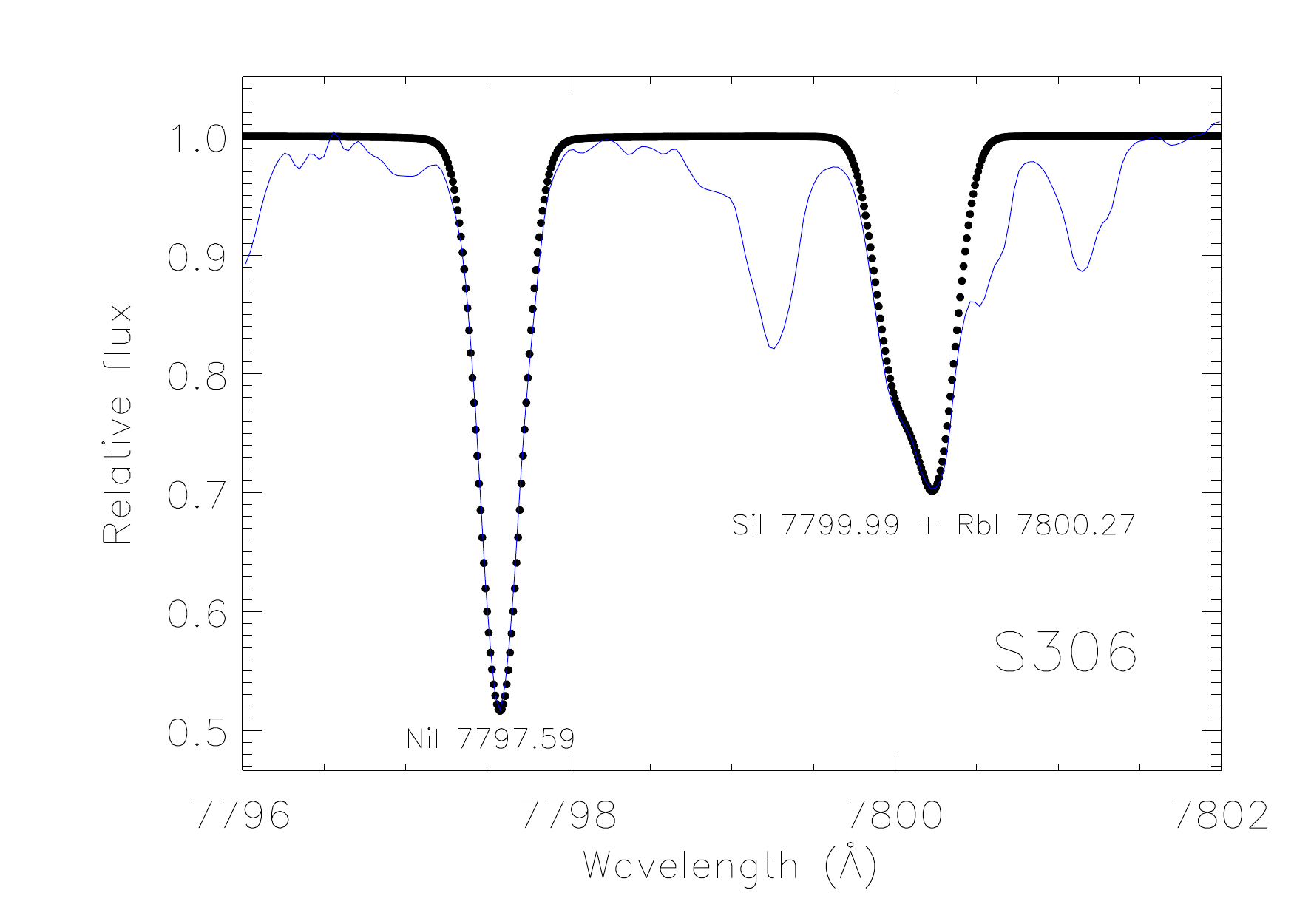}   %posición {h} me da problemas
  \caption{Example of spectral synthesis of the rubidium 7800\,\AA{} line for star 306. Note that the \ion{Si}{i} line at 7799\,\AA{} blends with the Rb line on its blue wing.} 
  \label{rb}  % ref para cuando la nombre, da igual lo que ponga, el programa las ordena fig1, fig2, ... pero es útil por si cambio la posición
\end{figure}

For lithium, we performed a classical analysis on the 6707.8\,\AA{} line. We measured its $EW$ by hand (in m\AA{}), using the {\scshape iraf} {\scshape splot} task. In some cases we had to correct 
the $EW($Li$)$ by taking into account the nearby \ion{Fe}{i} line at 6707.4\,\AA{}. The results are displayed in Table~\ref{Li}. We use the standard notation, where 
$A($Li$)= \log\,$[n(Li)/n(H)]\,+\,12. In the case of oxygen, we used the [\ion{O}{i}] 6300\,\AA{} line. This oxygen line is blended with a \ion{Ni}{i} feature; we corrected the $EW$s 
accordingly by using the  methodology and line atomic parameters described in \citet{bel15}. Finally we also derive rubidium abundances using stellar synthesis for the 
7800\,\AA{} \ion{Rb}{i} line, following the methodology in \citet{dor13}, as shown in Fig.~\ref{rb}.

In Table~\ref{Abund} we show the computed abundances for all the late-type stars in our sample.  From the analysis of thirteen evolved stars (316 was not analysed because of 
its binary nature) we derive a supersolar metallicity for the cluster. The weighted arithmetic mean (using the variances as weights) is [Fe/H]$=+0.19\pm0.05$~dex (see Table~\ref{Par}). We checked the consistency of our analysis by measuring [Fe/H] for each star using two alternative methods, equivalent widths and spectral synthesis. 
As shown in Fig.~\ref{check_feh}, both methods give measurements compatible within errors.

%\begin{landscape}
\begin{table*}
\caption{Chemical abundances, relative to solar abundances by \citet{As05}, measured on the cool stars.}
\begin{center}
\begin{tabular}{lccccc}   %longtable es para poner primero landscape
\hline\hline
Star & [O/H] &[Na/H] & [Mg/H] & [Si/H] & [Ca/H]\\
\hline
\multicolumn{6}{|c|}{Members}\\
\hline
240  &  0.08 $\pm$ 0.25    &  0.06 $\pm$ 0.44   &   ---              & 0.47 $\pm$ 0.27 & $-$0.13 $\pm$ 0.22\\
247  & $-$0.40 $\pm$ 0.28  &  0.24 $\pm$ 0.55   & $-$0.09 $\pm$ 0.10 & 0.33 $\pm$ 0.23 & $-$0.01 $\pm$ 0.21\\
261  & $-$0.25 $\pm$ 0.24  & $-$0.26 $\pm$ 0.09 & $-$0.09 $\pm$ 0.10 & 0.58 $\pm$ 0.23 &  0.48 $\pm$ 0.28\\
275  & $-$0.45 $\pm$ 0.33  & $-$0.17 $\pm$ 0.58 & $-$0.29 $\pm$ 0.22 & 0.26 $\pm$ 0.31 & $-$0.31 $\pm$ 0.31\\
276  &  0.03 $\pm$ 0.22    & $-$0.04 $\pm$ 0.69 &  0.02 $\pm$ 0.18   & 0.46 $\pm$ 0.21 & $-$0.11 $\pm$ 0.20\\
292  & $-$0.32 $\pm$ 0.24  &  0.16 $\pm$ 0.54   & $-$0.14 $\pm$ 0.09 & 0.40 $\pm$ 0.21 & $-$0.11 $\pm$ 0.18\\
297  &   ---               &  0.29 $\pm$ 0.38   & $-$0.12 $\pm$ 0.27 & 0.28 $\pm$ 0.13 & $-$0.49 $\pm$ 0.11\\
303  & $-$0.15 $\pm$ 0.22  &  0.03 $\pm$ 0.41   & $-$0.14 $\pm$ 0.10 & 0.32 $\pm$ 0.23 &  0.22 $\pm$ 0.31\\
306  & $-$0.09 $\pm$ 0.20  & $-$0.03 $\pm$ 0.55 & $-$0.06 $\pm$ 0.12 & 0.56 $\pm$ 0.28 & $-$0.22 $\pm$ 0.19\\
323  & $-$0.30 $\pm$ 0.16  &  0.16 $\pm$ 0.53   &  0.09 $\pm$ 0.06   & 0.49 $\pm$ 0.16 &  0.24 $\pm$ 0.15\\
329  & $-$0.02 $\pm$ 0.22  &  0.05 $\pm$ 0.56   & $-$0.08 $\pm$ 0.14 & 0.47 $\pm$ 0.19 & $-$0.22 $\pm$ 0.16\\
1294 &  0.08 $\pm$ 0.26    & $-$0.01 $\pm$ 0.37 &  0.06 $\pm$ 0.19   & 0.42 $\pm$ 0.23 & $-$0.19 $\pm$ 0.25\\
QZN  &  0.30 $\pm$ 0.48    &  0.56 $\pm$ 0.21   &  0.15 $\pm$ 0.58   & 0.48 $\pm$ 0.16 &  0.58 $\pm$ 0.15\\
\hline
Mean & $-$0.15 $\pm$ 0.07  & $-$0.08 $\pm$ 0.07 & $-$0.04 $\pm$ 0.03 & 0.42 $\pm$ 0.06 & $-$0.07 $\pm$ 0.05\\
\hline 
\multicolumn{6}{|c|}{Non-member}\\
\hline
229  &  0.32 $\pm$ 0.25    & -0.05 $\pm$ 0.29   &  0.23 $\pm$ 0.19   & 0.40 $\pm$ 0.20 & -0.12 $\pm$ 0.15\\ 
\hline
     &                     &                    &                    &                 &                 \\
     &                     &                    &                    &                 &                 \\
\hline\hline
Star & [Ti/H] & [Ni/H] & [Rb/H]& [Y/H] & [Ba/H]\\
\hline
\multicolumn{6}{|c|}{Members}\\
\hline
240  & $-$0.12 $\pm$ 0.35 &  0.11 $\pm$ 0.27 & $-$0.20  &   ---              & 0.49 $\pm$ 0.18\\
247  & $-$0.05 $\pm$ 0.28 &  0.03 $\pm$ 0.21 & $-$0.51  & $-$0.27 $\pm$ 0.62 & 0.54 $\pm$ 0.29\\
261  &  0.07 $\pm$ 0.30   &  0.46 $\pm$ 0.27 & $-$0.29  &  0.33 $\pm$ 0.97   & 0.72 $\pm$ 0.18\\
275  & $-$0.19 $\pm$ 0.25 &  0.04 $\pm$ 0.35 & $-$0.30  & $-$0.12 $\pm$ 0.78 & 0.20 $\pm$ 0.30\\
276  &  0.12 $\pm$ 0.23   &  0.34 $\pm$ 0.23 & $-$0.11  &  0.06 $\pm$ 0.23   & 0.53 $\pm$ 0.21\\
292  & $-$0.13 $\pm$ 0.25 &  0.02 $\pm$ 0.21 & $-$0.32  & $-$0.28 $\pm$ 0.55 & 0.34 $\pm$ 0.28\\
297  &  0.11 $\pm$ 0.29   &  0.03 $\pm$ 0.19 &   ---    & $-$0.08 $\pm$ 0.30 & ---  \\
303  & $-$0.21 $\pm$ 0.24 &  0.05 $\pm$ 0.24 & $-$0.27  & $-$0.10 $\pm$ 0.62 & 0.40 $\pm$ 0.21\\
306  & $-$0.11 $\pm$ 0.22 &  0.20 $\pm$ 0.21 & $-$0.34  &  0.04 $\pm$ 0.70   & 0.26 $\pm$ 0.23\\
323  &  0.06 $\pm$ 0.18   &  0.14 $\pm$ 0.11 & $-$0.15  & $-$0.14 $\pm$ 0.58 & 0.61 $\pm$ 0.19\\
329  & $-$0.06 $\pm$ 0.22 &  0.16 $\pm$ 0.21 & $-$0.36  &  0.07 $\pm$ 0.53   & 0.47 $\pm$ 0.26\\
1294 & $-$0.04 $\pm$ 0.24 &  0.11 $\pm$ 0.21 & $-$0.30  &  ---               & 0.69 $\pm$ 0.20\\
QZN  &  0.47 $\pm$ 0.28   &  0.43 $\pm$ 0.17 &   ---    &  0.28 $\pm$ 0.45   &  --- \\
\hline
Mean & $-$0.01 $\pm$ 0.07 & 0.17 $\pm$ 0.06  & $-$0.29 $\pm$ 0.11 & 0.00 $\pm$ 0.14 & 0.51 $\pm$ 0.07\\
\hline
\multicolumn{6}{|c|}{Non-member}\\
\hline
229  & -0.06 $\pm$ 0.26   &  0.18 $\pm$ 0.19 & -0.27    &  0.13 $\pm$ 0.48   & 0.39 $\pm$ 0.21\\
\hline
\end{tabular}

\label{Abund}
\end{center}
\end{table*}
%\end{landscape}

\begin{figure} 
  \centering         

  \includegraphics[width=\columnwidth]{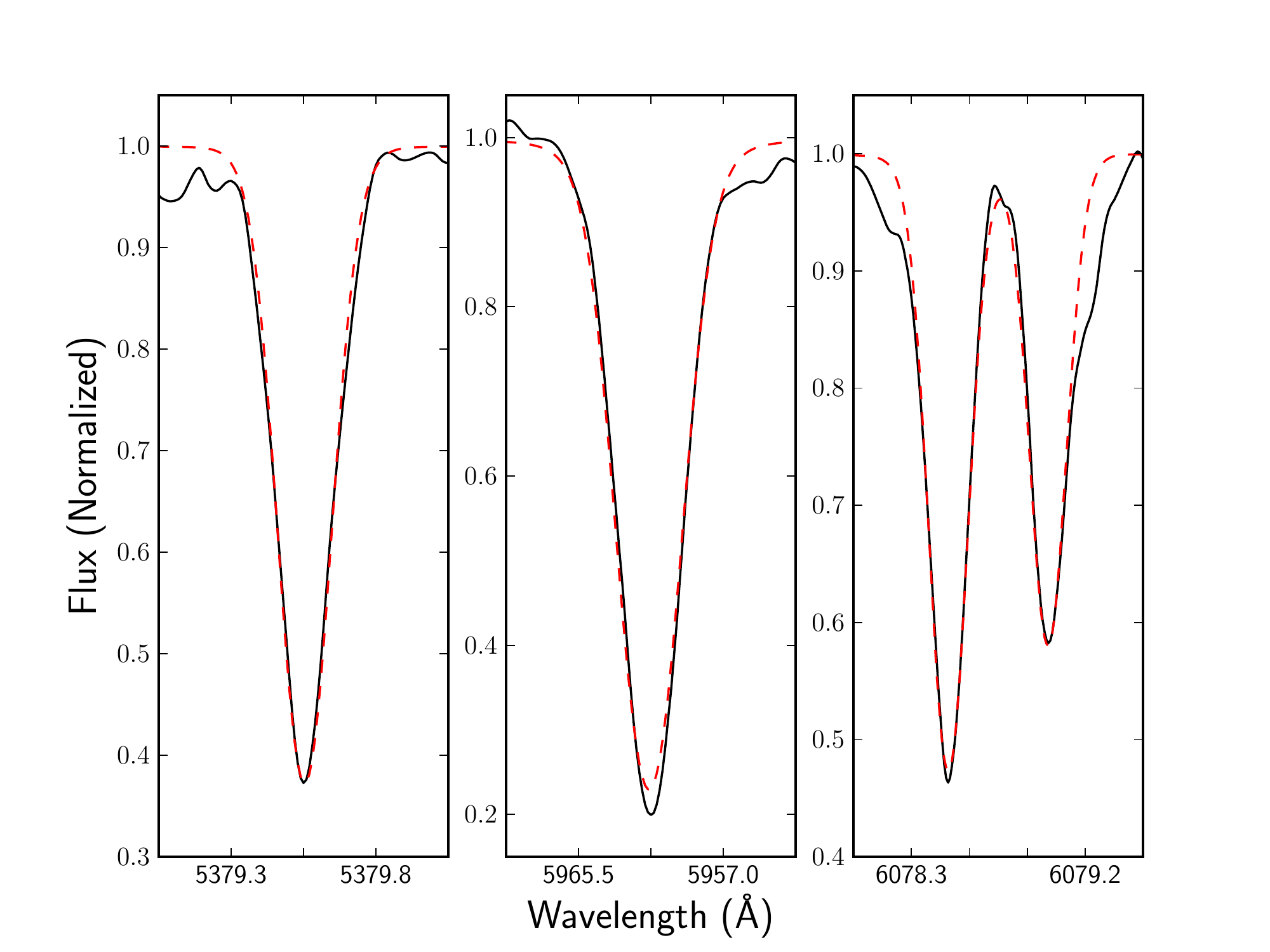}  
  \caption{Some examples of line fits in star 306. Three regions are shown centred on \ion{Fe}{i}~$\lambda$5379.6 (left), \ion{Fe}{i}~$\lambda$5956.7 (centre) and 
  \ion{Fe}{i}~$\lambda$6078.5 (right). The observed spectrum is the solid black line, whereas the synthetic one is the red dashed line.}

  \label{lineas_hugo}  
\end{figure}

\begin{table}
\caption{Lithium abundances measured on the cool stars.}
\begin{center}
\begin{tabular}{lccccc}   %longtable es para poner primero landscape
\hline\hline
Star & EW & EW(Fe) & EW(Li) & A(Li) & Note\\
\hline
\multicolumn{6}{|c|}{Members}\\
\hline
240  & 44.6  & 0.0  & 44.6  &  0.06 $\pm$ 0.18 & Li\\ 
247  & 174.7 & 46.3 & 128.4 &  1.21 $\pm$ 0.22 & Li+Fe\\
261  & 0.0   & 0.0  & 0.0   &  0.0             & NoLi\\
275  & 0.0   & 0.0  & 0.0   &  0.0             & NoLi\\
276  & 370.3 & 35.1 & 335.2 &  2.41 $\pm$ 0.12 & Li+Fe\\
292  & 19.1  & 0.0  & 19.1  & $-$0.06          & UL\\
297  & 0.0   & 0.0  & 0.0   &  0.0             & NoLi\\
303  & 179.1 & 34.4 & 144.7 &  1.15 $\pm$ 0.16 & Li+Fe\\
306  & 22.5  & 0.0  & 22.5  & $-$0.55          & UL\\
323  & 44.3  & 0.0  & 44.3  &  0.88 $\pm$ 0.11 & Li\\
329  & 107.1 & 0.0  & 107.1 &  0.61 $\pm$ 0.15 & Li\\
1294 & 0.0   & 0.0  & 0.0   &  0.0             & NoLi\\
QZN  & 0.0   & 0.0  & 0.0   &  0.0             & NoLi\\
\hline
\multicolumn{6}{|c|}{Non-member}\\
\hline
229  & 22.7  & 0.0  & 22.7  & -0.10            & UL\\
\hline

\end{tabular}

\label{Li}
\end{center}
\begin{list}{}{}
\item[]Legend for last column: \textbf{NoLi}: no lithium is found; \textbf{UL}: upper limit, only a little amount of lithium is seen; \textbf{Li}: direct measurement, 
lithium is deblended; \textbf{Li+Fe}: lithium is blended with iron, indirect measurement.
  \end{list}
\end{table}
%\end{landscape}

\begin{figure}
  \centering         
  \includegraphics[width=\columnwidth, height=8cm]{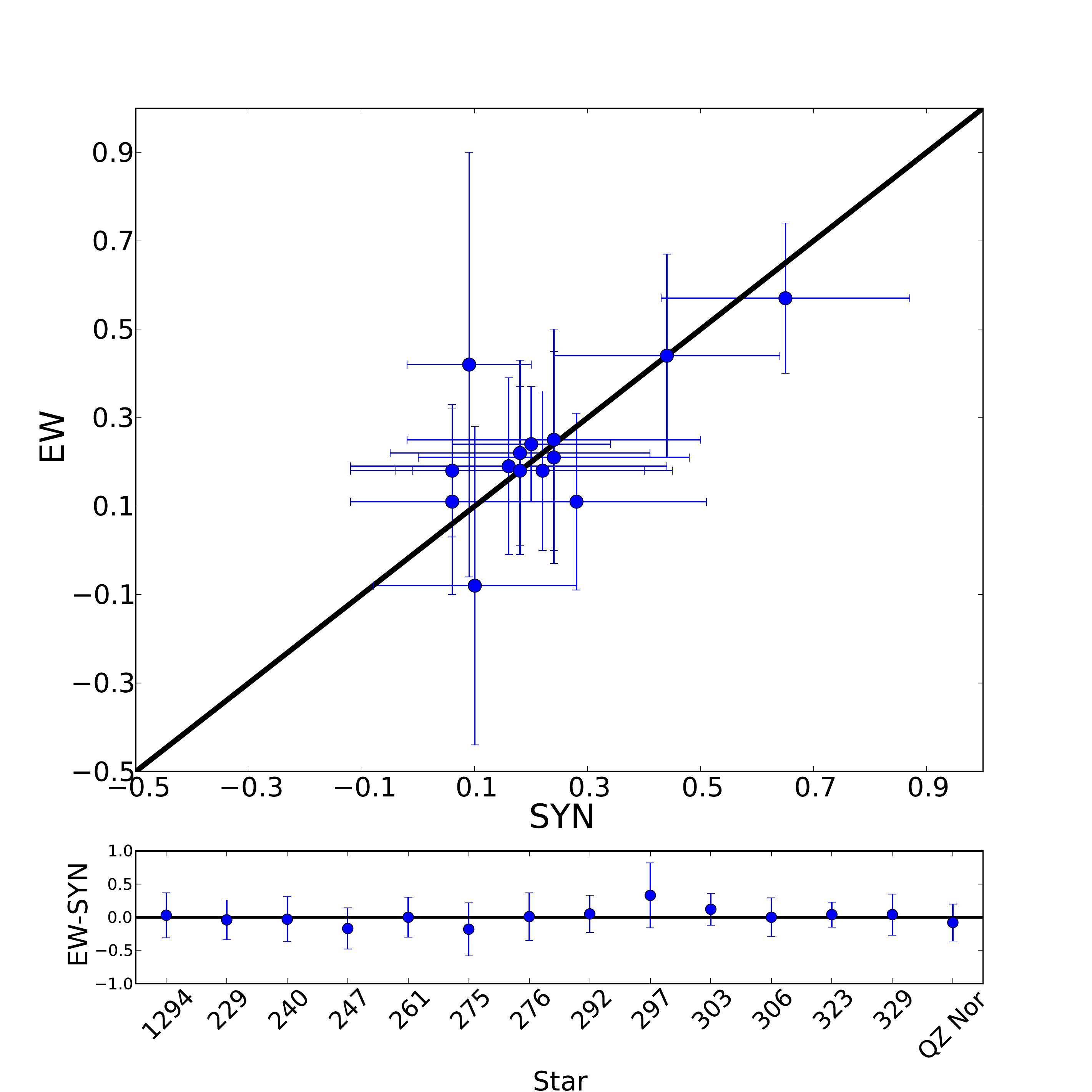}  
  \caption{Consistency test. Top: [Fe/H] determined via spectral synthesis (SYN) versus equivalent widths (EW). All the stars lie, within the errors, on the 1:1 line 
  (black line). Botton: Difference between the abundance measured with each method (EW\,$-$\,SYN) for each star. The black line marks the zero-point. No significant differences are 
  seen.\label{check_feh}}  
\end{figure}

\section{Discussion}\label{sec:4}

\subsection{Cluster mass}

We find a virial mass of $M_{\textrm{vir}}=6\,100^{+3\,200}_{-1\,700}\:$M$_{\sun}$ for NGC~6067. We also estimate the mass of the cluster using two different methods to count stars, both subject to large uncertainties. 
All three methods agree on a present-day mass around or slightly below 5\,000$\:$M$_{\sun}$. The corresponding initial mass, without taking into account dynamical losses, is roughly consistent with population 
synthesis models developed by \citet{Mes09}, who find an initial mass of $\approx7\,000\pm2\,000$ for a cluster of 50\,--\,100~Ma containing 15 evolved stars. This suggests that NGC~6067 is the descendent of a young massive
cluster, similar to those found close to the base of the Scutum arm \citep[e.g.][]{Neg11}. In fact, NGC~6067 should be (perhaps together with NGC~2354, whose age is not so well constrained) the most massive cluster among all those studied by 
\citet{Me08} in the 50\,--\,150 Ma range, since it hosts, by far, the largest number of evolved stars (15) compared with the average number (2) observed in other clusters 
(see Table~\ref{RG_OC}).

Another example of a moderate-age populous open cluster is M11, with a similar appearance to NGC~6067, though older \citep[$\tau= 250$\,--\,316~Ma;][]{GES_M11}. Both are situated in 
regions of high stellar density (Scutum and Norma Cloud respectively) and are very populous. 
%According to \citet{Trumpler}, NGC~6067 is classified as I2r whereas M11 is II2r: both are rich and 
%detached with a strong central concentration. 
Current mass estimates for M11 indicate an initial mass between $\sim5\,000$\:M$_{\sun}$ and more than $10\,000\:M_{\sun}$  \citep{SaM11,GES_M11}, with recent results by \citet{Bav16} strongly favouring the younger age and hence the higher mass value. The existence of these older populous clusters implies 
that such massive clusters have been forming regularly in the Galactic disc, and are not necessarily concentrated towards the tips of the long Bar. Both NGC~6067 and M11 are 
relatively nearby ($d\la2$~kpc) and located in low-extinction windows. However, they do not stand out strongly from their background, suggesting that similar clusters in 
higher-obscuration areas could be quite difficult to detect. 

%Because of its age and large population, NGC~6067 is an excellent laboratory to constrain the evolution of the most massive intermediate-mass stars (see Sect.~\ref{sec:parameters}). 
According to the isochrones, its red giants should
have masses $\approx6\:$M$_{\sun}$. M11, by comparison, has a clump of red giants with masses around $3.6\:$M$_{\sun}$ \citep{SaM11}. The only other example of a cluster of similar 
age with an equivalent population may be Mercer~13 \citep{Mes09}, but this cluster is heavily obscured, and its parameters are uncertain. On the contrary, in the Large Magellanic
Cloud (LMC), there are several young globular clusters with similar ages and more massive than NGC~6067, such as NGC~2136 or NGC 2157, or the much more massive NGC~1850 
\citep{Nie15}.

\subsection{Be stars} 

According to the standard definition \citep{Po03}, classical Be stars are non-supergiant B-type stars that show or had shown Balmer lines in emission at any time. Be stars are fast rotators \citep{Sl49} and their high rotation is believed to be connected with the formation of a gaseous circumstellar disc around the central star, where the observed emission originates
\citep{Po03}. However, based on extensive simulations of B-star populations, \citet{Granada_be} conclude that fast rotation alone cannot account for the existence of the Be phenomenon.  

The Be phenomenon has been observed in young open clusters, starting at an age $\sim10\:$Ma and up to the oldest ages at which B-type stars are present ($\sim300\:$Ma). \citet{Me82b}, by considering a sample of clusters covering a wide age range, found that the Be star fraction (defined as the number of Be stars divided by the total number of B-type stars) has two peaks, one at spectral types B1\,--\,B2 and the other at B7\,--\,B8. However, \citet{Sw05} analysed a large number of clusters in a homogeneous way, finding that the Be fraction presented a broad maximum for ages between 25 and $100\:$Ma. NGC 6067 is too old to host a population
in the B0\,--\,B2 range, but contains many stars in the B7\,--\,B8 range. In fact, most of the Be stars in the cluster present a spectral type B8 (see Table~\ref{Be}). If we look at Table~\ref{Par_cal}, we can see that many of the Be stars in NGC~6067 present very high projected rotational velocities, with three of them having values around $280\:$km\,s$^{-1}$, the highest measured in the cluster (note that the value measured for star 290 represents the width of the shell lines, and not the actual $v\,\sin\,i$). 

\citet{Zo97} estimated at $\sim 17$\% the Be star fraction among bright Galactic B stars.  Clusters with a high Be fraction (up to $\sim40$ per cent for early-B types), such as NGC~663 \citep{Pi01} or NGC~7419 
\citep{7419}, are much younger ($\tau \sim$\,15\,--\,25~Ma) than NGC~6067. In contrast, among 15 older clusters (ranging in age 35\,--\,$282\:$Ma), \citet{Sw05} found only one or 
two definite Be stars per cluster and up to five other candidate Be stars. In NGC~6067 we find a Be fraction around 21 per cent, detecting six Be stars. This value is biased because our
sample is not complete. We have only observed the brightest stars close to the MSTO (about one third of the total number of B-type stars in the cluster, since we have estimated the presence of $\approx$100 B-stars in Sect.~\ref{sec:lum_mass}). Therefore the real fraction of Be stars would probably decrease with increasing number of late-B faint stars, because Be stars are usually located around the MSTO \citep{Sw05}. According to \citet{Granada_be}, while early-type Be stars cannot be explained by critical rotation alone, this might be the dominant physical cause of the Be phenomenon for late Be stars. In their calculations, the fraction of critical rotators is $\sim5$\% for a cluster with an age around 100~Ma. The observed number of Be stars in NGC~6067 is compatible with this fraction under the reasonable expectation that there are very few or no Be stars among B8\,--\,B9 stars still on the main sequence. 
%is compatible with that expected from models. %, and the relatively high number of Be stars is a simple consequence of the richness of the cluster.

\begin{table}
\caption{Fraction of Be stars in NGC 6067.}
\begin{center}
\begin{tabular}{lccc}   %longtable es para poner primero landscape

\hline\hline
Sp T& B & Be & Fraction ($\%$)\\
%      &   &    &   Be/(B+Be) \\
\hline
B2 & 2 & 0 & 0.0\\
B3 & 0 & 0 & ---\\
B4 & 1 & 0 & 0.0\\
B5 & 0 & 1 & 100.0\\
B6 & 3 & 1 & 25.0\\
B7 & 9 & 1 & 10.0\\
B8 & 4 & 3 & 42.9\\
B9 & 3 & 0 & 0.0\\
\hline
Total & 22 & 6 & 21.4\\

\hline
\end{tabular}

\label{Be}
\end{center}
\end{table}
 
%______________________________________________________________________________________________________________________________________________________________________

\subsection{Stellar atmospheric parameters}\label{sec:parameters}

As seen in Figs.~\ref{BV} and \ref{JHK}, with the exception of Cepheids (see Sect.~\ref{sec:Cepheids} for a detailed discussion), Padova isochrones provide an excellent match to the position of both hot and cool stars 
in the observational CMD. However, when we plot these very same isochrones that fit best the CMDs in the $\log\,{g}$\,--\,$\log\,T_{\textrm{eff}}$ diagram against our derived stellar atmospheric parameters (Fig.~\ref{pHR}), 
the position of most of the cool stars seems incompatible with the isochrone. Their values of $\log\, g$ place them well above the isochrone. Red giants follow the isochrone very closely in the CMD, and seem to be well placed on the $T_{\textrm{eff}}$ axis of 
the Kiel diagram. Moreover, the temperatures obtained correlate well with spectral types and photometric colours. This suggests that the source of the discrepancy lies in the spectroscopic gravities. A search of the literature 
shows that this discrepancy is not peculiar to this cluster, but seems widespread \citep{Lu15}. Gravities derived from Fe ionisation balance
analyses are consistently lower (and thus luminosities are consistently higher) than those inferred from photometric calibrations or isochrones. This behaviour has been attributed to non-LTE effects \citep{All04}, but its ultimate
origin is unclear at the moment \citep{Lu15}. In addition to this well-known problem, we must not forget that the stars in NGC~6067 are very luminous compared with typical red giants, and the models used may be stretched to their 
limits, where NLTE effects and the consequences of extension start to be felt. 

\begin{figure}
  \centering         
  \includegraphics[width=\columnwidth]{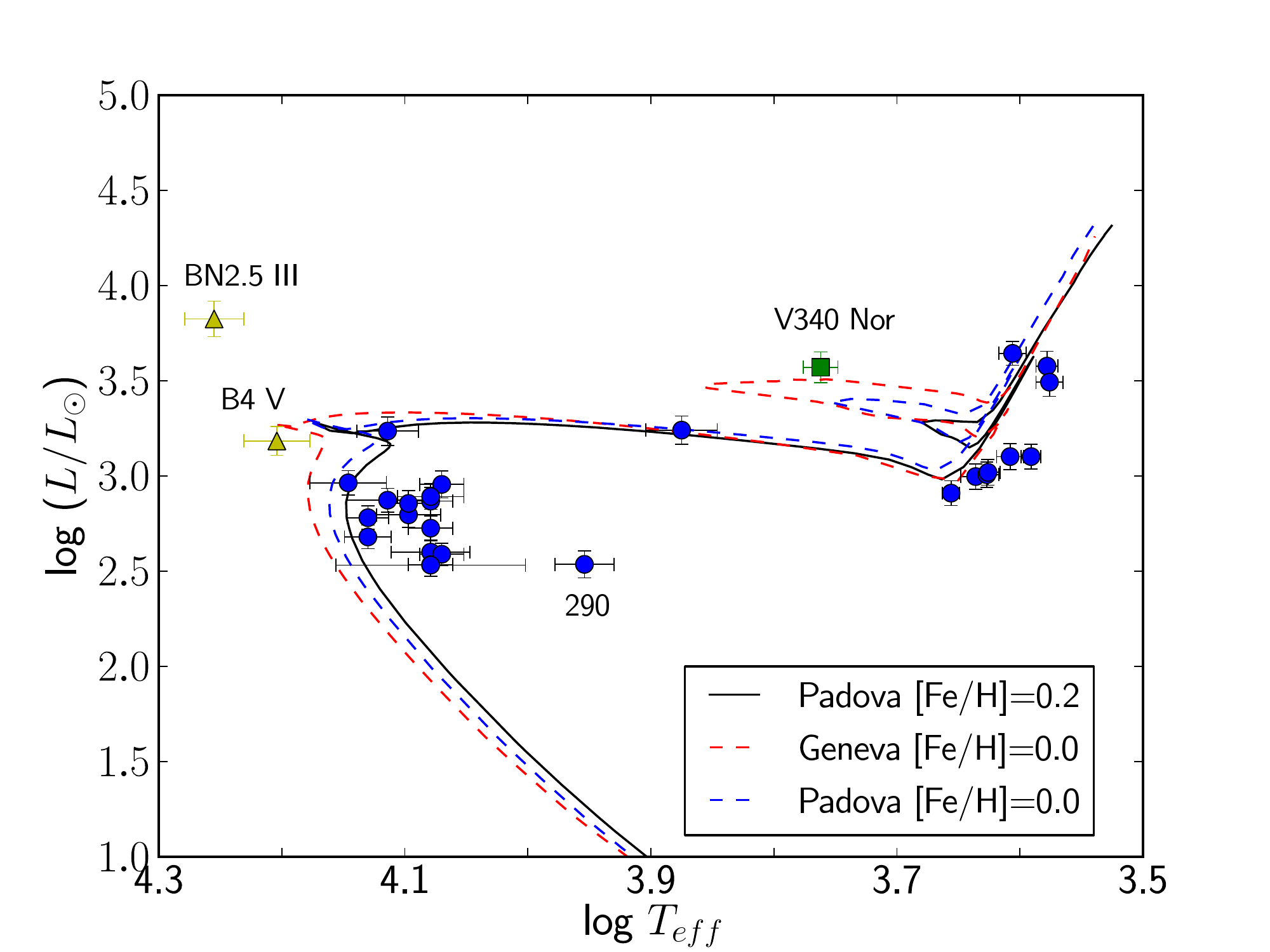}  
  \caption{HR diagram. Symbols and colours follow the same code as in Fig.~\ref{pHR}. Three isochrones with the age of the cluster (log\,$\tau$\,=7.95) are plotted.
  The solid line represents a Padova supersolar isochrone at the metallicity found in this work, whereas the dashed lines are solar isochrones for comparison: Padova (blue) and Geneva with average rotation,
  i.e. $\Omega/\Omega_{\textrm{crit}}=0.5$ (red).\label{luminos}} 
\end{figure}

As a test on this hypothesis, we plot in Fig.~\ref{luminos} the Hertzsprung--Russell (HR) diagram, where luminosity has been derived from photometry instead of gravity. From the $V$ magnitudes (see Table~\ref{Ph}), we calculated 
luminosities by using different bolometric corrections for the hot \citep{Ku94} and cool stars \citep{masana}. The overall fit to the isochrone is now improved, giving support to the idea that the 
spectroscopic gravities, as suspected from Fig.~\ref{pHR}, are somewhat lower than they should be. Even though the agreement between the two methods based on photometry is not surprising, 
the excellent match when the photometry is combined with the spectroscopic $T_{\textrm{eff}}$ strongly points to the spectroscopic gravities as the cause of the discrepancy.

\citet{Lu94} calculated temperatures, surface gravities and abundances for the two Cepheids and five other stars in the cluster. In Table~\ref{Comp_par}, the parameters for these
five stars are confronted with our results. In general, our temperatures are cooler than those of \citet{Lu94}. Only for star 298 (estimated with our hot grid) the temperature is
hotter, though compatible within the errors. In addition, for this star and for star 275, discrepancies in both $T_{\textrm{eff}}$ and $\log\,g$ are noticeable. We have to note, 
however, that 275 is the second brightest star in the cluster in $K_{\textrm {S}}$ (less than 0.1~mag fainter than 261), and, given the similar spectral type, the value of 
$\log\,g$ found by \citet{Lu94} seems too high. Moreover, the higher temperature found for 275 by \citet{Lu94} is at odds with its spectral classification and the similarity of 
its infrared magnitudes and colours to those of 276. Aside from these inconsistencies, differences between our and Luck's analysis can be explained by the different methodologies employed. 
Our analysis is done with a purely spectroscopy approach based on modern KURUCZ models \citep{mes12}, whereas \citet{Lu94} derives photometric stellar parameters and uses old MARCS atmospheric models. 
The underlying physics of the atmospheric models has been substantially revised since then, from atomic parameters to opacities \citep[a review of this topic is found in ][]{mes12}.

In Table~\ref{Cef}, we list the parameters found in the literature for the two Cepheids \citep{Lu94,Fry97,Ge14}. Since these stars are pulsating variables, their parameters 
depend on the moment of the observation (especially the effective temperature) reason for which the aim of Table~\ref{Cef} is only qualitative, to collect previous results. A 
proper comparison, nevertheless, is not possible because the Cepheids have been observed at different times of their variability cycle\footnote{V340\,Nor was observed at MJD=55693.3808
and QZ\,Nor at MJD=54189.2119.}. However, all authors when comparing both
Cepheids show the same trend: QZ~Nor has a higher temperature, gravity and iron abundance than V340 Nor. There are two columns for the [Fe/H]: the first one is the 
[Fe/H] as it appears in each paper, whereas in the second column, every [Fe/H] is rescaled to ours for a more adequate comparison (see next subsection). 

\begin{table}
\caption{Comparison of temperature and surface gravity with those of \citet{Lu94}. \label{Comp_par}}
\begin{center}
\begin{tabular}{l|cc|cc}   %longtable es para poner primero landscape
\hline\hline
\multirow{2}{*}{Star} & \multicolumn{2}{|c|}{Luck94$^{*}$} & \multicolumn{2}{c}{This work}\\
   & $T_{\textrm{eff}}$ (K)& $\log\, g$ & $T_{\textrm{eff}}$ (K)& $\log\, g$\\

\hline
261      & 4100  & $-0.15$ & 4036 $\pm$ 100 & 0.26$^{**}$ \\
275      & 4400  &  1.41   & 3782 $\pm$ 80  & 0.00$^{**}$\\
276      & 4200  &  1.00   & 3771 $\pm$ 92  & 0.72 $\pm$ 0.39\\
298      & 7000  &  2.50   & 7500 $\pm$ 500 & 1.90 $\pm$ 0.10\\
303      & 4500  &  0.50   & 4224 $\pm$ 94  & 0.70 $\pm$ 0.36\\

\hline

\end{tabular}
\end{center}
\begin{list}{}{}
\item[]$^{*}$ Typical uncertainties are $\pm$\,200\,K and $\pm$\,0.25\,dex.
\item[]$^{**}$ The tabulated value of gravity is an upper limit.
  \end{list}
\end{table}

\begin{table*}
\caption{Comparison of the stellar parameters derived by different authors for the Cepheids.}
\begin{center}
\begin{tabular}{l|cccc|cccc}   %longtable es para poner primero landscape
\hline\hline
\multirow{2}{*}{Author}   & \multicolumn{4}{|c|}{V340 Nor} & \multicolumn{4}{c}{QZ Nor}\\
  & $T_{\textrm{eff}}$ (K)& log $g$ & [Fe/H] & [Fe/H]$^{*}$ &  $T_{\textrm{eff}}$ (K)& log $g$ & [Fe/H] & [Fe/H]$^{*}$\\
\hline

\citet{Lu94}  & 5400 & 1.35 &  0.10   &  0.32   & 6000 & 1.48 &  0.24 &  0.46\\
\citet{Fry97} & 5450 & 1.50 & $-$0.18 & $-$0.10 & 5750 & 2.00 &  0.06 &  0.14\\
\citet{Ge14}  & 5685 & 0.60 &  0.07   &  0.12   & 5765 & 1.05 &  0.19 &  0.24\\
This work     & 5776 & 0.82 &  0.09   &  0.09   & 6031 & 1.21 &  0.65 &  0.65\\

\hline

\end{tabular}

\label{Cef}
\end{center}
\begin{list}{}{}
\item[]$^{*}$ Iron abundance referred to the solar value listed in \citet{As05}.
  \end{list}
\end{table*}

\subsection{Stellar chemical abundances}

In Table~\ref{metal}, we compare the iron abundance found in the literature to our results. Three studies
estimated it on the basis of a photometric calibration (Ph) whereas \citet{Lu94} and this work derived it by directly analysing the spectra (Sp). Our value for the cluster 
metallicity, [Fe/H]$=+0.19\pm0.05$, is higher than previous results. It must be noted, however, that \citet{Lu94} used as solar reference the abundances listed in \citet{Gr84}, 
$A($Fe$)=7.67$, while in this work we have taken $A($Fe$)=7.45$ \citep{As05}. If we rescale both values to the same solar abundance \citep{As05},
the two results differs by only 0.04~dex, and are compatible within their errors. It is noteworthy that the metallicity derived in the present work by analysing evolved stars is fully consistent with the Galactic gradient calculated by \citet{Ge14} by using classical Cepheids.

\begin{table}
\caption{Metallicities estimated for NGC 6067. \label{metal}}
\begin{center}
\begin{tabular}{lccc}   %longtable es para poner primero landscape
\hline\hline
Reference & [Fe/H] & $N^{*}$ & Data\\
\hline
\citet{Cla89}&  0.10 $\pm$ 0.06   & 6 & Ph\\
\citet{Lu94}$^{**}$ &  0.01 $\pm$ 0.12   & 7 & Sp\\
\citet{Pi95} & $-$0.01 $\pm$ 0.07 & 3 & Ph\\
\citet{Tw97} &  0.14 $\pm$ 0.06   & 5 & Ph\\
This work    &  0.19 $\pm$ 0.05   & 13& Sp\\
\hline
\end{tabular}
\end{center}
\begin{list}{}{}
\item[] $^{*}$ $N$ is the number of stars used.
\item[] $^{**}$ When the same solar abundance is used \citep{As05}, [Fe/H]=0.23, a value compatible with our estimation.       
  \end{list}
\end{table}

As mentioned, only \citet{Lu94} has previously determined chemical abundances for stars in NGC~6067. In Table~\ref{Lu94} we show the mean abundances, together with their rms, 
for the stars (6) and the chemical elements (9) that we have in common with him. On average, we do not find significant differences between the abundances obtained in both studies. 
%with the exceptions of sodium and magnesium. For these elements, especially for Mg, differences are important (0.3\,--\,0.4 dex).
The largest differences, up to 0.3\,--\,0.4~dex, are found for sodium and magnesium, but the values are still compatible within the errors.

\begin{table*}
\caption{Comparison of the chemical abundances, relative to solar abundances by \citet{As05}, derived in this work with those of \citet{Lu94}. Solid circles represent the stars used to calculate the abundance of each element.
}
\begin{center}
\begin{tabular}{l|cccccc|c|c}   %longtable es para poner primero landscape
\hline\hline
\multirow{2}{*}{Element} & \multicolumn{6}{|c|}{Star} & Luck94 & This work\\
   & 261 & 275 & 276 & 303 & V340 Nor & QZ Nor & [X/H] & [X/H]\\

\hline
O  & \textbullet & \textbullet & \textbullet & \textbullet &             & \textbullet & -0.08 $\pm$ 0.41  & $-$0.10 $\pm$ 0.28\\
Na &             &             &             & \textbullet & \textbullet & \textbullet &  0.67 $\pm$ 0.17  &  0.29 $\pm$ 0.27\\
Mg & \textbullet & \textbullet & \textbullet & \textbullet & \textbullet & \textbullet &  0.34 $\pm$ 0.22  & $-$0.08 $\pm$ 0.15\\
Si & \textbullet & \textbullet & \textbullet & \textbullet & \textbullet & \textbullet &  0.29 $\pm$ 0.12  &  0.39 $\pm$ 0.14\\
Ca &             & \textbullet & \textbullet & \textbullet & \textbullet & \textbullet & $-$0.11 $\pm$ 0.30& $-$0.02 $\pm$ 0.43\\
Ti & \textbullet & \textbullet & \textbullet & \textbullet & \textbullet & \textbullet &  0.03 $\pm$ 0.35  &  0.06 $\pm$ 0.25\\
Fe & \textbullet & \textbullet & \textbullet & \textbullet & \textbullet & \textbullet &  0.25 $\pm$ 0.12  &  0.26 $\pm$ 0.24\\
Ni & \textbullet & \textbullet & \textbullet & \textbullet & \textbullet & \textbullet &  0.14 $\pm$ 0.14  &  0.23 $\pm$ 0.21\\
Y  &             &             & \textbullet & \textbullet & \textbullet & \textbullet &  0.23 $\pm$ 0.07  &  0.04 $\pm$ 0.18\\

\hline
\end{tabular}

\label{Lu94}
\end{center}
\end{table*}

\begin{figure} 
  \centering         
  \includegraphics[width=\columnwidth]{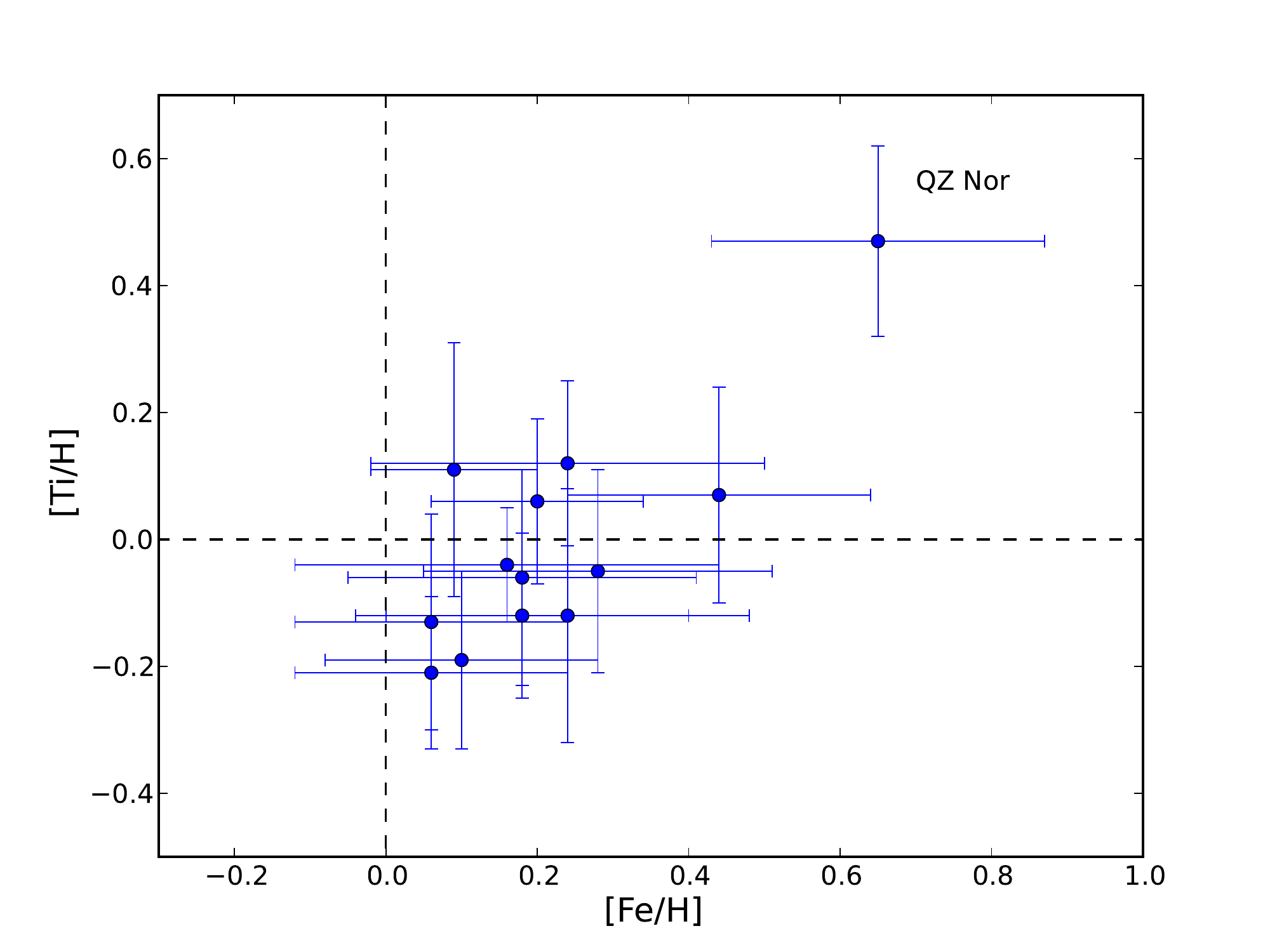}  
  \caption{Abundance ratio [Ti/H] vs [Fe/H]. All the stars, with the exception of the Cepheid QZ~Nor, have the same composition (within the errors), indicating chemical 
  homogeneity. As a reference, the dashed lines show  the solar values.\label{disp}}  
\end{figure}

NGC 6067 seems to be chemically homogeneous since the star-to-star scatter for each chemical element can be explained by the propagation of uncertainties in the determination of 
the abundances. All the stars are grouped around the average cluster value with the exception of the Cepheid QZ Nor that appears shifted to very high [Fe/H] (see Fig.~\ref{disp}). Although the individual errors are high, all the elements analysed show the same behaviour. At present, we cannot offer an explanation for these high abundances, but
note that the same trend is seen in the analysis of \citet{Lu94}, who observed it with an F6\,Iab spectral type. The only other star that presents very high abundances for a few 
elements is 261, the brightest red star. This again points to a failure of the models to reproduce the most luminous stars.

%We performed a least-squares fit ([O/Fe] vs [Na/Fe]) obtaining a correlation coefficient, $R^2<0.1$, thus discarding the Na-O anti-correlation. This trend, present in all globular clusters, is 
%not seen in young open clusters \citep{NaO,NaO2}.

We derived a roughly subsolar [Y/Fe] against a supersolar [Ba/Fe], which is in good agreement with the dependence on age and Galactic location found by \citet{Mi13} by comparing the
abundances of Y and Ba in different open clusters. In addition, this over-abundance of barium supports the enhanced ``$s$-process'' suggested by \citet{Ba} to explain the enrichment
of barium observed in young open clusters, in apparent conflict with the standard model (see their fig.~2).

\begin{figure*}  %usando el * ocupa las dos columnas, también para tablas / puedo usar para localizar la figura t(op), b(otton) o h(ere)  
  \centering         
%el pdflatex no compila .eps, pero sí funciona si lo compilo a mano acabo pasando epstopdf
  \includegraphics[width=19cm]{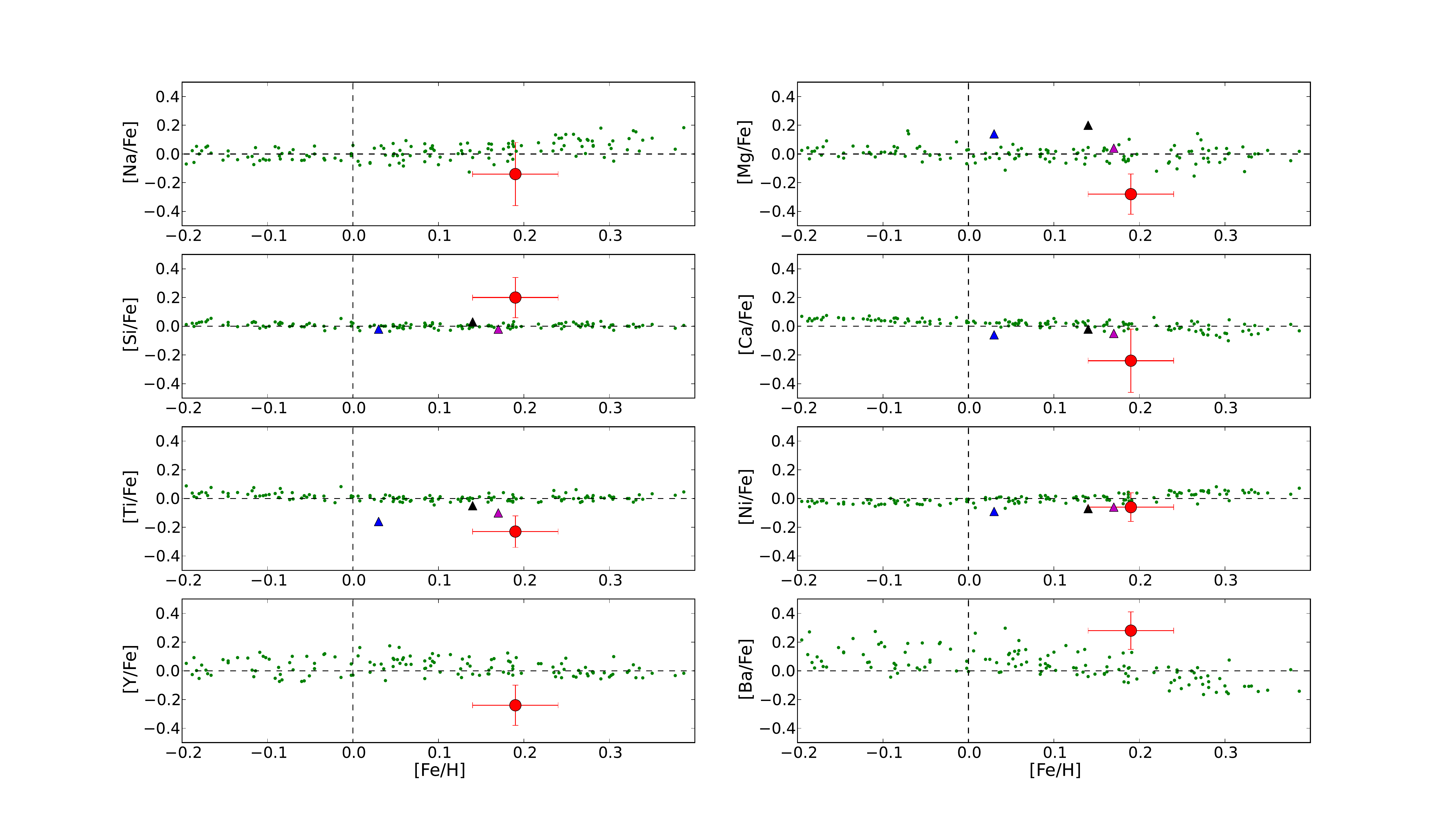}   %posición {h} me da problemas
  \caption{Abundance ratios [X/Fe] versus [Fe/H]. The green dots represent the galactic trends for the thin disc \citep{Jon10,Jon13}. The
  much higher quality of their observational data ($R=110\,000$ and $S/N\approx850$ on average) is reflected in the low scatter. NGC~6067 is the red circle whereas other open clusters
  studied in the $Gaia$-ESO Survey \citep{GES_comp} are represented by coloured triangles: M11 (black), Tr20 (magenta), and NGC 4815 (blue). Clusters are represented by their 
  mean values. The error bars for NGC 6067 show the standard deviation. The dashed lines indicates the solar value.}
  \label{trends}  % ref para cuando la nombre, da igual lo que ponga, el programa las ordena fig1, fig2, ... pero es útil por si cambio la posición
\end{figure*}

Finally, in Fig.~\ref{trends}, we compare our abundances with the Galactic trends ([X/Fe] vs [Fe/H]) obtained by \citet{Jon10,Jon13} for each element. For Na and Ni the relative abundance
found is solar; for Si and Ba, it is supersolar; and for the rest of elements (Mg, Ca, Ti, and Y) we found a subsolar value. In all cases our results are compatible, within errors, 
with those expected for a cluster with this age and position in the Galaxy.

\subsubsection{Lithium}

According to canonical models \citep{ib67,ib67_2,sod93}, lithium (Li) depletion in giant stars is a natural consequence of stellar evolution. 
%Once stars leave the main sequence, theyexpand and cool becoming red giants and increase the depth of the convective envelope. 
During the first dredge-up this external envelope reaches internal regions removing the inner
material outwards. One consequence is the overall dilution of Li, decreasing its abundance down to an upper limit of A(Li)=1.5 dex \citep{Ch00}. In good agreement with this scenario, 
Li is not found, or just in a small amount, in most of the stars studied. Only stars 303, 247, and especially 276 show a supersolar abundance. In particular, star 276 shows
a high lithium abundance, $A($Li$)=2.41$, which represents a value 23 times greater than solar. \citet{Lu94} gives Li abundances for 276 and 303. For the latter, his rescaled abundance, $A($Li$)=1.12$, is almost identical to our value (1.15), whereas for star 276 his abundance, $A($Li$)=1.30$, is significantly smaller than ours\footnote{This difference could be explained taking into account the possibility of a misidentification of stars 276 (Li-rich) and 1294 (Li-normal), given their proximity (firstly noted as a binary system by \citetalias{Th62}) and similarity in spectral type in previous works (K3\,Ib and K3\,II respectively).} (2.41). Interestingly, the Li-rich star 276 has the latest spectral type and is the coolest 
star in the cluster (star 261 is equally cool within the errors). It is also the star with the highest Rb abundance in the cluster, although it shows a subsolar value.
%Star 276 is not an outlier, but one more Li-rich giant. 
These unexpected stars, around one per cent of all the giants studied, have been discovered in isolation \citep{Br89}, as well
as within open clusters \citep{De16}. Different internal and external scenarios have been proposed to explain this overabundace. The \citet{CFM} mechanism (CFM) can produce 
Li via the hot botton burning \citep{hbb} in intermediate-mass stars during the AGB phase as well as during the RGB phase in low-mass stars \citep{sac99}, although it requires an 
extra mixing process. Other mechanisms proposed to explain the presence of extra Li are the engulfment of a sub-stellar companion \citep{Si99,Ag16} or interstellar medium enrichment caused by a SN explosion \citep{Woo95}. As the cluster chemical composition (and that of star 276) is on the galactic average, the SN explosion scenario can be discarded. Given its very low abundance of $s$-process elements, such as Rb, and its position on the CMD, 276 should not be an AGB star, and so the CFM is also rejected. After discarding all other possibilities, the engulfment of a planet or a brown dwarf by the star could be a plausible explanation, but our data cannot confirm or discard it. To shed some light on this issue, we should 
determine the $^6$Li/$^7$Li ratio, for which a spectrum with rather higher resolution is needed. 

\begin{figure}  %usando el * ocupa las dos columnas, también para tablas / puedo usar para localizar la figura t(op), b(otton) o h(ere)  
  \centering         
%el pdflatex no compila .eps, pero sí funciona si lo compilo a mano acabo pasando epstopdf
  \includegraphics[width=\columnwidth]{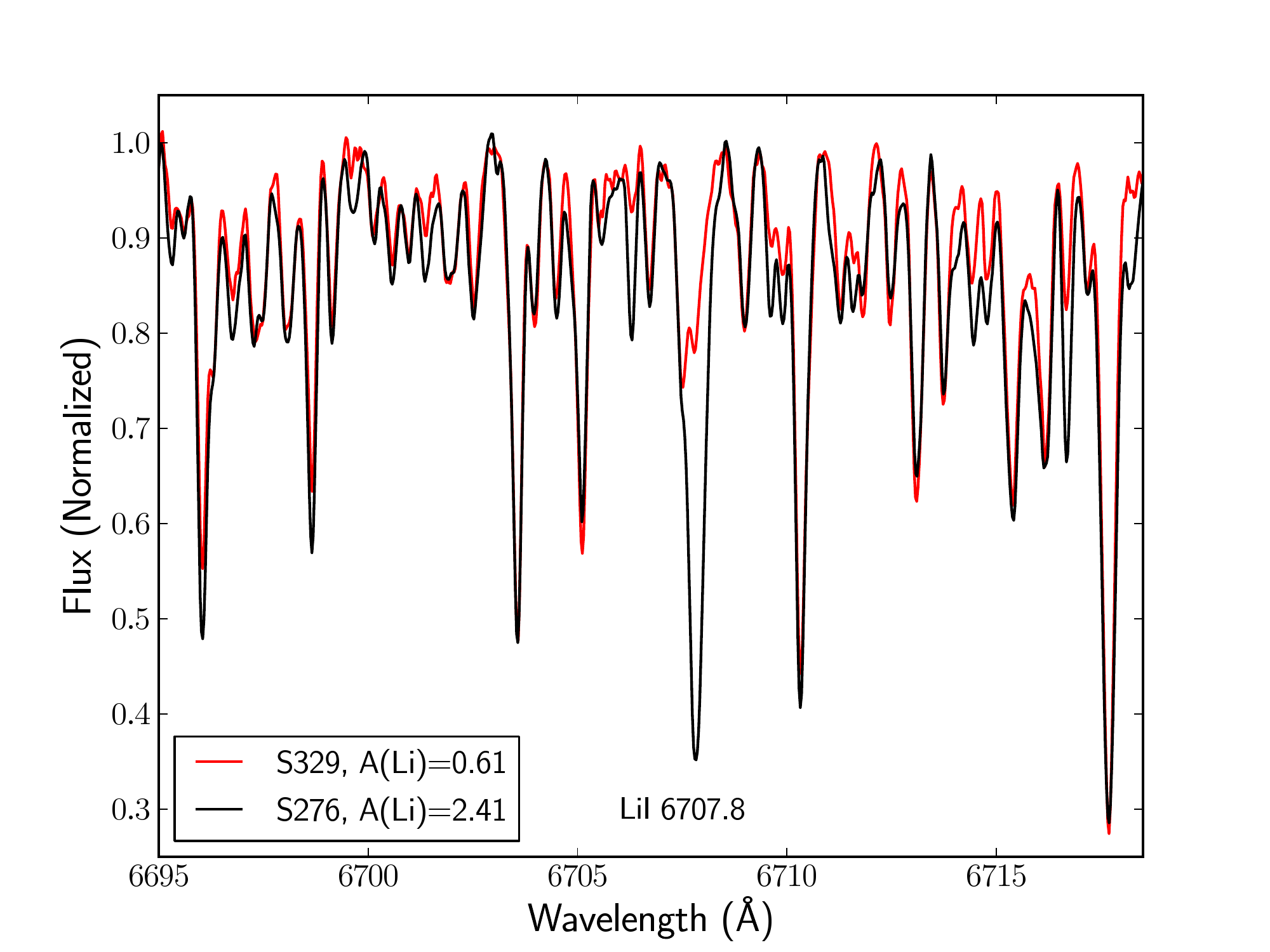}   %posición {h} me da problemas
  \caption{Spectra around Li line at 6707 \AA{}. Li-rich giant, star 276, is represented versus star 329, a Li-normal giant with similar stellar parameters.}
  \label{litio}  % ref para cuando la nombre, da igual lo que ponga, el programa las ordena fig1, fig2, ... pero es útil por si cambio la posición
\end{figure}

\subsection{Cepheids}

NGC 6067
%, as has been mentioned throughout the paper, 
hosts two well-studied classical Cepheids: V340~Nor and QZ~Nor. Cepheids in clusters are useful calibrators and thus several recent papers have analysed in detail membership criteria \citep{Ma13,An13,Chen15}. All of them confirm that both Cepheids belong to NGC\,6067, based on many parameters: radial velocities, Wesenheit distances and radial velocity-distance gradient in \citet{Ma13}; spatial, kinematic and population-specific membership constraints in \citet{An13}.

Both Cepheids present quite similar spectra, appearing as early-G supergiants in our data. They have different
pulsation periods ($P$) inferred from well-studied light curves \citep{lan94}: V340~Nor is a fundamental mode pulsator ($P=11.30$~d), whereas QZ~Nor is an overtone mode pulsator ($P=3.78$~d). 
According to these periods, \citet{Ma13}, by using different 
period-age relations \citep{efremov,bono,turner},
estimated the age of the  Cepheids (see their table~1). The mean age of QZ~Nor ($89\pm13$~Ma) is in good agreement with that obtained in this work for the cluster. Contrariwise, V340~Nor (with 
an average $52\pm10$~Ma) appears younger than the cluster itself. However, as recently shown by \citet{An16}, ages derived from pulsation periods are subject to large uncertainties, 
sometimes close to 100 per cent, due to a number of physical effects. Among them, the initial rotation of the star has a strong impact on its observed properties as a Cepheid. The relation 
between initial rotation and pulsation period is not monotonic, and the strongest effects are seen for initial rotational velocities close to those typically observed in this mass range 
\citep[the ``average rotation'' case in][]{An16}. Using their period relations for this case and solar metallicity, we would obtain ages of 71 and $110\,$Ma for V340~Nor and QZ~Nor respectively. 
These ages are compatible with that of the cluster, within the errors estimated in this work.
%: the age of V340\,Nor is just at the lower limit whereas QZ\,Nor is at the upper one.} 
However, the two Cepheids occupy different positions in the CMDs (Figs.~\ref{BV} and ~\ref{JHK}) and the Kiel diagram (Fig.~\ref{pHR}). 
%In all of them V340~Nor sits well above the corresponding isochrone. 

By using the Period-Mass relations from \citet{and14} and \citet{An16}, we derive masses for the Cepheids: $6.1\pm0.5\,$M$_{\sun}$ for V340~Nor and $5.3\pm0.4\,$M$_{\sun}$ for QZ~Nor (having taken 
into account that it is an overtone pulsator). According to the isochrone, the most evolved stars have masses very slightly below $6\,$M$_{\sun}$. Thus, the masses of the Cepheids, estimated with 
calibrations from models including the effect of rotation, are compatible with those expected for the cluster members. Unfortunately, the period evolution, which could greatly constrain their 
evolutionary stage, cannot be determined yet for these two Cepheids \citep{Ma13}.

Our abundance analysis results in a chemical composition for V340\,Nor fully compatible with the cluster average. However, QZ~Nor is much more metal-rich, a result also found by \citet{Lu94}. Given that all existing analyses confirm membership for QZ~Nor, we can offer no explanation for this discrepancy. We do not expect stars with such a high metallicity at this Galactocentric distance, while the observed properties of QZ~Nor ($P$, reddening, etc.) are incompatible with the much higher distance at which this metallicity is expected according to the Galactic gradient. Perhaps an unresolved binary companion could lead to artificially high abundances. Accurate \textit{Gaia} distances will soon definitely resolve this issue.

\subsubsection{Cepheids/RG ratio}\label{sec:Cepheids}

Cepheids are yellow supergiants in the mass range 3\,--\,$10\:$M$_{\sun}$. These intermediate-mass stars evolve from the main sequence, where they are B-type dwarfs, and cross the instability strip (IS), where they are subject to pulsation and become Cepheid variables, three times. The first crossing is during the H-shell burning (before the first dredge-up) and very short \citep[for a typical $\sim5\:$M$_{\sun}$ Cepheid, it lasts less than one per cent of the lifetime in the Cepheid phase;][]{and14}. They cross twice more, during He-core burning, as they trace the blue loop. 

Comparing the observed number of stars in two regions of the HR diagram, red giants (RG) and Cepheids (Cep),  provides a very stringent test on different stellar evolution models. Based on Geneva
stellar models at solar metallicity \citep{Eks12}, we calculated the theoretical ratio $t_{\textrm{loop}}/t_{\textrm{red}}$, i.e.\ the time spent by the star in the
blue loop versus that in the RG branch. For this purpose, we drew a line parallel to the Hayashi limit, shifted blueward by $\log\,\Delta\,T_{\textrm{eff}}=0.045$. This 
value was chosen in order to retain all the stars in the red clump (equivalent, in Fig.~\ref{JHK}, to $(J-K_{\textrm S})_0\approx0.6$). The lifetime to the right of this line is considered as $t_{\textrm{red}}$, while the time spent to its left is adopted as $t_{\textrm{loop}}$. In order to avoid an overlap, we considered a small gap of $\log\,\Delta\,T_{\textrm{eff}}=0.005$ between both regions. With these definitions, for a $6\:$M$_{\sun}$ star we find a $t_{\textrm{loop}}/t_{\textrm{red}}=0.60$ (models with 
no rotation) or 0.66 (for an initial rotation rate, $V$/$V_{\textrm{crit}}=0.4$) in good agreement with the first case from \citet{Matraka82}. They calculated $t_{\textrm{loop}}/t_{\textrm{red}}=0.64$ 
(with overshooting) and 1.64 (no overshooting). As shown by \citet{An16}, introducing the effects of initial rotation does not significantly affect the extent of the instability strip, but impacts on 
the luminosity of stars and the extent of the blue loops. The numbers observed in NGC\,6067, 2 Ceps per 12 RGs, i.e.\ a ratio of 0.17, differ significantly from this prediction. To check this discrepancy, we searched for similar clusters in order to estimate this ratio in a sufficiently significant sample.

In the whole Milky Way, besides NGC~6067, there are only two other open clusters containing more than one Cepheid\footnote{Recently, \citet{Chen15} have claimed 
membership of the Cepheids CN~Sct and TY Sct in the open cluster Dolidze~34. Unfortunately, this cluster is too poorly characterised to provide any meaningful analysis of its population}. 
The first one is NGC~7790 that hosts three:  CEab~Cas (a binary system in which both components are Cepheids) and CF~Cas \citep{kr58}. The second cluster is NGC~129. Traditionally it 
was thought to contain only one Cepheid, DL~Cas \citep{Kra57}, but recently \citet{An13} suggested the membership of V379~Cas, confirmed by \citet{Chen15}. This addition creates an 
interesting similarity to NGC~6067: DL~Cas is a fundamental pulsator located in the cluster core, whereas V379~Cas is an overtone pulsator in the halo. These three clusters have roughly the 
same age, but present very different Cepheids/RG ratios.

There are a few other clusters containing one Cepheid at a similar age. Photometric data available in the WEBDA database were used to derive their age (as done in 
Sect.~\ref{sec:isochrones}). The number of confirmed evolved members was taken from \citet{Me08}. In Table~\ref{ratio}, we show the ratio of the number of Cepheids 
to the number of RGs in all these clusters. This sample, however, is obviously biased, as the clusters have been selected precisely because they contain Cepheids. 
To obtain a more representative sample, we have taken all the clusters in the age range 50\,--\,$150\:$Ma present in the sample studied by \citet{Me08}, using for this task the ages 
provided by the WEBDA database. This latter list provides a more suitable sample of clusters containing evolved members for this comparison. We display in Table~\ref{RG_OC} the number 
of red giants present in every cluster. At solar metallicity, tracks for moderately massive stars ($7\:$M$_{\sun}$ and above) present the hot side of the loop at warmer temperatures than 
the blue edge of the IS. This means that the number of F-type (super)giants may be higher than the number of Cepheids. \citet{Me08} do not separate yellow and red (super)giants, but we have marked in Table~\ref{RG_OC} all the cluster members whose published spectral types and colours identify as yellow (super)giants.

In total, in this age range we find 53 clusters containing 23 yellow (super)giants, not only Cepheids, and 112 RGs. The overall ratio, i.e. YSGs/RGs\,=\,0.21, is smaller than one, 
in disagreement with the ratios calculated above for solar metallicity. 
This ratio is indeed very sensitive to the extension of the blue loop, which not only depends on the amount of overshooting but is also very sensitive to metallicity and 
input physics \citep{Matraka82,Eks12}. %In this sense, NGC~6067 is especially important because its very high population of evolved stars 
%strengthens the statistical significance of its ratio, the smallest among Milky Way clusters with Cepheids, but more similar to the overall ratio.

\begin{figure}  %usando el * ocupa las dos columnas, también para tablas / puedo usar para localizar la figura t(op), b(otton) o h(ere)  
  \centering         
%el pdflatex no compila .eps, pero sí funciona si lo compilo a mano acabo pasando epstopdf
  \includegraphics[width=\columnwidth]{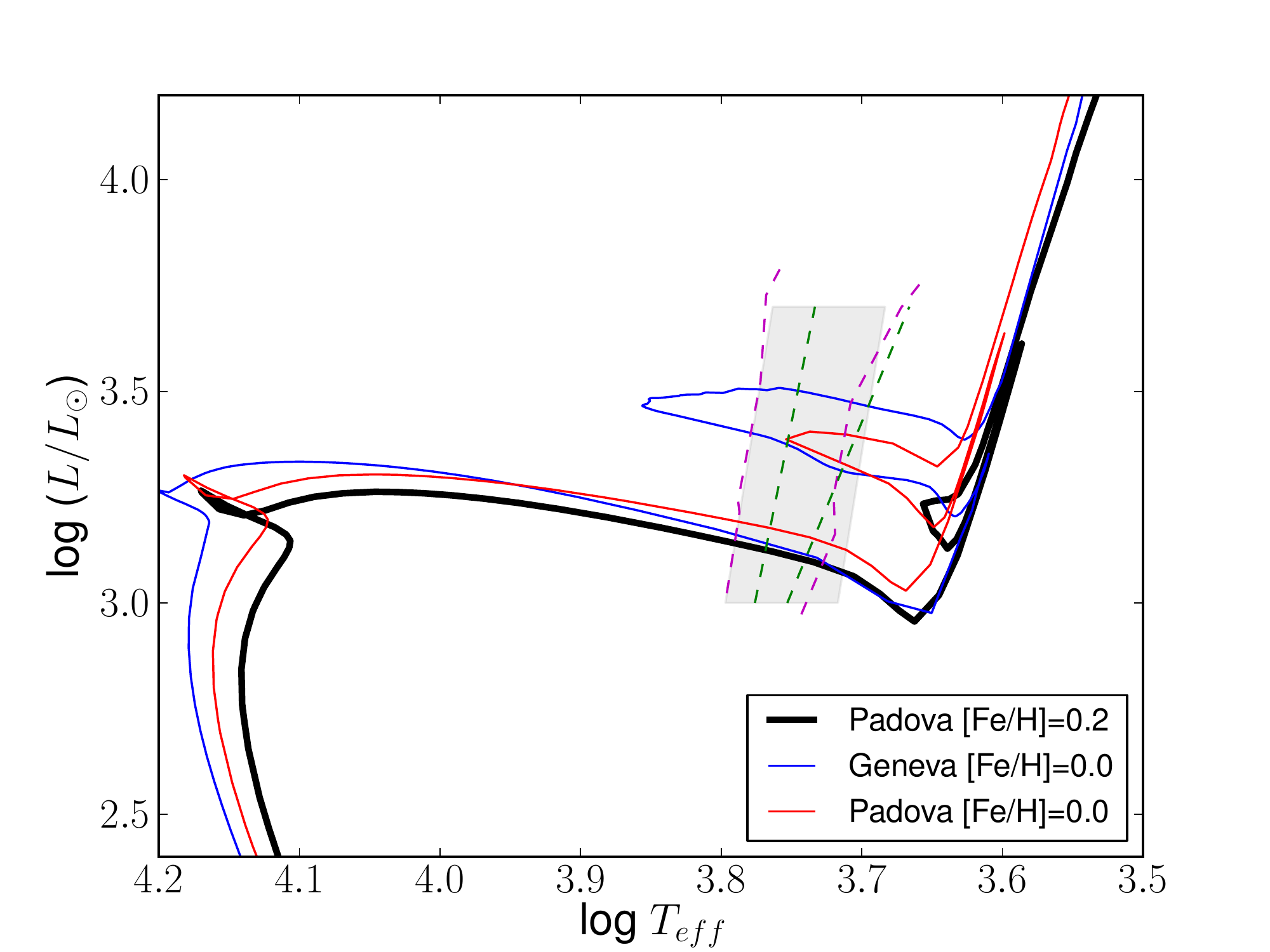}   %posición {h} me da problemas
  \caption{HR diagram from Padova (log $\tau$=7.95) with different metallicities: solar (black) and supersolar (red) and Geneva solar (blue) isochrones. The regions for the
  IS after \citet{Bo00}, green dashed lines, \citet{An16}, magenta lines, and \citet{Tam03}, grey shaded area, are shown.}
  \label{iso795} % ref para cuando la nombre, da igual lo que ponga, el programa las ordena fig1, fig2, ... pero es útil por si cambio la posición
\end{figure}

Since increased metallicity  decreases looping \citep{Eks12,Wal15}, in the supersolar Padova isochrones He-burning happens away from the IS and, therefore, stars do not 
reach the locus of Cepheids. At the metallicity that we find for NGC~6067 (see Fig.~\ref{iso795}) no Cepheids are predicted at this age. Indeed at this metallicity, 
only stars in the $\approx8$\,--\,$10\:$M$_{\sun}$ range have pronounced blue loops, in contradiction with the masses that we find for our Cepheids. Moreover, by $Z=0.45$, 
the loops are suppressed at all masses. However, \citet{Ge15} find a high number of Cepheids with metallicities $>0.4$. All these high-metallicity Cepheids have $P>10$~d and
are thus relatively massive, but this could be a selection effect, because only the most luminous Cepheids can be observed towards the inner regions of the Galaxy 
(at high distance and beyond moderate extinction), where these high metallicities are found. Therefore, we conclude that the supersolar tracks suppress the Cepheid loops 
more efficiently than the observations suggest. Unfortunately, there are no supersolar Geneva models available yet for a proper comparison to our data. Both sets of isochrones,
Padova \citep{Gi00,Marigo} and Geneva \citep{Eks12,Ge13}, are quite different regarding the solar composition (Y=0.273, Z=0.019 and Y=0.266, Z=0.014, respectively) or 
the input physics: Padova models include more overshooting (in terms of $d_{\textrm{over}}/H\textrm{p}$) than Geneva models (0.25 vs. 0.10) and do not take into account rotation, as Geneva models do. 
For comparison, in Fig.~\ref{iso795} we also include a Geneva isochrone at solar metallicity (as this is the highest metallicity available in Geneva models) and an average rotation
($\omega=\Omega/\Omega_{\textrm{crit}}=0.5$). We see that the extension of the blue loop in the Geneva model is larger than in the Padova one.
Moreover, \citet{An16} find that the extent of blue loops also decreases with increasing metallicity for the Geneva models. It is thus quite possible that supersolar Geneva isochrones, could still reach the IS at the cluster metallicity, since they seem to have wider loops than Padova models at all metallicities. 

On the other hand, the fact that NGC~6067 has the lowest Cepheid/RG ratio of all the clusters listed in Table~\ref{ratio} suggests that high metallicity really inhibits the 
formation of Cepheids. The two clusters in the Perseus arm, which are expected to have the lowest metallicity in the whole Galactic sample, have ratios 3/0 and 2/2, in contrast 
with the clusters towards the inside of the Galaxy, which typically have more RGs than Cepheids. The LMC young globular cluster NGC~1866, which has a metallicity around 
$0.5\;$Z$_{\sun}$ has about the same number of blue and red giants, including 24 Cepheids \citep{Mu16}. Although it has an age similar to NGC~6067 (around 100~Ma), NGC 1866 is 
much richer \citep[$\approx$\,5\,x\,10$^4$ M$_{\sun}$;][]{Muc11}, and again provides a statistically significant sample on its own.

\begin{table*}
\caption{Cepheid/RG ratio in several young open clusters. \label{ratio}}
\begin{center}
\begin{tabular}{lcccccccc}   %longtable es para poner primero landscape
\hline\hline
Cluster & $\log\,\tau$ & $R_{\textrm{GC}}^{*}$ & Cep  & RG &  Ratio & Cepheid & Period (d)$^{**}$ & Reference\\
\hline

NGC 0129 & 7.90 & 9.0  & 2 (+1) & 0  & --- & DL Cas, V379 Cas & 8.00, 4.31 & a\\
NGC 5662 & 7.95 & 7.5  & 1 & 2  & 0.50 & V Cen & 5.50 & b,c \\
NGC 6067 & 7.95 & 6.5  & 2 & 12 & 0.17 & V340 Nor, QZ Nor & 11.30, 3.78 & d\\
NGC 6087 & 8.00 & 7.3  & 1 & 1  & 1.00 & S Nor & 9.75 & e\\
NGC 6649 & 7.90 & 6.4  & 1 (+1)$^{***}$ & 3  & 0.33 & V367 Sct & 6.29 & f\\
NGC 6664 & 7.70 & 5.2  & 1 & 4  & 0.25 & EV Sct & 3.09 & g,h \\
NGC 7790 & 7.90 & 10.0 & 3 & 0  & ---  & CEab Cas, CF Cas & 5.14, 4.48, 4.88 & i\\
IC 4725  & 7.97 & 7.40 & 1 & 2  & 0.50 & U Sgr & 6.75 & e \\
Trumpler 35 & 7.86&6.96& 1 & 2  & 0.50 & RU Sct & 19.72 & b,c\\
vdBergh 1& 8.03 & 9.52 & 1 & 0  & --- & CV Mon & 5.38 & h \\
\hline
\end{tabular}
\end{center}
\begin{list}{}{}
\item[]$^{*}$ A value of $R_{\sun}$\,=\,8.0~kpc is used.
\item[]$^{**}$ The values have been taken from \citet{An13}.
\item[]\textbf{Ref}: \textbf{a}: \citet{vdb57}, \textbf{b}: \citet{Tsa66}, \textbf{c}: \citet{Tu82}, \textbf{d}: \citet{Eg83}, \textbf{e}: \citet{Ir55},
\textbf{f}: \citet{Ro63}, \textbf{g}: \citet{Kr57}, \textbf{h}: \citet{vdb57}, \textbf{i}: \citet{Sa58}. Here are cited the authors who suggested for the first time the cluster membership for each Cepheid.
The reader will find a more complete set of references in \citet{An13}.
\item[]$^{***}$ NGC 6649 contains a second F (super) giant that is not a Cepheid.
  \end{list}
\end{table*}

%___________________________________________________________________________________________________________________________________________________________________________

%___________________________________________________________________________________________________________________________________________________________________________

\subsection{Blue Straggler candidates}

BSSs are stars that seem much younger than the rest of the cluster population. They lie above and bluewards of the turn-off point in the CMD, where no stars are expected on the 
basis of standard stellar evolution, under the assumption that all cluster stars are coeval. Several channels have been suggested to produce these objects, but those most accepted are: 
(i) the merger of two stars induced by stellar collision and (ii) coalescence or mass-transfer between the companions in a binary system \citep{Perets}. The first mechanism is 
only believed to work effectively at very high stellar densities, never reached in Galactic open clusters.
 
In this work we found three likely BSSs. \citet{Ah95} gave a list of seven BSS candidates in this cluster. When we examine spectral types and position on the CMDs (see Figs.~\ref{BV} and~\ref{JHK}), Kiel (Fig.~\ref{pHR}) 
and HR diagrams (Fig.~\ref{luminos}) only two of their objects can be considered good BSS candidates: 267 and 264. Their five other candidates are normal B-type stars on the main sequence. Star 264, with spectral 
type B4\,V, is only a weak candidate for a BSS, slightly to the left of the MSTO, a location that can be easily explained if, for example, we assume that the star was initially a faster rotator than the average, by comparing 
isochrones from \citet{Ge13} with different rotation rates. Star 267 is a much more obvious BSSc, much earlier and brighter than all the other blue cluster members, and had already been proposed as such by 
\citet{Me82} and, more recently, \citet{DeMa06}. The third BSSc, nominated here for the first time, is HD\,145304. It is very similar to star 267, both in spectral type and 
parameters, but lies far away ($\approx9\arcmin$) from the cluster core, and thus its membership had not been previously noticed. The spectral types and positions in the CMD of these
two stars suggest that they have $\approx10\:$M$_{\sun}$.

The chemical composition of these two stars can give us some hints as to the most likely mechanism to have generated them. The collisional channel predicts no chemical signatures \citep{Lo95}, whereas 
the mass transfer mechanism may be accompanied by anomalies in the CNO abundances, because of the mixing with material coming from stellar regions where CNO burning occurs \citep{Sa96}. According to
this and after consulting the abundances obtained for stars 267 and HD\,145304 (see Table~\ref{abund_cal}), we infer that these BSSs probably have a mass transfer origin. They present
low abundances of carbon and, at the same time, a strongly enhanced nitrogen. In fact their N/C ratios are 61.7 and 3.9 respectively, much higher than that expected for non-evolved B stars, around 0.3
\citep{nieva}. The latter abundance could also be caused by strong rotational mixing in a very fast rotator, but, according to the
tracks by \citet{Ge13}, fast rotation alone cannot explain the brightness of these stars or their high temperatures, especially in the case of star 267.

In addition to these objects, which fulfill the conditions to be considered good BSS candidates (a blueward position in the CMD, an evolved chemical composition and an earlier spectral type 
than MSTO), there are two other stars in the cluster whose spectral types and luminosities suggest that they
are more massive than their peers.  One of them is the A-type supergiant HD\;145324. The parameters of this object are quite similar to those of star 298, which
falls right on the cluster isochrone, but is more than one magnitude fainter (see Fig.~\ref{JHK}). HD\;145324 lies more than $8\arcmin$ away from the cluster centre, but its radial velocity confirms that
it as a halo member. The other object is the K supergiant 261. It is the brightest star in the infrared, and the only red star that is decidedly a morphological supergiant, and it 
is located at the cluster core.

These objects are likely to be the result of the evolution of BSSs because the difference in brightness with other members 
of similar colours and spectral type seem too large to be explained by rotation alone. According to \citet{Me82} or \citet{DeMink14} these mass gainers
are expected to be among the brightest stars in young clusters, often at the cluster core where the stellar density is higher.

%__________________________________________________________________________________________________________________________________________________________________________
\section{Conclusions}\label{sec:5}

We have performed the most complete analysis to date of NGC~6067, by combining archival photometry (optical and 2MASS) and high-resolution spectroscopy.
Many of the stars forming our sample have been observed in this work for the first time. We provide spectral classifications and parameters for them. We also notice the 
presence of several SB2, Be stars and BSSs.  We derive stellar atmospheric parameters for over 40 cluster members, using tailored models for stars across the whole HR 
diagram. We find that NGC~6067 is located at a distance of $\approx1.8$~kpc, and has low reddening, $E(B-V)$\,=\,0.35~mag.

We confirm the cluster age derived from isochrone fitting, around $90\:$Ma, using two different approaches: the spectral type around the MSTO (B6\,V) and the theoretical HR 
diagram, built from our stellar parameters. This age is compatible with the parameters of the two classical Cepheids contained in the cluster. There are two obvious BSS candidates, one of which, star
HD\;145304 was found for the first time in this work. There are two other members (stars 261 and HD\;145324) that seem too bright for the cluster age, most likely testimony to
the consequences of mass transfer in interacting binaries.

From star counts, we estimate a present cluster mass in the range 4\,000\,--\,5\,000$\:$M$_{\sun}$, compatible with the virial mass, which corresponds to an initial mass up to 7\,500$\:$M$_{\sun}$, 
comparable to that of the populous cluster M11. NGC~6067 is the most massive Galactic cluster in the 50\,--\,150~Ma range that has been characterised and, 
in consequence, contains the largest population of evolved massive intermediate-mass stars (around $6\:$M$_{\sun}$) in the Milky Way. From their analysis, we find a supersolar metallicity, 
with an average value of [Fe/H]\,=\,+0.19 dex.  We also calculated abundances of O, Li, Na, some $\alpha$-elements (Mg, Si, Ca, and Ti), Ni (Fe-group element) 
and some $s$-elements (Rb, Y, and Ba) for a sample of thirteen cool stars.  We deduce a homogeneous chemical composition for NGC~6067, compatible with abundance gradients in the Milky Way.
We identify a Li-rich star (276, with A(Li)\,=\,2.41), which also happens to be the coolest cluster member and have the highest,
 although still subsolar, Rb abundance. Our results support the enhanced ``$s$-process'' model because of the over-abundance of barium. 

 The parameters derived by the {\sc fastwind} analysis for the hot stars agree very well with the fits to the optical and infrared isochrones, which manage to fit very well the position of the MSTO and
 the red giants. The parameters derived for the cool stars reproduce the temperatures very well, but there are significant discrepancies between the luminosities inferred from  photometry and isochrone fitting 
 and those spectroscopically derived from effective gravities. In addition to this effect, the location of the Cepheids, is not well matched. Supersolar metallicity Padova models do not predict the existence 
 of Cepheids at this age. The low ratio of Cepheids to red giants in NGC~6067 when compared with clusters in lower-metallicity environments gives support to a strong dependency of the blue loop characteristics
 on metal content. It seems, however, that current models overestimate this effect.

%________________________________________________________________________________________________________________________________________________________________________________
\section*{Acknowledgements}

We thank the referee for many valuable comments and suggestions that have resulted in a clear improvement of this paper. This research is partially supported by the Spanish Government Ministerio de Econom\'{\i}a 
y Competitividad under grants BES 2013-065384 and AYA2015-68012-C2-2-P (MINECO/FEDER). JAS thanks John Pritchard for his advice on installing and using the {\scshape feros-drs} pipeline. AM acknowledges support
from the Ministerio de Educaci\'{o}n, Cultura y Deporte through grant PRX15/00030. This research has made use of the Simbad database, operated at CDS, Strasbourg (France). This publication also made use of data
products from the Two Micron All Sky Survey, which is a joint project of the University of Massachusetts and the Infrared Processing and Analysis Center/California Institute of 
Technology, funded by the National Aeronautics and Space Administration and the National Science Foundation.

%%%%%%%%%%%%%%%%%%%%%%%%%%%%%%%%%%%%%%%%%%%%%%%%%%



\bibliographystyle{aa}
\bibliography{ref} % if your bibtex file is called example.bib

%\begin{thebibliography}{99}
%\end{thebibliography}

%%%%%%%%%%%%%%%%%%%%%%%%%%%%%%%%%%%%%%%%%%%%%%%%%%

%%%%%%%%%%%%%%%%% APPENDICES %%%%%%%%%%%%%%%%%%%%%
\appendix

\section{Additional tables}

\begin{landscape}

\begin{table*}
\caption{Photometric magnitudes of the stars observed spectroscopically. As mentioned in Sect.~\ref{sec:2.2}, optical values are taken from \citet{An07}. The typical 
photometric error is 0.025 mag. $JHK_{\textrm{S}}$ magnitudes are selected from the 2MASS catalogue. \label{Ph}}
\begin{center}
\begin{tabular}{lccccccc}   %longtable es para poner primero landscape
\hline\hline
NGC 6067 & CPD   &  Spectral type  & $B$ & $V$ & $J$ & $H$ & $K_{\textrm{S}}$\\
\hline
 HD\;145139 &  CPD~$-$53 7270  & B9.5\,III	       &	&	 &  9.313$\pm$0.021 &  9.265$\pm$0.023 &  9.216$\pm$0.022\\
 229	    &  CPD~$-$53 7288  & K0\,III	       &	&	 &  6.222$\pm$0.026 &  5.561$\pm$0.027 &  5.371$\pm$0.018\\
 240	    &  CPD~$-$53 7308  & K2\,II 	       & 11.690 & 9.990  &  7.121$\pm$0.020 &  6.372$\pm$0.042 &  6.161$\pm$0.027\\
 244	    &  CPD~$-$53 7317  & B7\,III-IV	       &	&	 &  9.542$\pm$0.021 &  9.417$\pm$0.023 &  9.402$\pm$0.022\\
 247	    &  CPD~$-$53 7324  & G8\,Ib-II	       & 11.620 & 10.080 &  7.367$\pm$0.020 &  6.739$\pm$0.026 &  6.544$\pm$0.017\\
 254	    &  CPD~$-$53 7334  & B7\,IV 	       & 10.960 & 10.700 & 10.127$\pm$0.023 & 10.006$\pm$0.022 & 10.050$\pm$0.025\\
 257	    &  CPD~$-$53 7339  & B6\,IV 	       &	&	 & 10.482$\pm$0.023 & 10.398$\pm$0.025 & 10.412$\pm$0.026\\
 260	    &  CPD~$-$53 7343  & B6\,IV 	       & 10.660 & 10.420 &  9.959$\pm$0.024 &  9.854$\pm$0.025 &  9.893$\pm$0.028\\
 261	    &  CPD~$-$53 7344  & K2\,Iab-Ib	       & 11.792 & 10.737 &  5.728$\pm$0.021 &  4.998$\pm$0.023 &  4.737$\pm$0.023\\
 264	    &  CPD~$-$53 7347  & B4\,V  	       & 10.470 & 10.340 & 10.020$\pm$0.023 &  9.984$\pm$0.022 & 10.010$\pm$0.023\\
 HD\;145304 &  CPD~$-$53 7350  & B2\,III	       &	&	 &  8.500$\pm$0.021 &  8.454$\pm$0.023 &  8.472$\pm$0.026\\
 267	    &  CPD~$-$53 7353  & BN2.5\,III	       & 9.202  & 9.027  &  8.608$\pm$0.020 &  8.557$\pm$0.029 &  8.531$\pm$0.022\\
 271	    &  CPD~$-$53 7360  & B8\,III + B8\,V       & 10.851 & 10.631 & 10.056$\pm$0.023 &  9.972$\pm$0.023 &  9.957$\pm$0.023\\
 272	    &  CPD~$-$53 7361  & B8\,IV\,shell         & 11.300 & 11.090 & 10.342$\pm$0.023 & 10.222$\pm$0.023 & 10.199$\pm$0.023\\
 273	    &  CPD~$-$53 7362  & B7\,IV 	       & 11.469 & 11.254 & 10.746$\pm$0.023 & 10.694$\pm$0.026 & 10.690$\pm$0.028\\
 274	    &  CPD~$-$53 7363  & B8\,III	       & 11.192 & 10.927 & 10.296$\pm$0.027 & 10.216$\pm$0.030 & 10.240$\pm$0.032\\
 275	    &  CPD~$-$53 7364  & K3\,Ib-II	       & 10.959 & 9.134  &  5.896$\pm$0.021 &  5.084$\pm$0.021 &  4.823$\pm$0.024\\
 276$^{*}$  &  CPD~$-$53 7366N & K4\,II 	       & 11.560 & 9.690  &  6.080$\pm$0.019 &  5.251$\pm$0.018 &  4.974$\pm$0.018\\
 1294	    &  CPD~$-$53 7366S & K2\,II 	       &	&	 &  6.370$\pm$0.079 &  5.619$\pm$0.036 &  5.440$\pm$0.024\\
 277	    &  CPD~$-$53 7370  & B9\,III\,Si	       & 11.277 & 11.067 & 10.579$\pm$0.024 & 10.498$\pm$0.025 & 10.522$\pm$0.025\\
 279	    &  CPD~$-$53 7372  & B8\,III	       & 11.102 & 10.917 & 10.475$\pm$0.023 & 10.398$\pm$0.023 & 10.417$\pm$0.025\\
 285	    &  CPD~$-$53 7383  & B7\,IV 	       & 11.201 & 10.941 & 10.280$\pm$0.026 & 10.200$\pm$0.028 & 10.138$\pm$0.028\\
 286	    &  CPD~$-$53 7384  & B7\,IIIe	       & 10.330 & 10.150 &  9.537$\pm$0.023 &  9.430$\pm$0.023 &  9.404$\pm$0.022\\
 287	    &  CPD~$-$53 7385  & B7\,IV 	       &	&	 & 10.628$\pm$0.030 & 10.544$\pm$0.028 & 10.572$\pm$0.022\\
 288	    &  CPD~$-$53 7388  & B7\,III-IV	       & 10.505 & 10.550 & 10.058$\pm$0.042 &  9.947$\pm$0.058 &  9.973$\pm$0.062\\
 290	    &  CPD~$-$53 7390  & B5\,shell	       & 10.840 & 10.690 & 10.035$\pm$0.021 &  9.868$\pm$0.022 &  9.715$\pm$0.023\\
 291	    &  CPD~$-$53 7392  & B9\,III\,Si	       & 11.501 & 11.261 & 10.637$\pm$0.030 & 10.539$\pm$0.037 & 10.534$\pm$0.035\\
 292	    &  CPD~$-$53 7393  & K0\,Ib-II	       & 11.610 & 10.090 &  7.355$\pm$0.023 &  6.700$\pm$0.033 &  6.489$\pm$0.024\\
 293	    &  CPD~$-$53 7395  & B7\,V  	       &	&	 & 10.977$\pm$0.021 & 10.915$\pm$0.023 & 10.949$\pm$0.022\\
 294	    &  CPD~$-$53 7397  & B8\,IVe	       &	&	 & 10.489$\pm$0.021 & 10.450$\pm$0.023 & 10.431$\pm$0.022\\
 295	    &  CPD~$-$53 7398  & B7\,IV 	       & 11.390 & 11.180 & 10.637$\pm$0.024 & 10.564$\pm$0.023 & 10.586$\pm$0.028\\
297 (V340 Nor)& CPD~$-$53 7400p& G2\,Iab	       & 9.445  & 8.340  &  6.156$\pm$0.020 &  5.731$\pm$0.018 &  5.526$\pm$0.023\\
 298	    &  CPD~$-$53 7400f & A5\,II 	       & 9.500  & 9.000  &  7.773$\pm$0.019 &  7.560$\pm$0.033 &  7.475$\pm$0.018\\
 299	    &  CPD~$-$53 7402  & B7\,III + B8?         & 10.460 & 10.230 &  9.675$\pm$0.021 &  9.570$\pm$0.022 &  9.572$\pm$0.025\\
 303	    &  CPD~$-$53 7416  & G8\,II 	       & 11.520 & 9.970  &  7.391$\pm$0.023 &  6.731$\pm$0.026 &  6.544$\pm$0.020\\
1006	    &  ---	       & A3\,V  	       & 11.836 & 10.076 & 10.051$\pm$0.023 &  9.976$\pm$0.026 &  9.921$\pm$0.026\\
 306	    &  CPD~$-$53 7419  & K2\,II 	       & 11.345 & 10.400 &  6.981$\pm$0.019 &  6.214$\pm$0.018 &  5.993$\pm$0.020\\
 310	    &  CPD~$-$53 7426  & B8\,III	       & 11.076 & 10.776 & 10.022$\pm$0.023 &  9.922$\pm$0.027 &  9.873$\pm$0.025\\
 316	    &  CPD~$-$53 7442  & K2\,Ib + B	       & 9.790  & 8.860  &  6.263$\pm$0.018 &  5.576$\pm$0.029 &  5.313$\pm$0.021\\
 320	    &  CPD~$-$53 7449  & B6\,V  	       &	&	 & 10.727$\pm$0.021 & 10.651$\pm$0.023 & 10.673$\pm$0.023\\
 323	    &  CPD~$-$53 7454  & G8\,II 	       & 11.780 & 10.240 &  7.587$\pm$0.021 &  7.008$\pm$0.031 &  6.817$\pm$0.031\\
 324	    &  CPD~$-$53 7456  & B7\,IV 	       &	&	 & 10.359$\pm$0.023 & 10.286$\pm$0.026 & 10.251$\pm$0.023\\
 325	    &  CPD~$-$53 7458  & B8\,IIIe	       & 10.680 & 10.360 &  9.501$\pm$0.021 &  9.422$\pm$0.023 &  9.350$\pm$0.023\\
 7467	    &  CPD~$-$53 7467  & B8\,IIIp	       &	&	 & 10.261$\pm$0.023 & 10.286$\pm$0.027 & 10.215$\pm$0.023\\
 QZ Nor     &  CPD~$-$54 7159  & G1\,Iab	       &	&	 &  7.004$\pm$0.027 &  6.700$\pm$0.051 &  6.544$\pm$0.021\\
 HD\;145324 &  CPD~$-$54 7226  & A5\,Ib-II	       &	&	 &  6.268$\pm$0.019 &  6.135$\pm$0.033 &  5.995$\pm$0.021\\
 329	    &  CPD~$-$54 7244  & K0\,Ib 	       &	&	 &  7.069$\pm$0.023 &  6.394$\pm$0.023 &  6.153$\pm$0.018\\
1796	    &  ---	       & B6\,Ve 	       & 11.231 & 11.011 & 10.324$\pm$0.035 & 10.202$\pm$0.037 & 10.195$\pm$0.033\\
\hline
\end{tabular}
\end{center}
\begin{list}{}{}
\item[] $^{*}$ For this star, there is a typographic error in the reference paper. It has been corrected in this work by using directly the values from \citetalias{Th62}.
  \end{list}
\end{table*}

\end{landscape}

\begin{table*}
\caption{Number of red giant stars (and yellow supergiants in brackets) in open clusters, containing no Cepheids, with ages between 50 and 150 Ma \citep[from][]{Me08}. The age of every cluster has been
 taken from the WEBDA database.\label{RG_OC}}
\begin{center}
\begin{tabular}{lcc|c|lcc}   %longtable es para poner primero landscape
\hline\hline
Cluster & $\log\,\tau$ & $N_{\textrm{RG}}$ & & Cluster & $\log\,\tau$ & $N_{\textrm{RG}}$\\
\hline
NGC 0225 & 8.11 & 0 &  &  NGC 5617 & 7.92 & 3 \\
NGC 0436 & 7.93 & 1 (1) &  &  NGC 5749 & 7.73 & 0 \\
NGC 1647 & 8.16 & 2 &  &  NGC 6124 & 8.15 & 8 \\
NGC 1778 & 8.16 & 1 &  &  NGC 6192 & 8.13 & 5 \\
NGC 2168 & 7.98 & 2 (1) &  &  NGC 6405 & 7.97 & 1 \\
NGC 2186 & 7.74 & 1 &  &  NGC 6416 & 8.09 & 0 \\
NGC 2232 & 7.73 & 0 &  &  NGC 6520 & 7.72 & 2 (1) \\
NGC 2323 & 8.10 & 1 &  &  NGC 6546 & 7.89 & 1 \\
NGC 2345 & 7.85 & 5 &  &  NGC 6694 & 7.93 & 2 \\
NGC 2354 & 8.13 & 13&  &  NGC 6709 & 8.18 & 2 \\
NGC 2422 & 7.86 & 0 &  &  NGC 6755 & 7.72 & 3 \\
NGC 2516 & 8.05 & 4 &  &  NGC 7031 & 8.14 & 1 \\
NGC 2546 & 7.87 & 2 &  &  NGC 7063 & 7.98 & 0 \\
NGC 2669 & 7.93 & 1 &  &  NGC 7654 & 7.76 & 0 (1) \\
NGC 2972 & 7.97 & 3 &  &  Collinder 258 & 8.03 & 1 \\
NGC 3033 & 7.85 & 1 &  &  IC 2488  & 8.11 & 3 \\
NGC 3114 & 8.09 & 6 (1) &  &  Melotte 20  & 7.85 & 0 (1) \\
NGC 3228 & 7.93 & 1 &  &  Melotte 101 & 7.89 & 1 \\
NGC 3247 & 8.08 & 1 &  &  Trumpler 2 & 8.17 & 1 \\
NGC 4609 & 7.89 & 1 &  &  Trumpler 3  & 7.83 & 2 \\
NGC 5138 & 7.99 & 4 &  &  Trumpler 9  & 8.00 & 0 (1) \\ 
NGC 5168 & 8.00 & 0 &  &  &  & \\
\hline
\end{tabular}
\end{center}
\end{table*}



%%%%%%%%%%%%%%%%%%%%%%%%%%%%%%%%%%%%%%%%%%%%%%%%%%

% Don't change these lines
\bsp	% typesetting comment
\label{lastpage}
\end{document}